\documentclass[cits]{PoS}
\title{Critical examination of RHIC paradigms---mostly high $\mathbf{p_T}$}

\ShortTitle{Critical examinations of RHIC paradigms---mostly high $p_T$}

\author{\speaker{Michael J.  Tannenbaum}%
         \thanks{Supported by the U.S. Department of Energy, Contract No. DE-AC02-98CH1-886.}\\
         Physics Dept., 510c, Brookhaven National Laboratory, Upton, NY 11973-5000,USA\\
        E-mail: \email{mjt@bnl.gov}}

\abstract{A critical examination of RHIC paradigms is presented. Topics include: search for a critical point with a low energy scan; the lack of understanding of radiative processes in a medium in QCD compared in detail to examples from QED; the reason why some physicists started to measure particles at large $p_T$ in the 1960's; a review of the discovery of hard-scattering in p-p collisions in the 1970's via single-inclusive and two-particle correlations and application of these techniques at RHIC. Several paradigms in both soft and hard physics which are popular at RHIC are discussed and challenged. }

\FullConference{Workshop on Critical Examination of RHIC Paradigms\\
		April 14-17, 2010\\
		Austin, Texas, U.S.A.}

\def\lsim{\raise0.3ex\hbox{$<$\kern-0.75em\raise-1.1ex\hbox{$\sim$}}}
\def\gsim{\raise0.3ex\hbox{$>$\kern-0.75em\raise-1.1ex\hbox{$\sim$}}}

\def\mean#1{\left<#1\right>}
\def\Journal#1#2#3#4{ {\it{#1}} {\bf #2}, #3 (#4)}
\def\IJMPA{{Int. J. Mod. Phys.}~{\rm A}}

\def\EPJC{{Eur. Phys. J.}~{\rm C}}

\def\JPG{{J. Phys.}~{\rm G}}
\def\JPCS{{J. Phys: Conf. Series\ }}

\def\NPA{{Nucl. Phys.}~{\rm A}}
\def\NPB{{Nucl. Phys.}~{\rm B}}
\def\PLB{{Phys. Lett.}~{\rm B}}

\def\PR{Phys. Rev.\ } 
\def\PRL{Phys. Rev. Lett.\ }
\def\PRD{{Phys. Rev.}~{\rm D}}
\def\PRC{{Phys. Rev.}~{\rm C}}
\def\PPNP{{Prog. Part. Nucl. Phys.\ }}
\def\ZPC{{Z. Phys.}~{\rm C}}
\def\ARNPS{{Ann. Rev. Nucl. Part. Sci.\ }} 
\def\RMP{Rev. Mod. Phys.\ }

\def\QGP{\Red Q\Blue G\Green P\Black}
\def\QCD{\Red Q\Green C\Blue D\Black}
\begin{document}

\section{The Quark Gluon Plasma (\QGP)}
\begin{figure}[h]
\begin{center}
a)\includegraphics[width=0.47\linewidth]{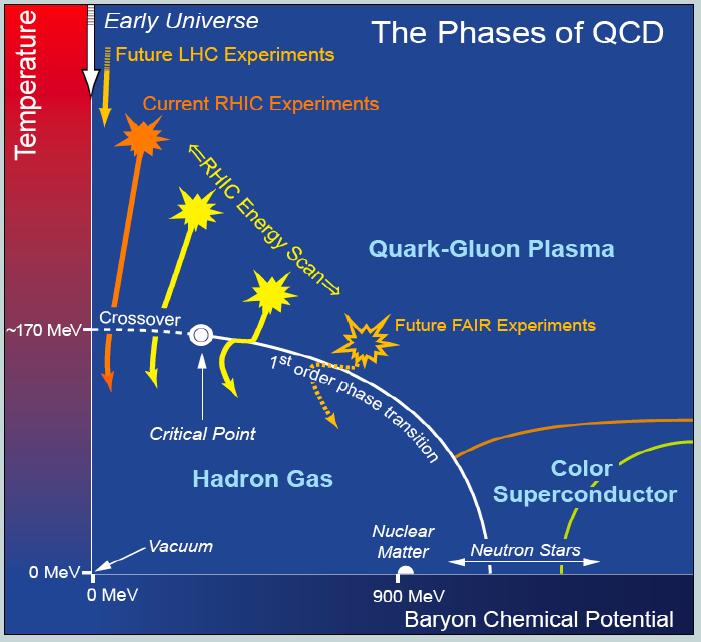}
b)\includegraphics[width=0.47\linewidth]{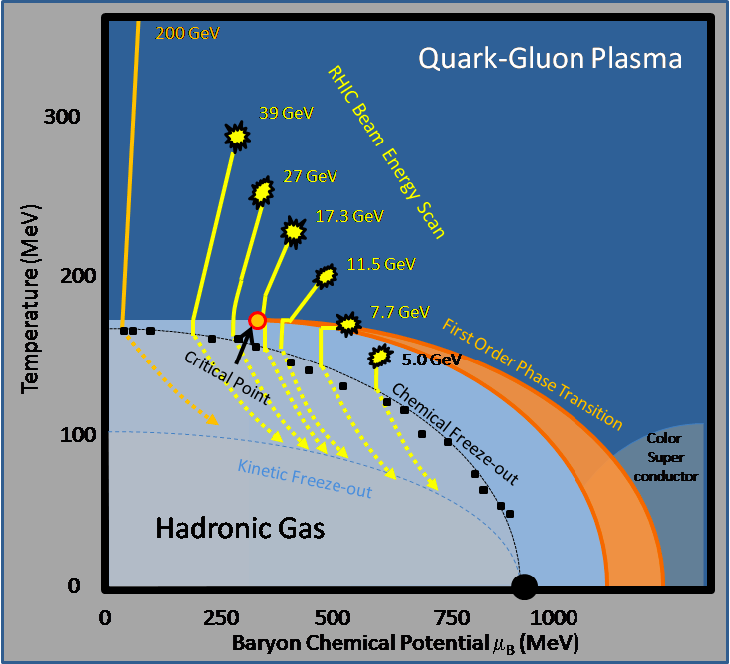}
\end{center}
\caption[]{a) A schematic representation of the QCD Phase Diagram~\cite{LRP07}. b) The location of the
critical point, the separation between the 1st-order transition and chemical freeze-out, and
the focusing of the event trajectories towards the critical point, are not based on specific
quantitative predictions, but are all chosen to illustrate plausible possibilities~\cite{STAR09}}
\label{fig:phase_boundary}
\end{figure}

Let me start out with the famous schematic plots of the phase diagram of nuclear matter~Fig.~\ref{fig:phase_boundary}~\cite{LRP07,STAR09}. 
The presumed trajectories of the evolution of the medium for collisions at RHIC and LHC c.m. energies are shown in Fig.~\ref{fig:phase_boundary}a, where the axes are the temperature $T$ vs. the baryon chemical potential $\mu_B$. The temperature for the transition from the Quark Gluon Plasma (\QGP) to a hadron gas is taken as 170 MeV for $\mu_B=0$ and the phase boundary is predicted to be a smooth crossover down to a critical point below which the phase boundary becomes a first order phase transition. 
Also shown are idealized trajectories for the RHIC c.m. energy scan and future experiments at FAIR which are being performed in order to find the QCD critical point.  

\begin{figure}[h]
\begin{center}
a)\includegraphics[width=0.47\linewidth]{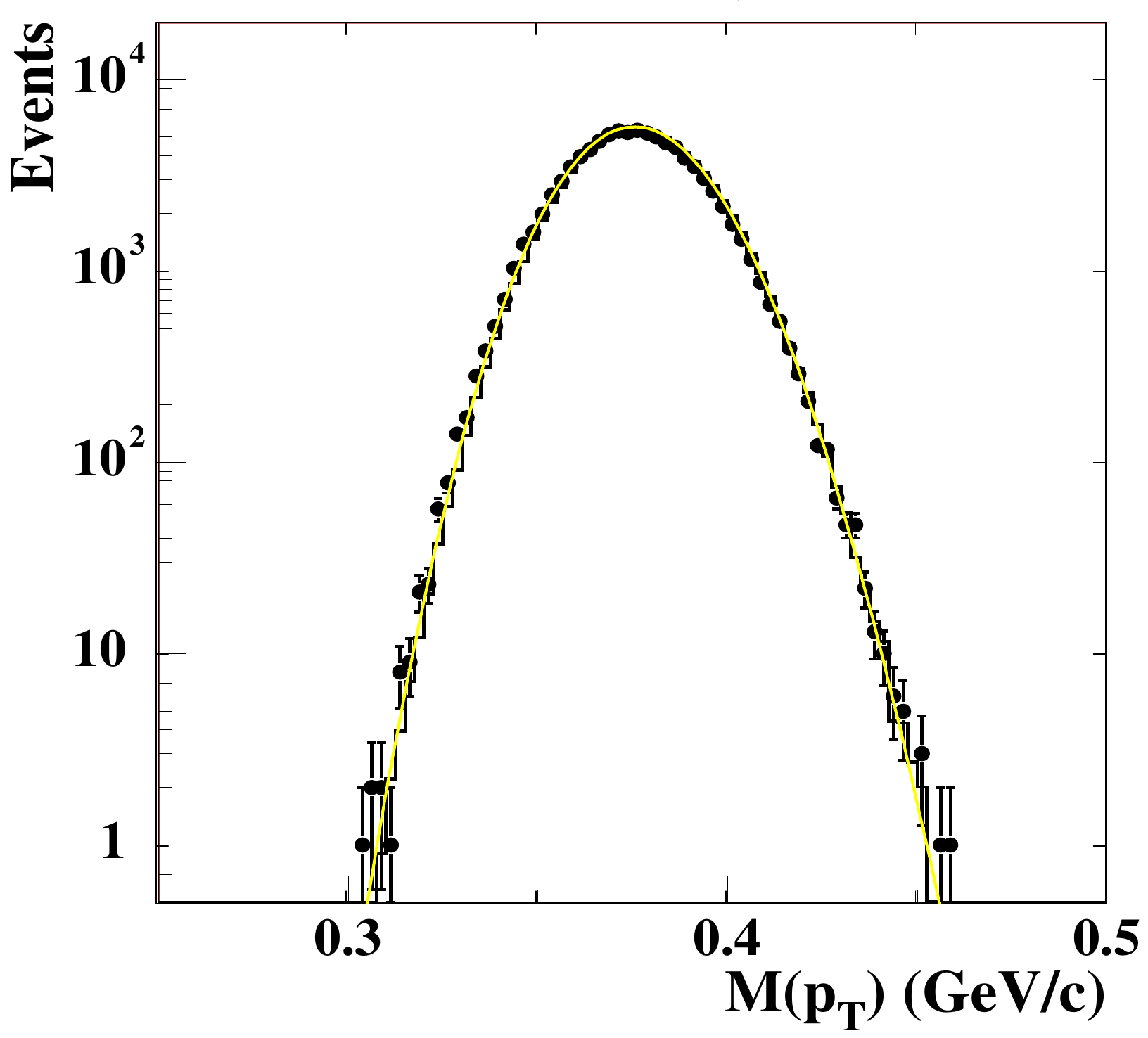}
b)\includegraphics[width=0.47\linewidth]{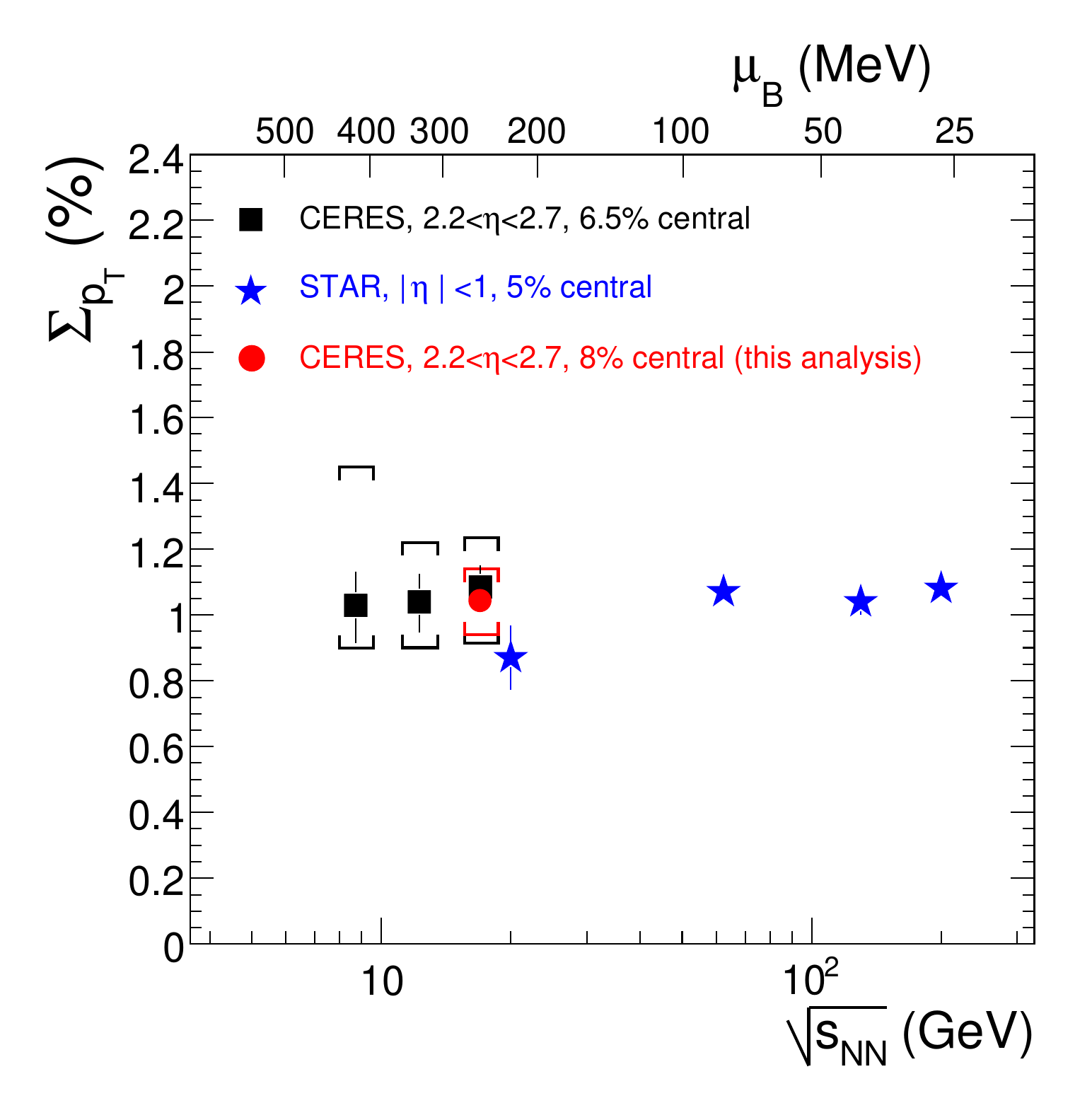}
\end{center}
\caption[]{a) Distribution of event-by-event average $p_T$, $M_{p_T}=\overline{p_T}={1\over n} \sum_{i=1}^n p_{T_i}$, where $n$ is the number of particles on a given event, in 158 GeV/nucleon Pb+Pb collisions (data points); histogram is from mixed events~\cite{NA49}. Line is calculated random convolution of inclusive $p_T$ spectrum~\cite{NA49MJT}.  b) rms/mean of $M_{p_T}$ distribution with random effect subtracted, $\Sigma_{p_T}=\sqrt{{ \sigma^2_{M_{p_T}} \over {\mean{p_T}^2} } - \left ( { \sigma^2_{M_{p_T}} \over {\mean{p_T}^2} }\right )_{\rm random}}$ as a function of $\sqrt{s_{NN}}$~\cite{CERES08}. }
\label{fig:NA49MJT}
\end{figure}

One paradigm that I don't believe is that one can find the critical point by an energy scan at RHIC. Luciano Moretto at WWND2010 said, ``critical fluctuations depend on a first order phase transition below the critical point whose large fluctuations drive the critical fluctuations'', leading me to ask again whether all the interesting physics just outside the E802 aperture?  What I mean by this is that since no fluctuations, other than from known correlations such as Bose-Einstein interference, were observed at the AGS fixed target Au+Au program at c.m. energy $\sqrt{s_{NN}}=4.8$ GeV (see below), or from the CERN Pb+Pb fixed target program at c.m. energy $\sqrt{s_{NN}}=17.2$ GeV (Fig.~\ref{fig:NA49MJT}a)~\cite{NA49,NA49MJT}  there is no first order phase transition to drive the critical fluctuations. 
Specifically, from AGS to RHIC energies, the non-random fluctuations of $M_{p_T}$, the event-by-event average $p_T$, are about 1\% of the $\mean{p_T}$, averaged over all events, independent of $\sqrt{s_{NN}}$ (Fig.~\ref{fig:NA49MJT}b)~\cite{Koch08,CERES08} and show no evidence of fluctuations from a first-order phase transition. The only escape is if the QGP can not be reached in A+A collisions below the critical point, i.e. the endpoints of the trajectories in Fig.~\ref{fig:phase_boundary}b for $\sqrt{s_{NN}}=17.2$ GeV and below lie between the Chemical Freeze-out line and the phase-boundary. In principle, this is possible, but in my opinion it is very unlikely given the constancy of $\Sigma_{p_T}$ as a function of $\sqrt{s_{NN}}$ in Fig.~\ref{fig:NA49MJT}b.  
\section{Given the Quark Gluon Plasma (\QGP), what is my interest?}
\begin{figure}[h]
\begin{center}
\includegraphics[width=0.62\linewidth]{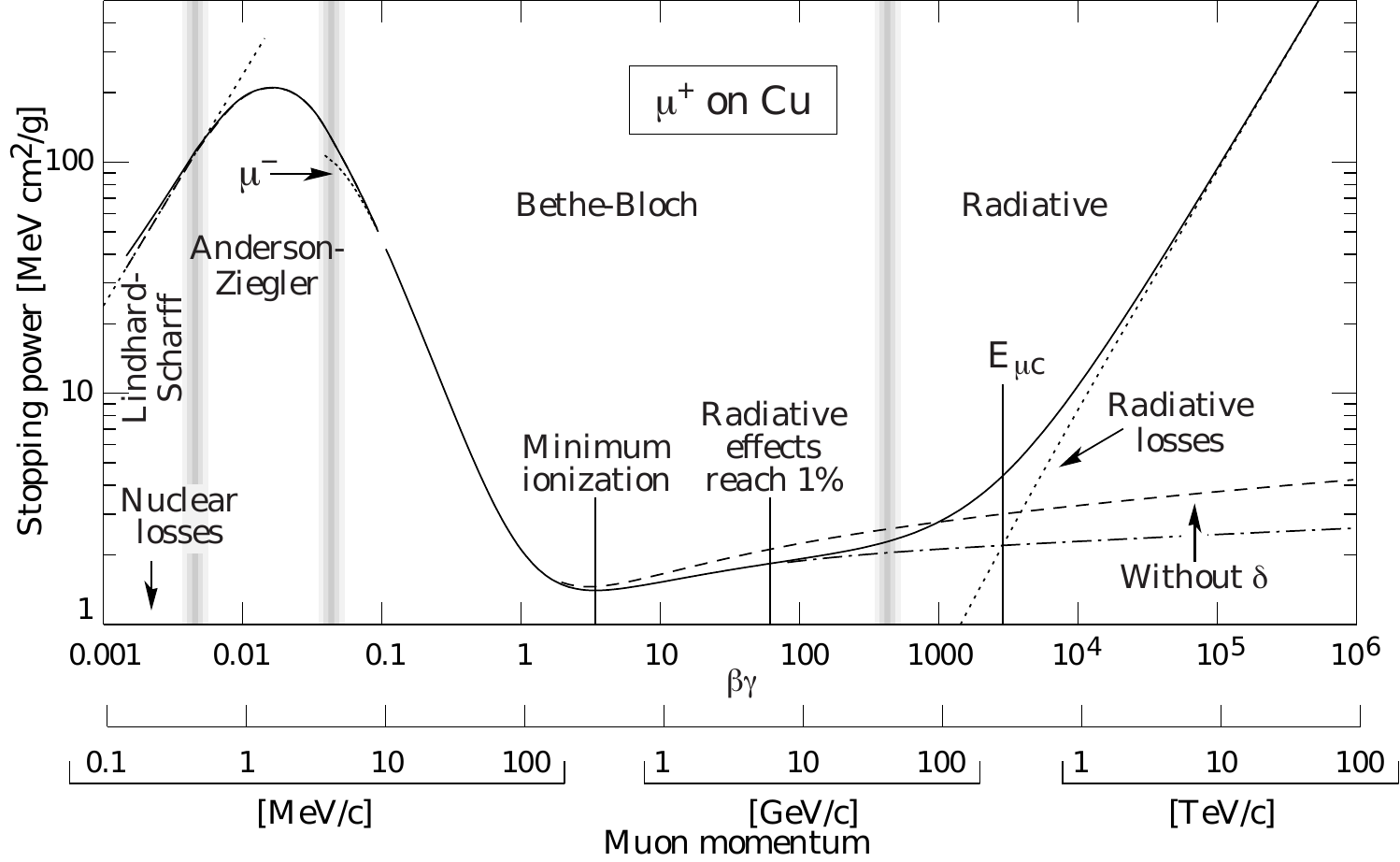}
\end{center}\vspace*{-0.25in}
\caption[]{$dE/dx$ of a $\mu^+$ in Copper as a function of muon momentum~\cite{PDG}.}
\label{fig:muCu}
\end{figure}

   The {\QGP} is the only place in the universe where we can in principle and in practice understand {\QCD} for color-charged quarks in a color-charged medium. How long will it take before we understand a quark in a {\QCD} medium as well as we understand the passage of a muon through Copper in QED (Fig.~\ref{fig:muCu})? 
\subsection{QED Bremsstrahlung in matter and crystals}
\begin{figure}[h]
\begin{center}
\begin{tabular}{cc}
\begin{tabular}[b]{c}
a)\hspace*{-1pc}\includegraphics[width=0.45\linewidth]{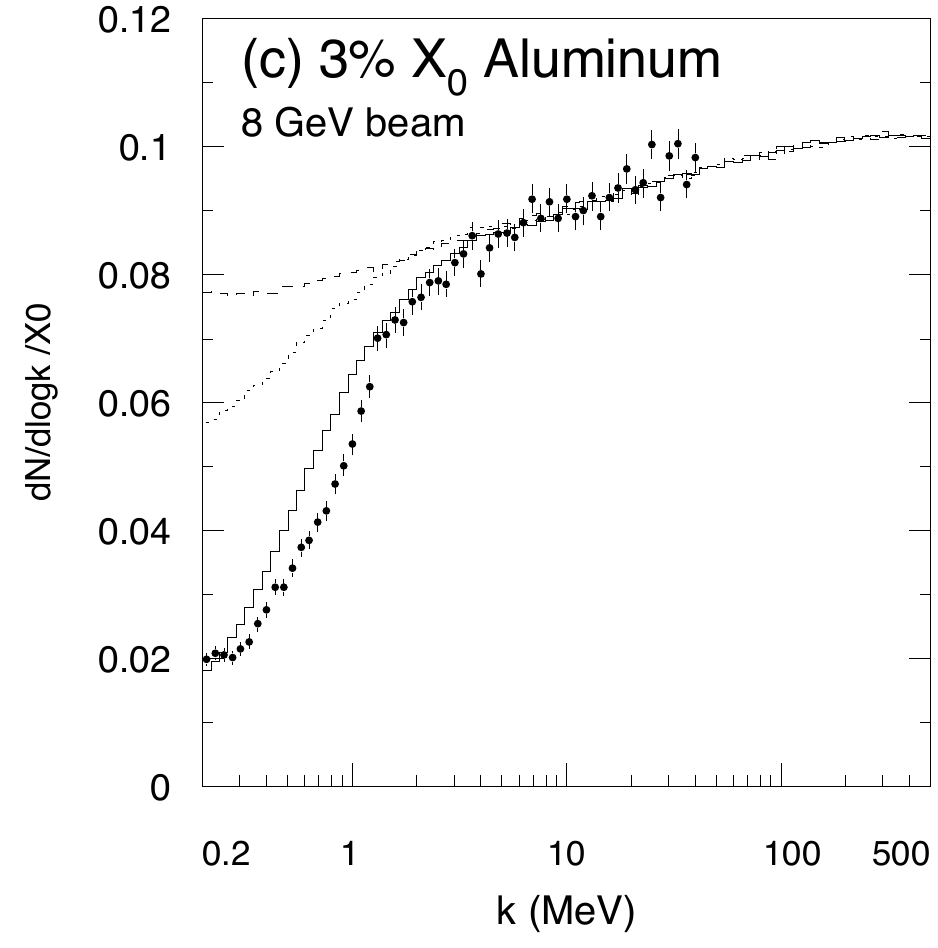}\hspace*{1pc}\cr
\end{tabular}
b)\begin{tabular}[b]{c}
\hspace*{-2pc}\includegraphics[angle=-0.9,width=0.50\linewidth]{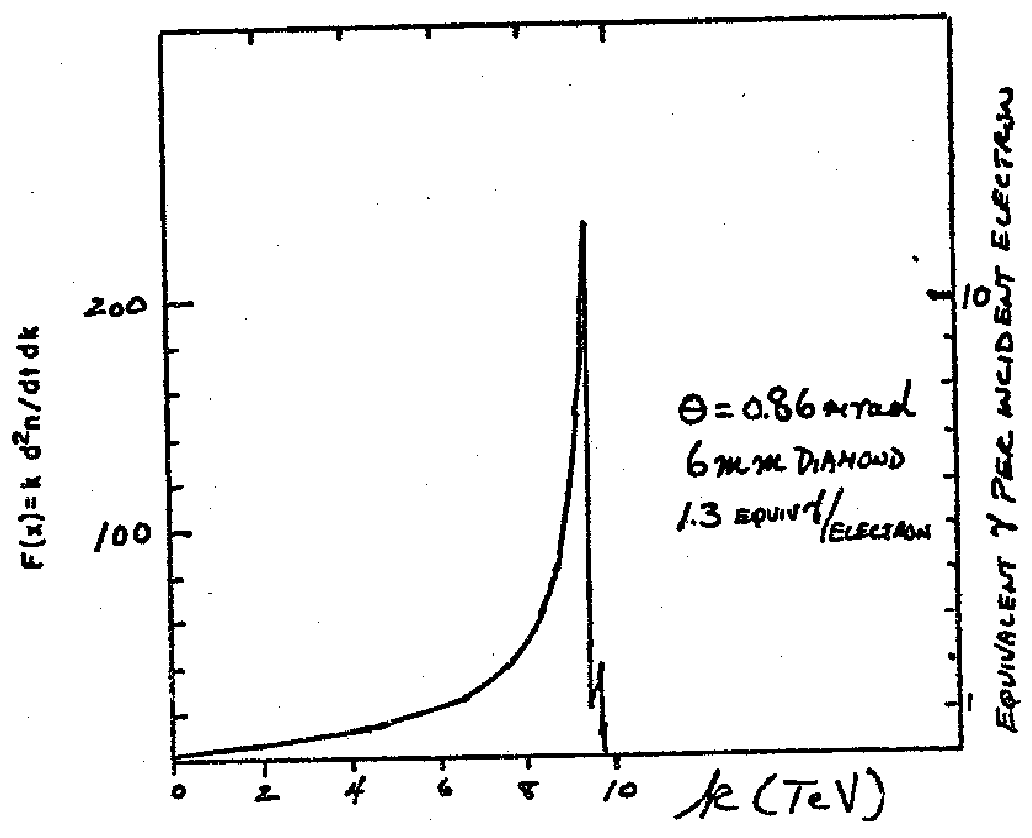}\cr
\begin{picture}(50,15)
\end{picture}

\end{tabular}
\end{tabular}

\end{center}
\vspace*{-1pc}
\caption[]
{a) SLAC measurement of equivalent photon spectrum $kdN/dk$ in 3\% $X_o$ Al target~\cite{LPMSLAC} illustrating LPM and dielectric suppression at low $x=k/E$.   b) Calculation of coherent bremsstrahlung in a diamond crystal with a peak at 9.48 TeV for a proposed 10 TeV electron beam at the SSC~\cite{MJTSSC}.
\label{fig:effects} }
\end{figure}

Even for such a seemingly simple reaction as muon bremsstrahlung, there is no simple formula~\cite{seeAtlas}. For bremsstrahlung of an electron of energy $E$ in a solid, apart from the standard Bethe-Heitler radiation ($k dN/dk=d/X_o$, where $k$=energy of radiated photon and $d/X_o$=the thickness in radiation lengths, $X_o$), there are also very interesting effects~\cite{LPMSLAC}: the Landau-Pomeranchuk-Migdal (LPM) effect, which decreases the radiation at low $x=k/E$; transition radiation at low $x$; and a medium effect on the forward outgoing photon from Compton scattering off the atomic electrons which if inside the coherence length kills the $dN/dk$ divergence. The coherence length is set by $t_{\rm min}$. For bremsstrahlung of a photon with energy $k$ by a particle with mass $m$, energy $E$:
\begin{equation}
\frac{1}{L_{coh}}=q_L=\sqrt{t_{\rm min}}=\frac {m^2 k}{2 E (E-k)}
\label{eq:coh}
\end{equation}
where $q=\sqrt{|t|}$ is the 4-momentum transfer to the target. 
Figure~\ref{fig:effects}a~\cite{LPMSLAC} shows the bremsstrahlung spectrum of equivalent photons, $k dN/dk$, from 8 GeV electrons incident on a 3\% radiation length ($X_o$) Aluminum target. The pure Bethe-Heitler spectrum, $k dN/dk=d/X_o$ is shown by the dashed line. Inclusion of the 
LPM suppression gives the dotted line. Further inclusion of the dielectric suppression gives the histogram, which agrees with the measurement. For crystals, there is also coherent bremsstrahlung when the momentum transfer vector $\vec{q}$ equals a characteristic momentum of the reciprocal lattice (Bragg condition) which greatly increases the radiation~\cite{DiambriniPalazzi} as shown in Fig.~\ref{fig:effects}b~\cite{MJTSSC}. 

   I hope that the discussion in this section has illustrated the point that the understanding of $dE/dx$ for quarks and gluons in a {\QGP}, which is much more complicated than for leptons and photons in solids in QED, is in its infancy and has barely scratched the surface.
   \section{Why were some people studying ``high $\mathbf{p_T}$'' physics in the 1960's}
     The quick answer is that they were looking for a `left handed' intermediate boson $W^\pm$, the proposed carrier of the weak interaction~\cite{LYW60}.
    
\begin{figure}[!b]
\hspace*{0.05\linewidth}\begin{minipage}{0.5\linewidth}
\includegraphics[width=\linewidth]{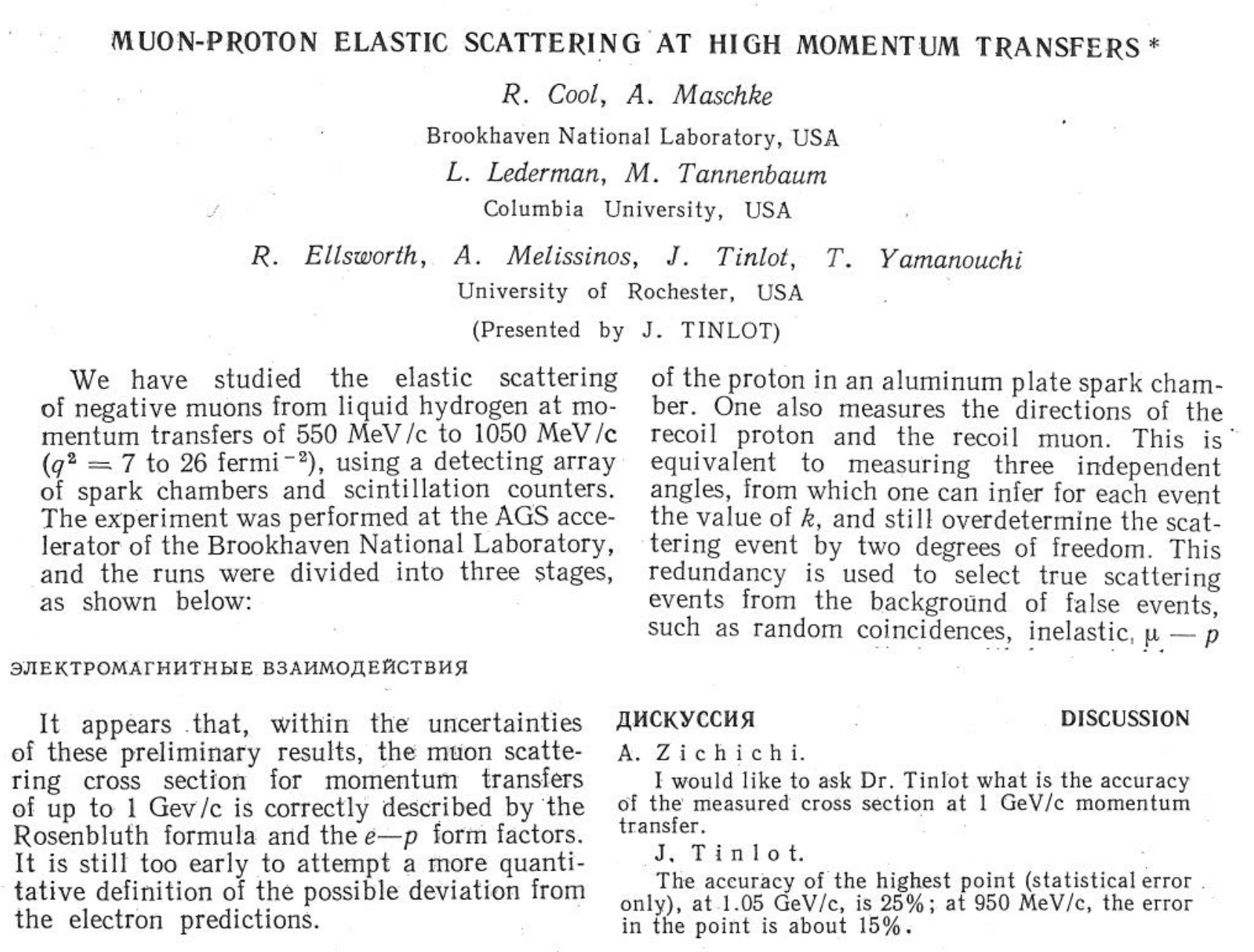}
\includegraphics[width=\linewidth]{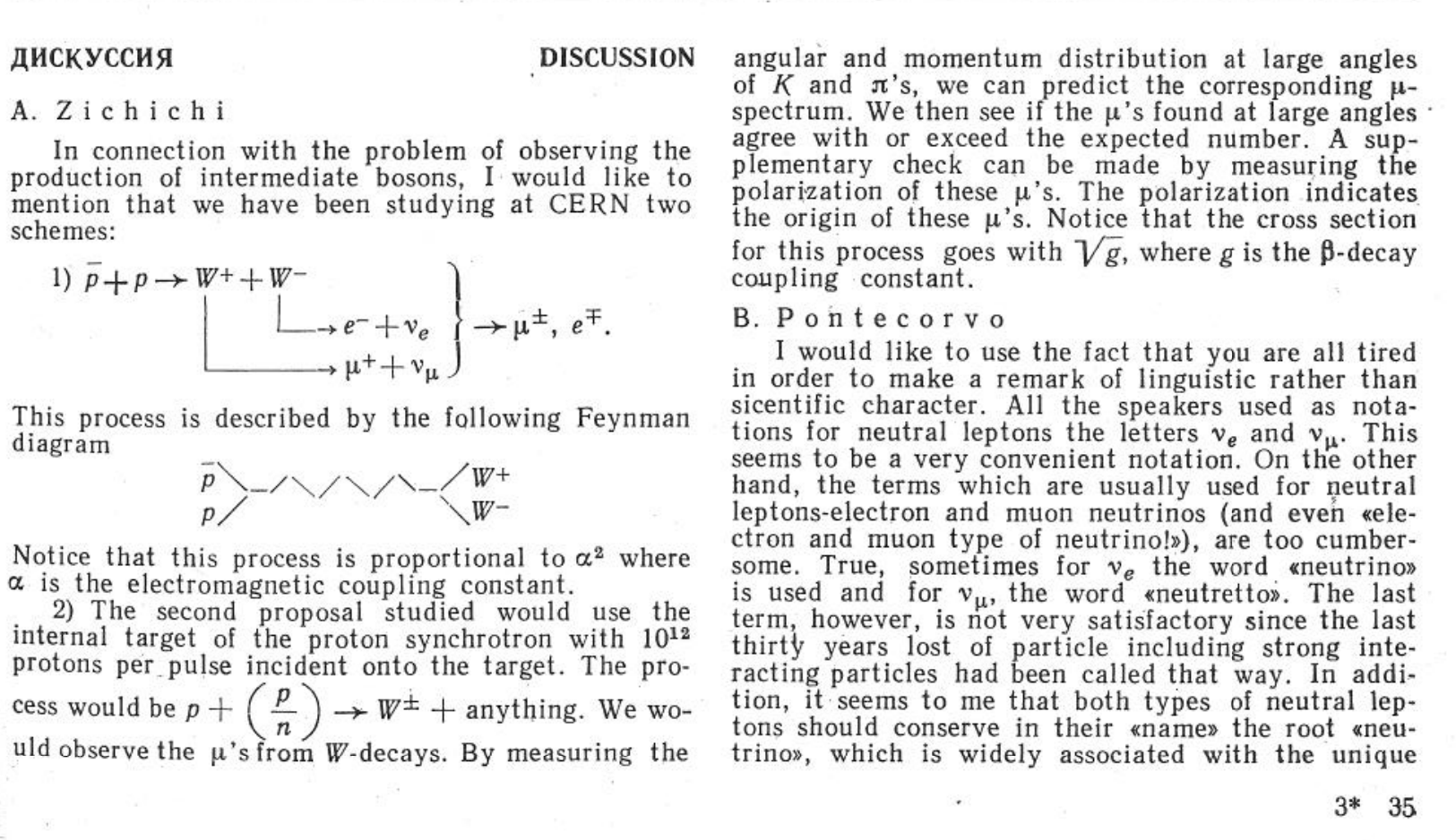}
\end{minipage}\hspace{1pc}%
\begin{minipage}{0.4\linewidth}
\includegraphics[width=\linewidth]{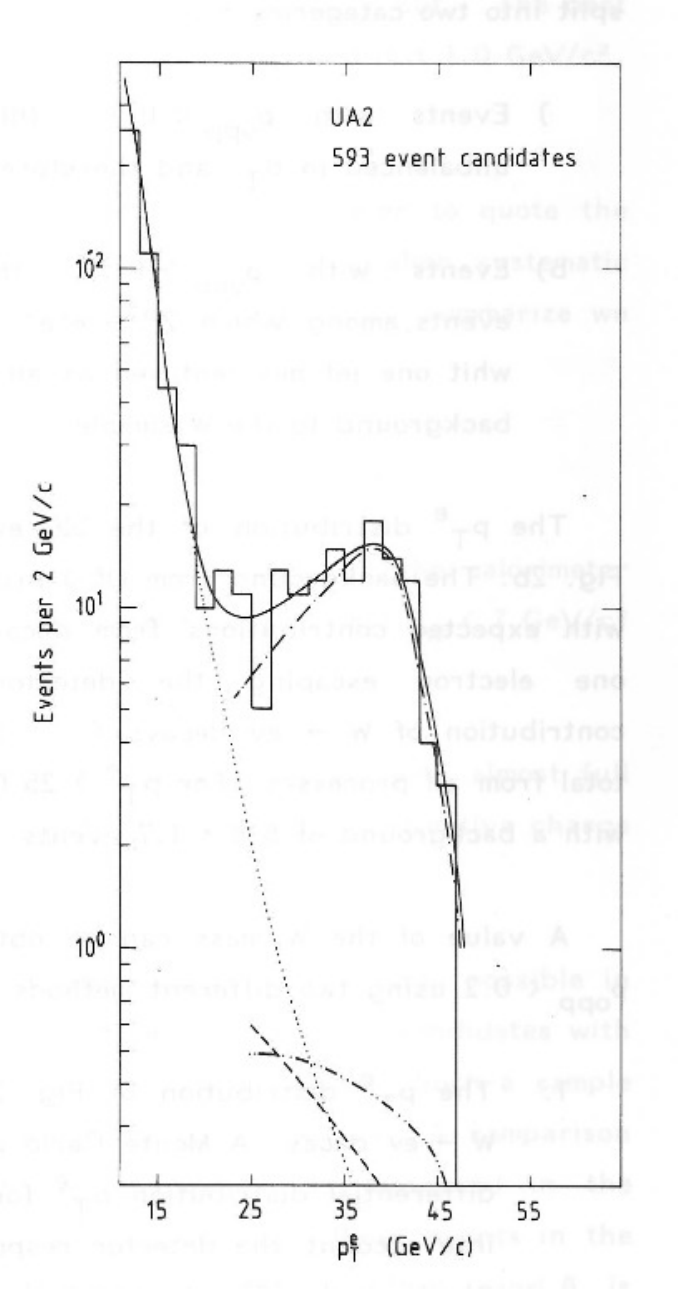}
\end{minipage} \vspace*{-1.0pc}
\caption[]{a)(left)Zichichi ICHEP 1964~\cite{MuPDubna64,NinoWDubna64} and b)(right) $W\rightarrow e +\nu$ from UA2~\cite{UA2WZ86} 1982-1984 runs, about 1/4 of the final sample~\cite{UA2WZ92}\label{fig:ZichichiW}}
\end{figure}

\begin{figure}[h]
\begin{center}
a)\includegraphics[width=0.57\linewidth]{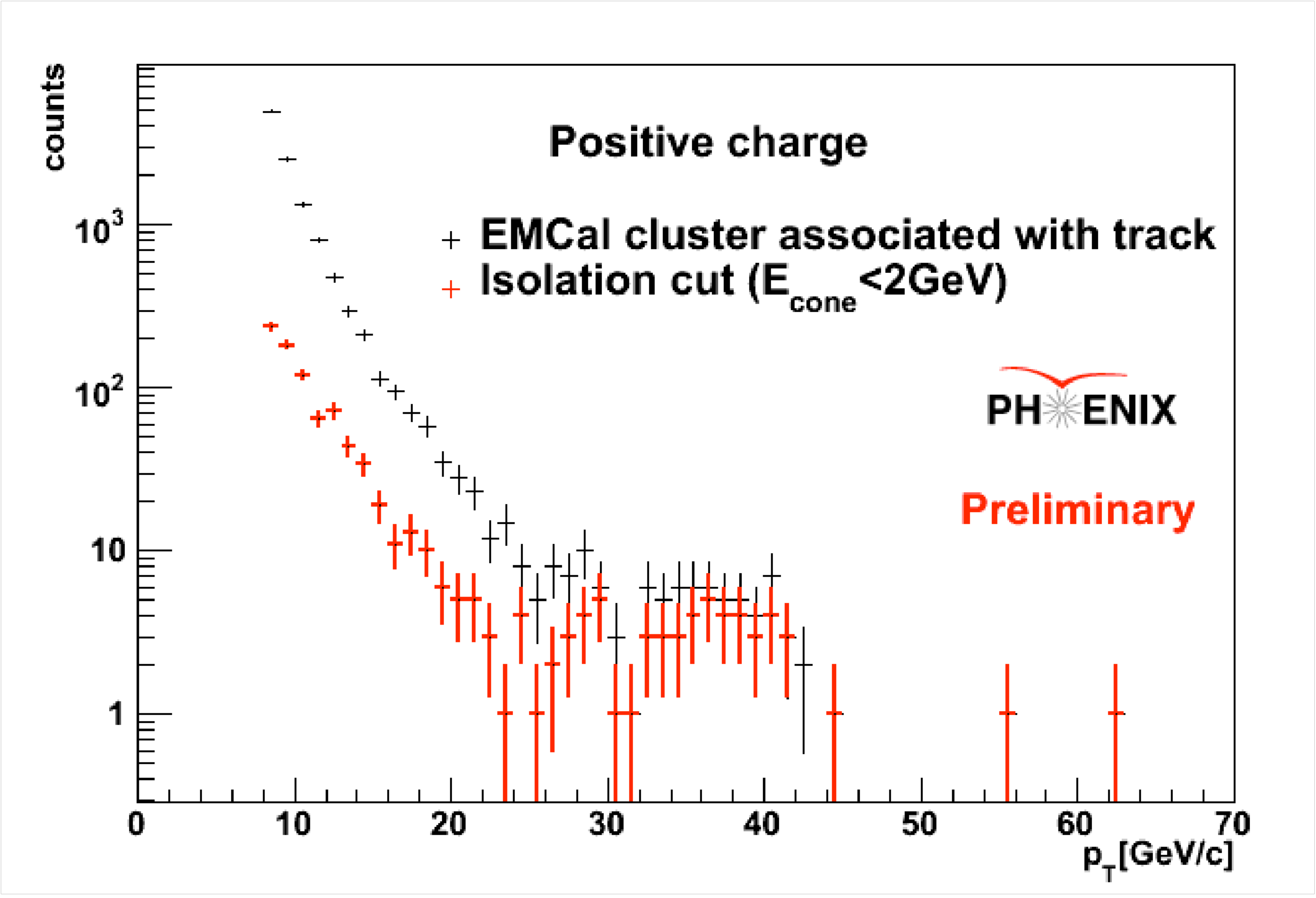}
b)\includegraphics[width=0.38\linewidth]{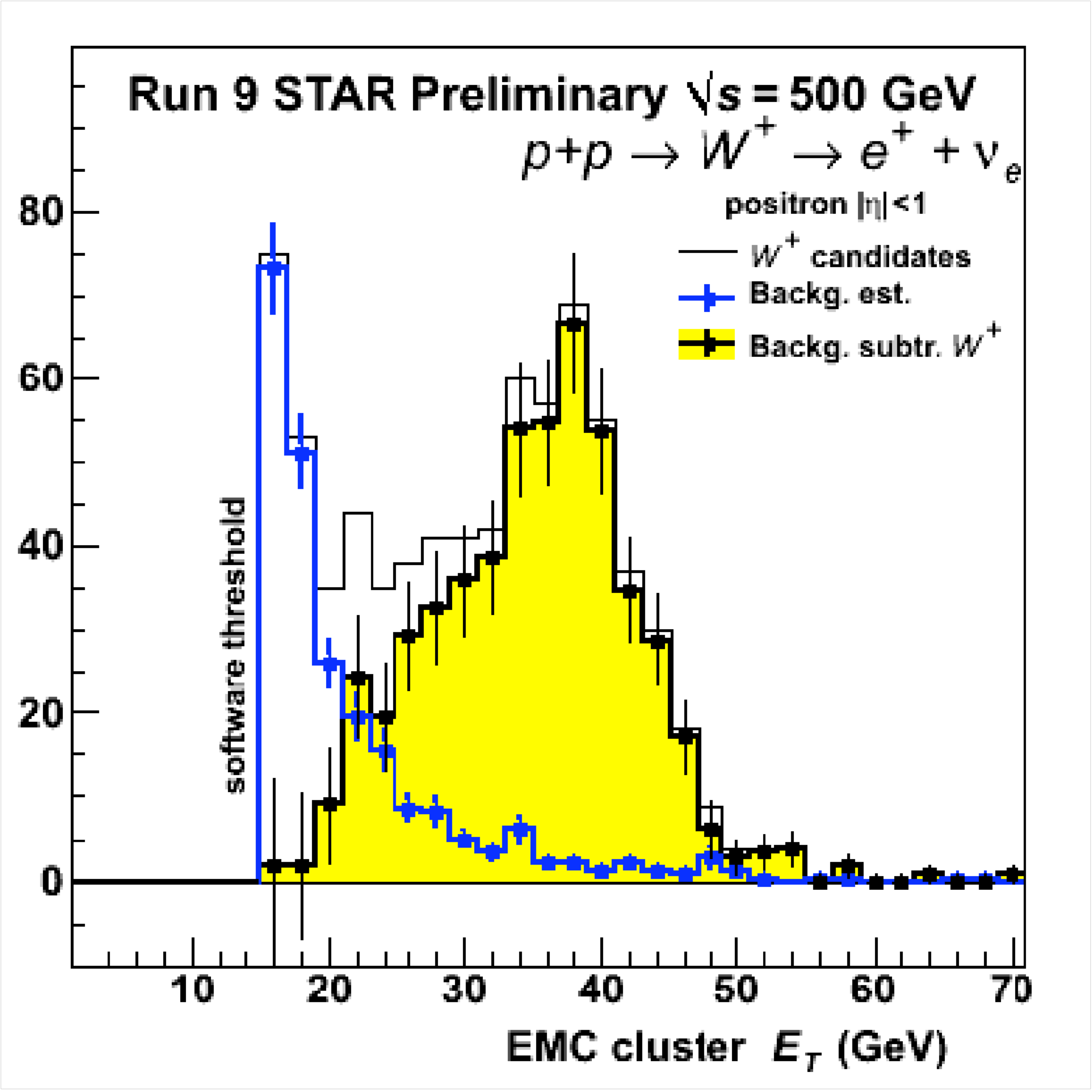}
\end{center}
\caption[]{PHENIX (a) and STAR (b) preliminary results~\cite{WAPSApr2010} for  $W^+\rightarrow e^+ + X$ from a short run at RHIC of polarized p-p collisions at $\sqrt{s}=500$ GeV in 2009. The parity violation in real $W^+$ production has been measured for the first time with this data to $\sim 4.0\sigma$ significance, combined. }
\label{fig:WRHIC}
\end{figure}

    The first opportunity to study weak interactions at high energy was provided by the development of neutrino beams at the new accelerators in the early 1960's, the CERN-SPS and the BNL-AGS~\cite{MS60} where the $\mu$-neutrino was discovered~\cite{Danbyetal62}. 
However, it was soon recognized that the intermediate (weak) boson, $W^{\pm}$ might be more favorably produced in p-p collisions~\cite{LYW60PR}. The canonical method for discovering the $W$-boson in p-p collisions was described by Nino Zichichi in a comment at the 1964 ICHEP, which deserves a verbatim quote because it was exactly how the $W$ was discovered at CERN 19 years later~\cite{Wdiscovery1,Wdiscovery2}: ``We would observe the $\mu$'s from W-decays. By measuring the angular and momentum distribution at large angles of K and $\pi$'s, we can predict the corresponding $\mu$-spectrum. We then see if the $\mu$'s found at large angles agree with or exceed the expected numbers.'' I became aware of this ``Zichichi signature'' in my graduate student days when I checked the proceedings of the 12th ICHEP in Dubna, Russia in 1964 to see how my thesis results were reported~\cite{MuPDubna64} and I found several interesting questions and comments by an ``A. Zichichi'' printed in the proceedings (Fig.~\ref{fig:ZichichiW}). 
The $W$ is still being measured by this method at RHIC (Fig.~\ref{fig:WRHIC})~\cite{WAPSApr2010} except now we use polarized p-p collisions and can measure the parity violation in $W$ production which will enable us to measure flavor identified spin structure functions of $\bar{u}$ and $\bar{d}$ quarks~\cite{spin95}. Now back to the mid 1960's
\subsection{The absence of high $\mathbf{p_T}$ single leptons leads to lepton-pair measurements.}
	Proton-beam-dump experiments at the ANL-ZGS~\cite{Lamb65} and BNL-AGS~\cite{Burns65,Wanderer69} looking for ``large angle'' muons didn't find any. Then the big question became, ``How do you know how many $W$'s should have been produced?''. Chilton Saperstein and Shrauner~\cite{Chilton66} emphasized the need to know the timelike form factor of the proton in order to calculate the $W$ production rate; and Yamaguchi~\cite{Yama66} proposed that the timelike form factor could be found by measuring the number of lepton pairs ($e^+ e^-$ or $\mu^+ \mu^-$), ``massive virtual photons'', of the same invariant mass as the $W$ ({but} also noted that the individual leptons from these electromagnetically produced pairs might mask the leptons from the $W$). This set off a spate of single and di-lepton experiments, notably the discovery by Lederman {\it et al.} of ``Drell-Yan'' pair production at the BNL-AGS~\cite{DYdiscovery,DYexplanation} followed by the experiments E70 and CCR at the new FNAL-Tevatron and CERN-ISR machines, respectively.  
	
\begin{figure}[h]
\begin{center}
a)\includegraphics[width=0.30\linewidth]{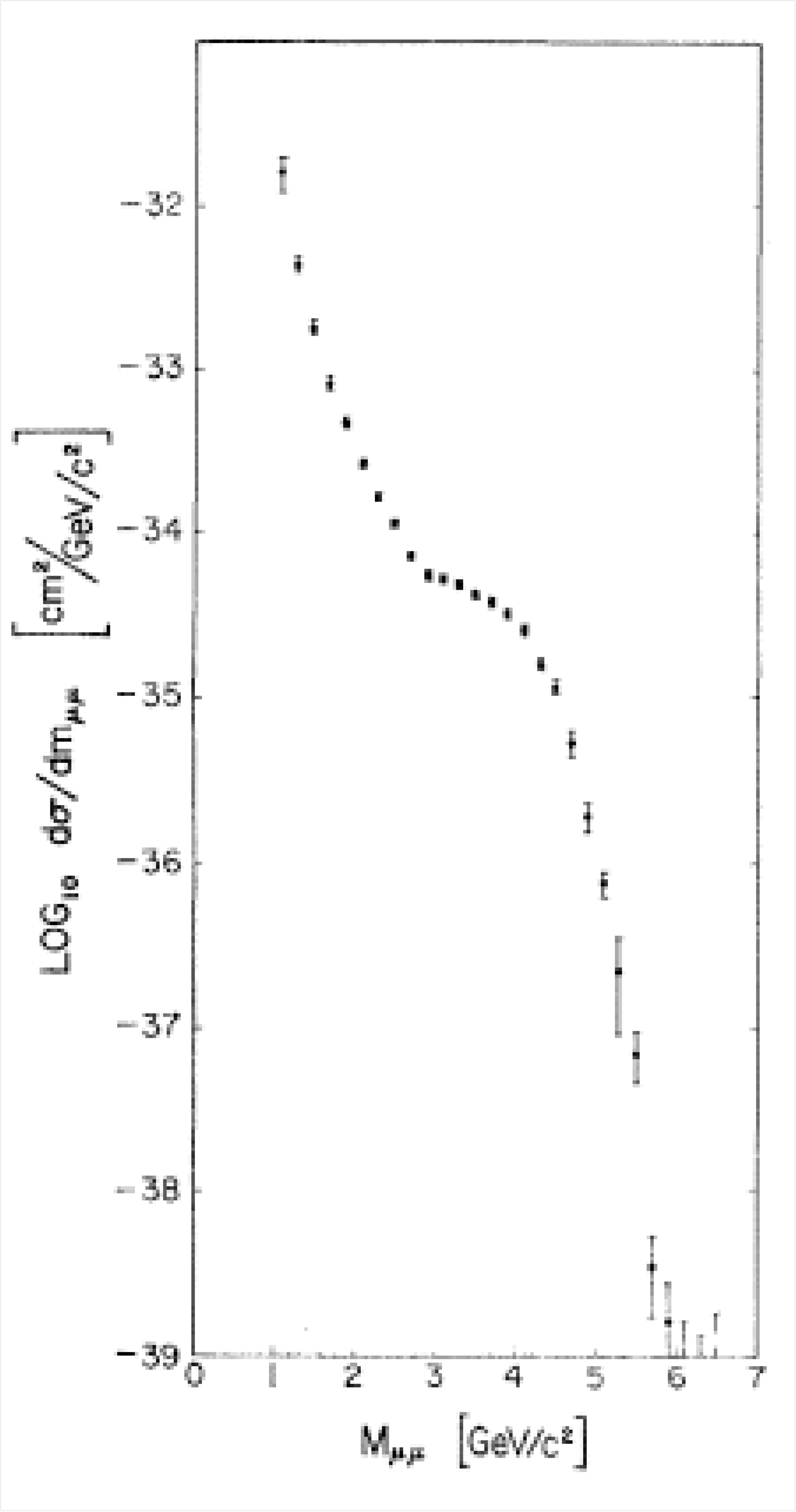}\hspace*{2pc}
b)\includegraphics[width=0.38\linewidth]{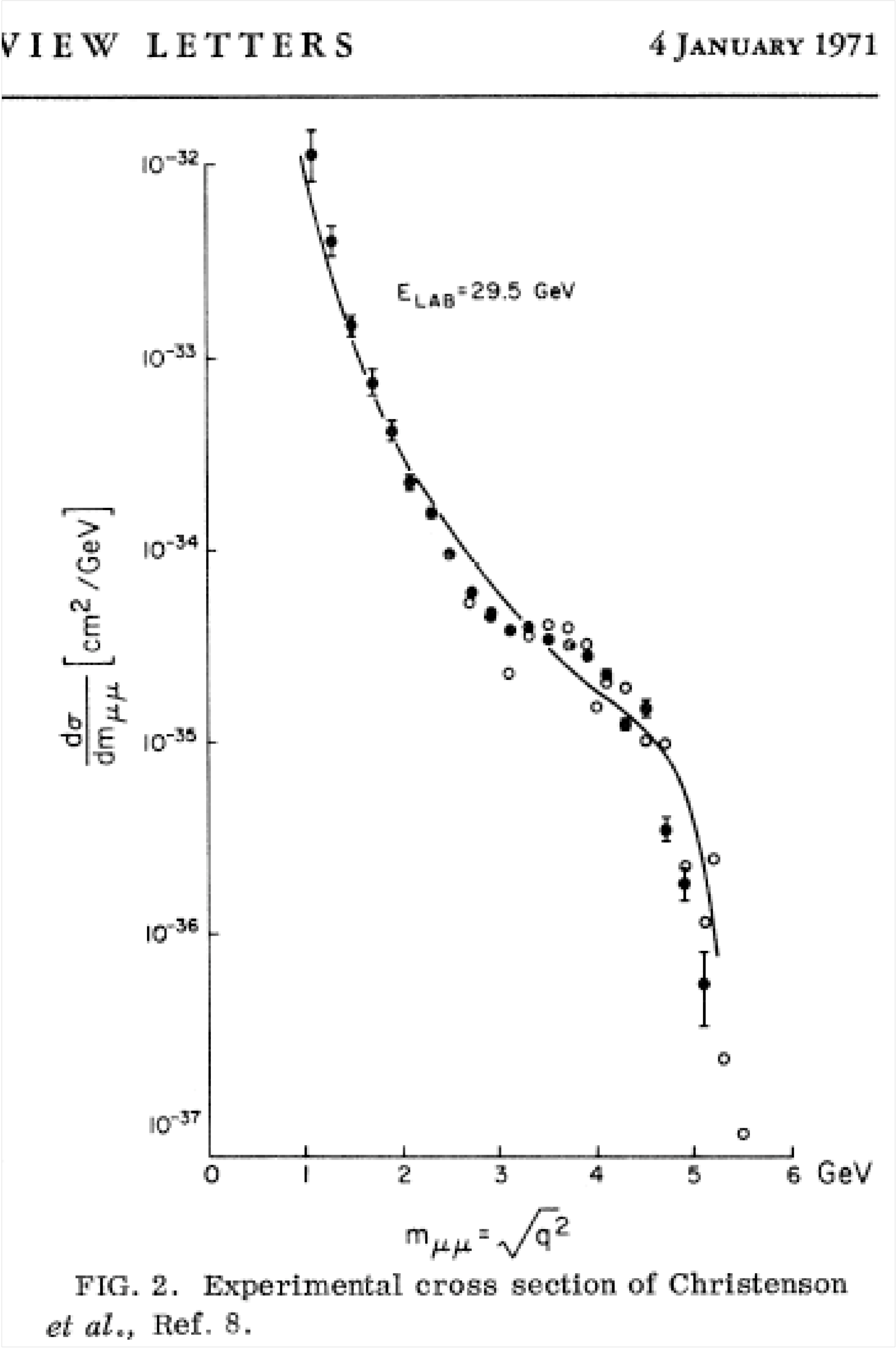}
\end{center}\vspace*{-1pc}
\caption[]{a) Dimuon invariant mass spectrum $d\sigma/dm_{\mu \mu}$~\cite{DYdiscovery};  b) theoretical prediction~\cite{ABP71} for $d\sigma/dm_{\mu \mu}$.}
\label{fig:DrellYan}\vspace*{-1pc}
\end{figure}

	The discovery of ``Drell-Yan'' pairs at the AGS proved to be seminal in future Relativistic Heavy Ion Physics as well as providing an interesting lesson. Figure~\ref{fig:DrellYan}a) shows the di-muon invariant mass spectrum $d\sigma/dm_{\mu \mu}$ from the collisions of 29.5 GeV protons in a thick Uranium target. There was definitely a dispute in the group about the meaning of the shoulder or ``bump'' or ``??'' for $2.5<m_{\mu \mu}<4.0$ GeV/c$^2$, which was apparently resolved adequately by the long forgotten theory paper~\cite{ABP71} which produced the curve that beautifully agreed with the data (Fig.~\ref{fig:DrellYan}b) with the explanation,``The origin of the shoulder comes from an interplay between the phase-space control of the integration region and the $q^2$ dependence of the coefficient.'' The important lesson that I learned from Fig.~\ref{fig:DrellYan} is to NEVER be influenced by theoretical curves which ``explain'' your data. It is only when the curves fail to explain the data that you learn something definitive: the theory is wrong. 

\begin{figure}[!b]
\begin{center}
\includegraphics[width=0.80\linewidth]{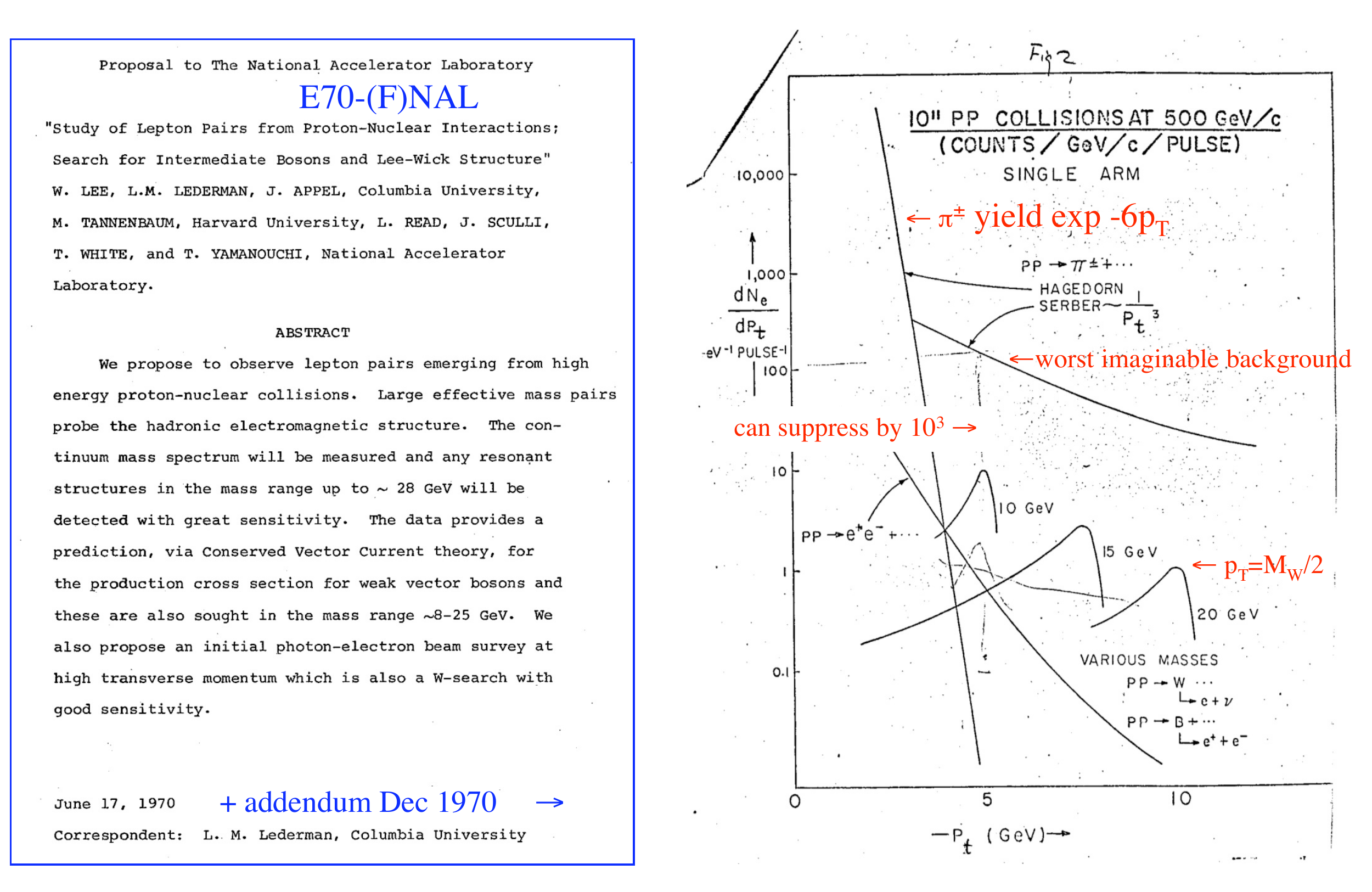}
\end{center}\vspace*{-2pc}
\caption[]{a) (left) Proposal for E70 Fermilab, June 17, 1970; b) (right) Cross section and background calculation for $W^{\pm}\rightarrow e^{\pm} +X$ from Addendum~\cite{E70Ad}.  }
\label{fig:E70}
\end{figure}

Leon Lederman was very excited in 1970 to be in the possession of the di-muon continuum mass spectrum, $d\sigma/dm_{\mu \mu}$, because by combining this result at $\sqrt{s}=7.4$ GeV with the newly found Bjorken Scaling~\cite{BjScaling} as used by Drell and Yan~\cite{DYexplanation}, he could calculate the $W$ cross section at any $\sqrt{s}$, and hence the sensitivities of his two proposals E70~\cite{E70Ad} and CCR~\cite{CCRprop} (see Fig.~\ref{fig:E70}).
Details worthy of note from Fig.~\ref{fig:E70}b are the $e^{-6p_T}$ pion yield, the line with the worst imaginable background and the Jacobean peaks at $p_T=M_{W}/2$ from $W\rightarrow  e+ X$ for various $W$ masses.
\section{The November Revolution and Birth of a {\QGP} Paradigm}
\begin{figure}[hbt]
\begin{center}
\begin{tabular}{cc}
\begin{tabular}[b]{c}
\includegraphics[width=0.35\linewidth]{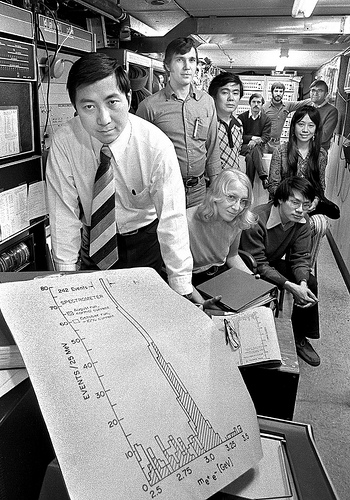}\hspace*{0.014in}\cr
\begin{picture}(50,5)
\end{picture}
\end{tabular}
\begin{tabular}[b]{c}
\includegraphics[width=0.50\linewidth]{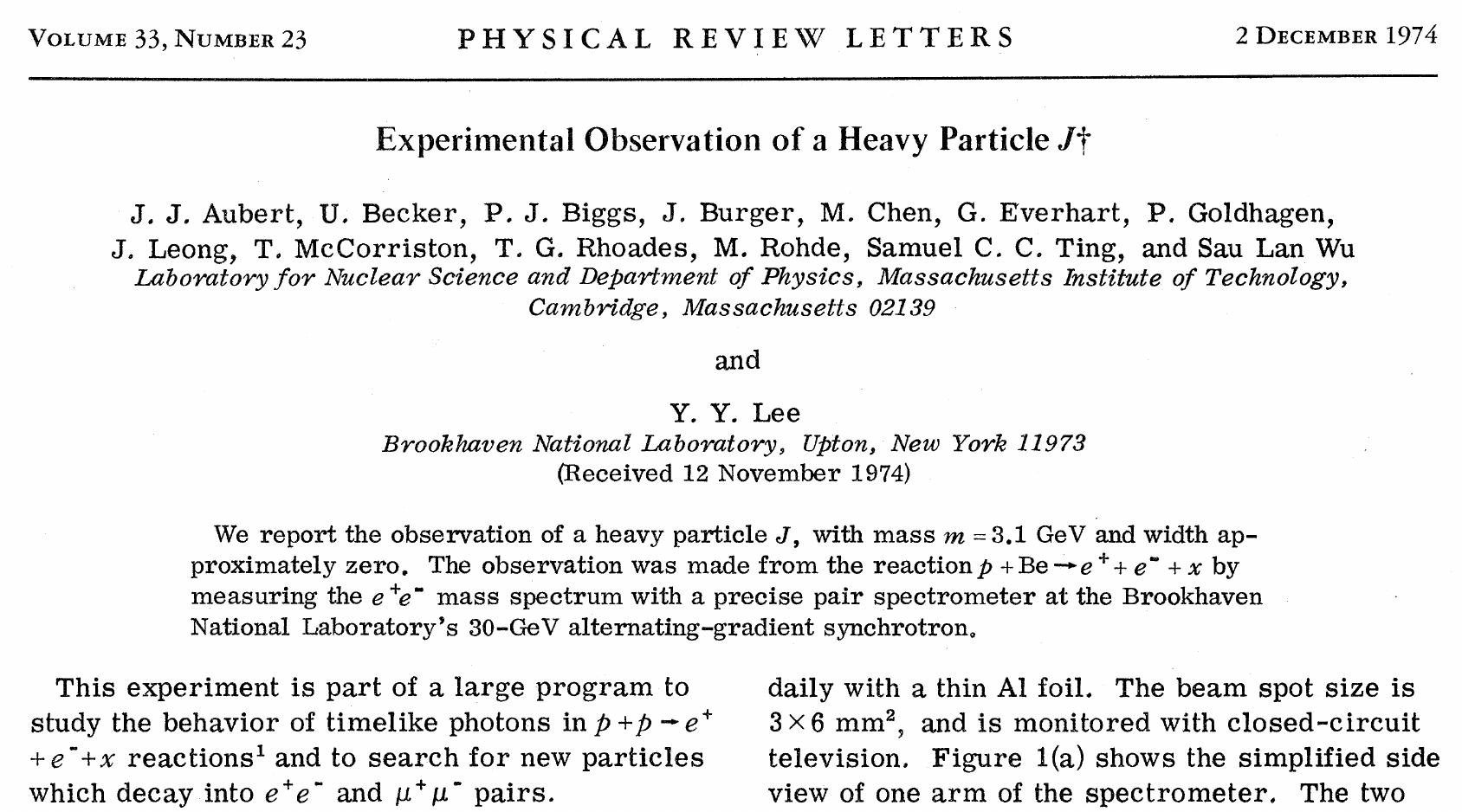}\cr
\hline\\
\includegraphics[width=0.50\linewidth]{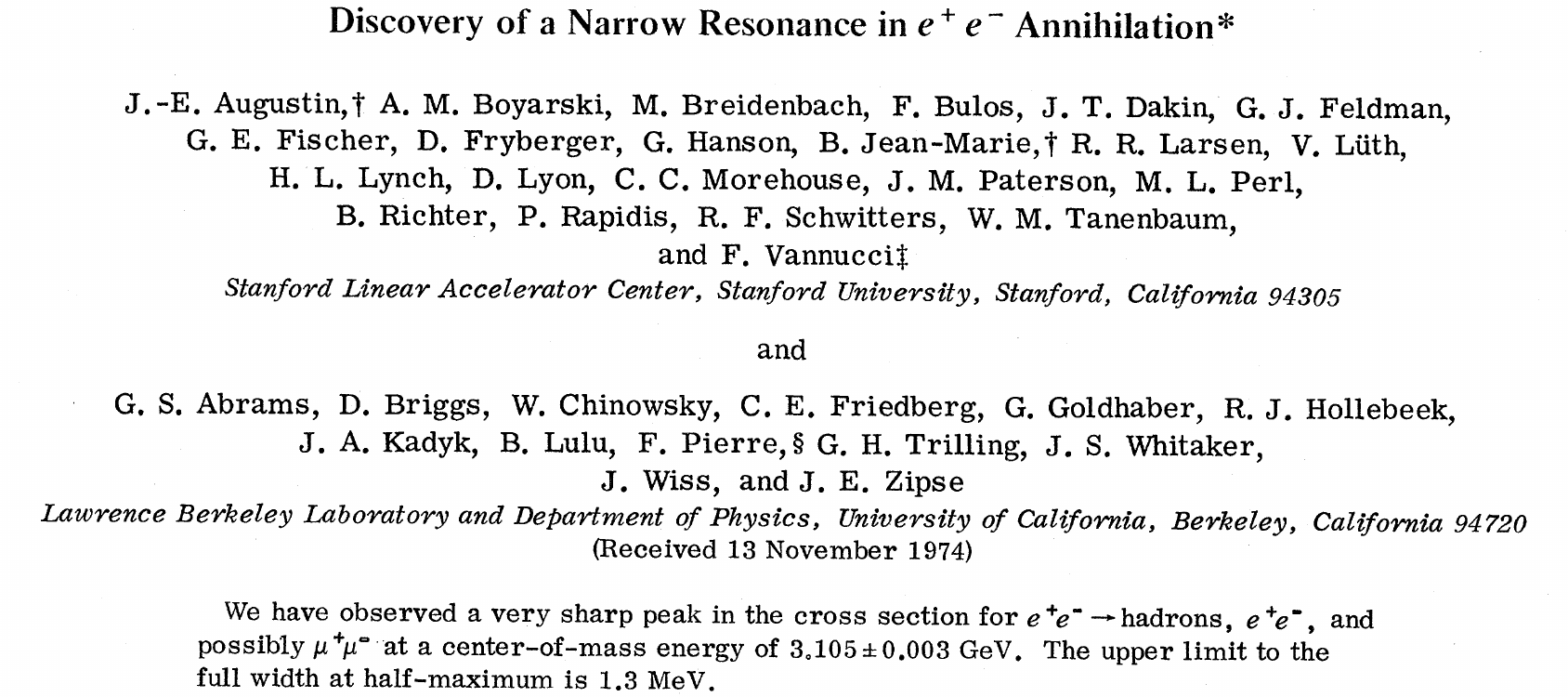}
\end{tabular}
\end{tabular}

\end{center}
\vspace*{-0.12in}
\caption[]
{ (left) Sam Ting with $J$ discovery at BNL; (right) $J$~\cite{TingJ} and $\psi$~\cite{RichterPsi} discoveries in PRL
\label{fig:JPsi} }
\end{figure}

\begin{figure}[h]
\begin{center}
a)\includegraphics[width=0.33\linewidth]{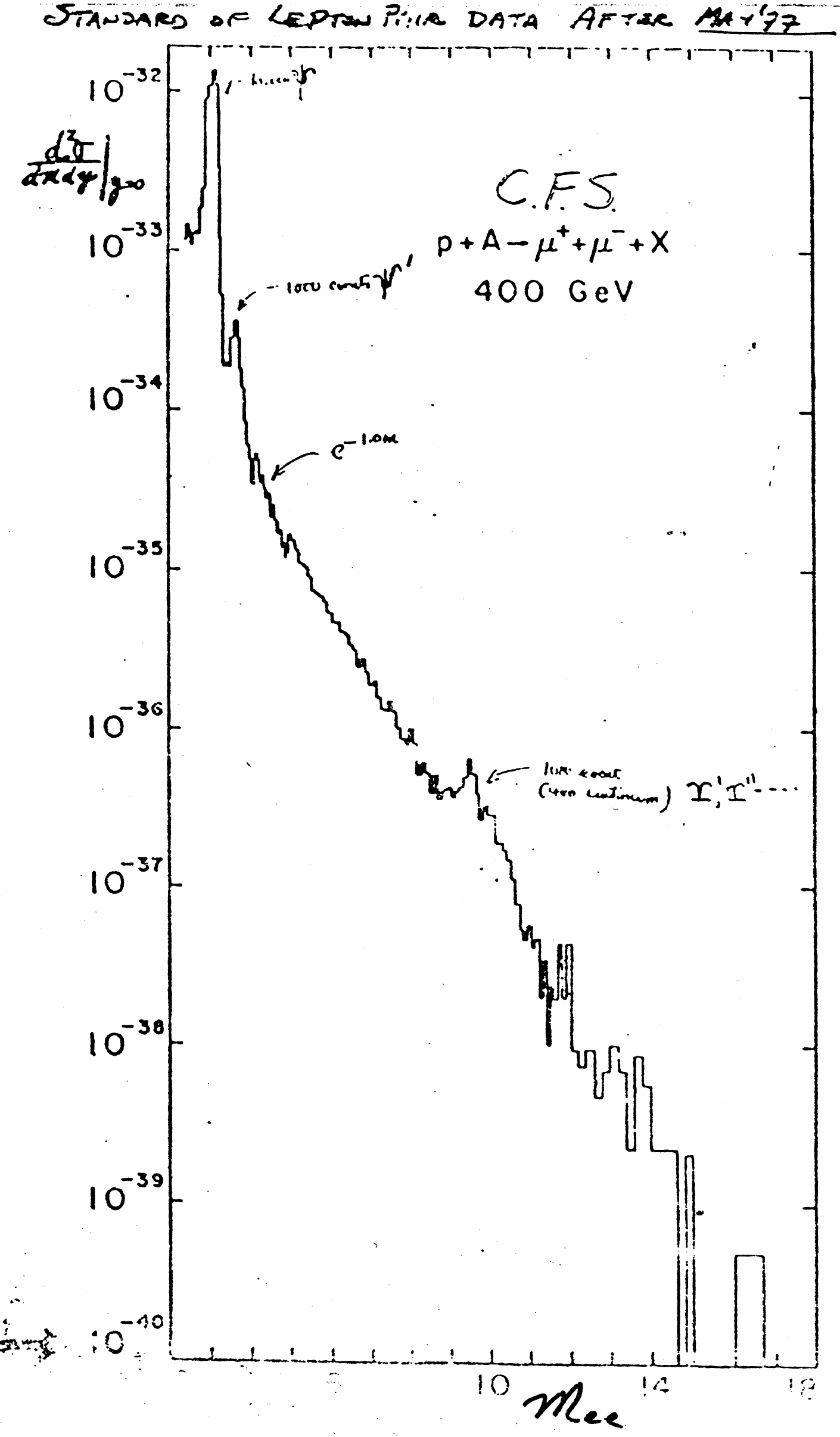}
b)\includegraphics[width=0.47\linewidth]{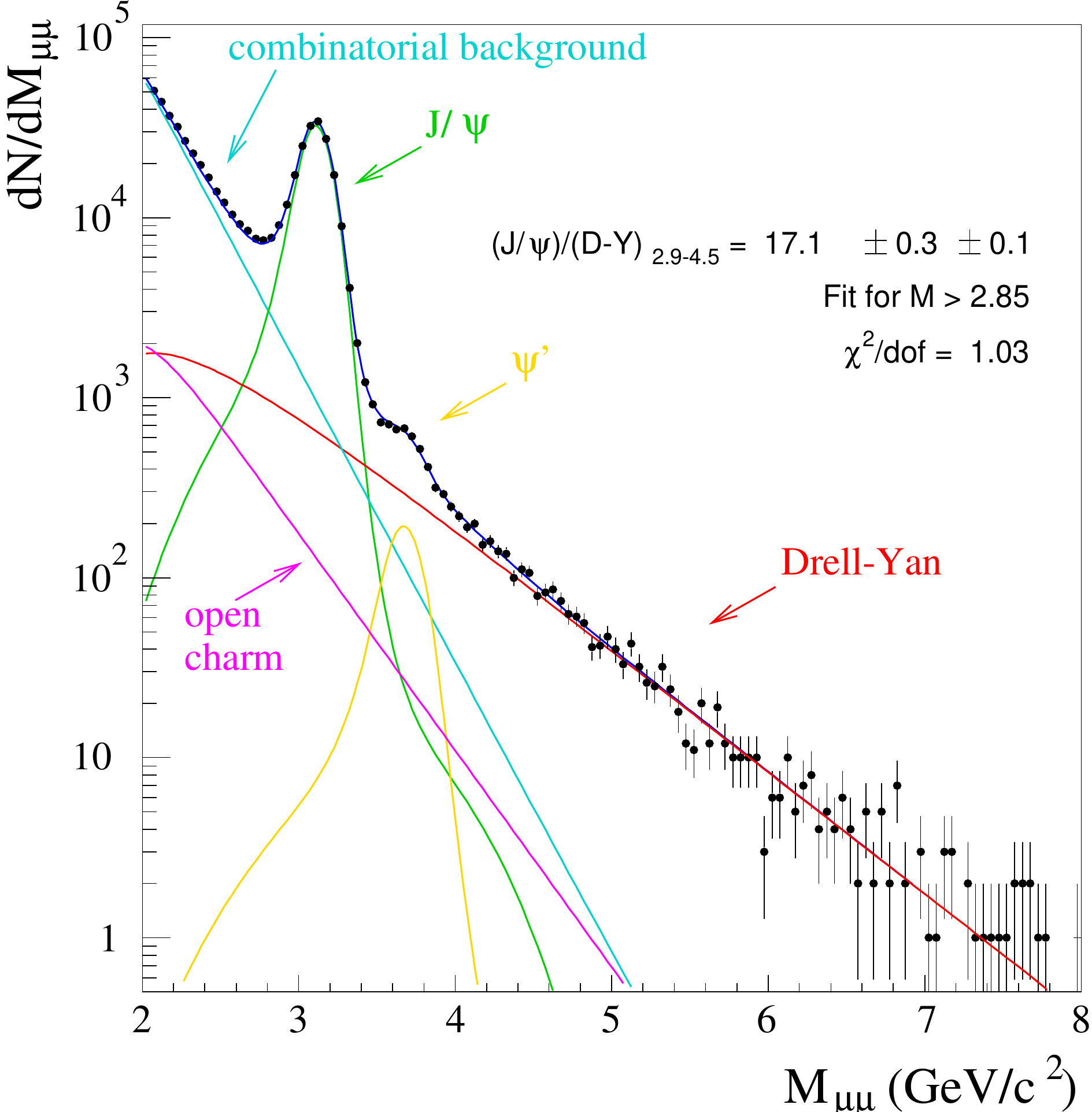}
\end{center}\vspace*{-1pc}
\caption[]{a) CFS di-muon spectrum at $\sqrt{s_{NN}}$=27.4 GeV~\cite{E70PRL39}, with $J/\psi$, $\psi{\, '}$, $\Upsilon$ clearly seen along with the Drell-Yan continuum, $d\sigma/dm\propto e^{-1.0m}$.    
b) NA50 di-muon spectrum in 158A GeV Pb+Pb collisions~\cite{Prino}
\label{fig:dimu}}
\end{figure}

   Lederman's shoulder was explained in November 1974 by the discovery of the $J/\psi$ at the BNL-AGS~\cite{TingJ} and at SLAC~\cite{RichterPsi} (Fig.~\ref{fig:JPsi}). This discovery revolutionized high energy physics since it was 
a heavy vector meson with a very narrow width (it decayed slowly) which implied a new conservation law, similar to the discovery of strange particles in cosmic rays~\cite{Peyrou}. The $J/\psi$ was quickly understood to be a bound state of heavy $c-\bar{c}$ quarks (charmonium)~\cite{bound2}---the hydrogen atom of {\QCD}. This was clear evidence for a ``2nd'' generation of quarks, and made all physicists believe in quarks and {\QCD}. This discovery also changed the paradigm of di-lepton production from the measurement of background for $W$ production to the search for new resonances. Greatly improved di-muon measurements by E70 in 1977 in p+A collisions at Fermilab~\cite{E70PRL39} set the standard for di-muon production (Fig.~\ref{fig:dimu}a) and were rewarded with the discovery of an even heavier ``3rd'' generation of quark, the $b$-quark~\cite{E70PRL39}. At CERN, di-muon measurements were extended to h+A collisions by NA3~\cite{NA3PLB86} and NA10~\cite{NA10PLB158} and then to Pb+Pb collisions by NA50 (Fig.~\ref{fig:dimu}b)~\cite{Prino,NA50PLB450}.  

\subsection{$\mathbf{J/\psi}$ suppression as the first paradigm for detecting the {\QGP}}
\begin{figure}[!b]
\begin{center}
\includegraphics[width=0.48\linewidth]{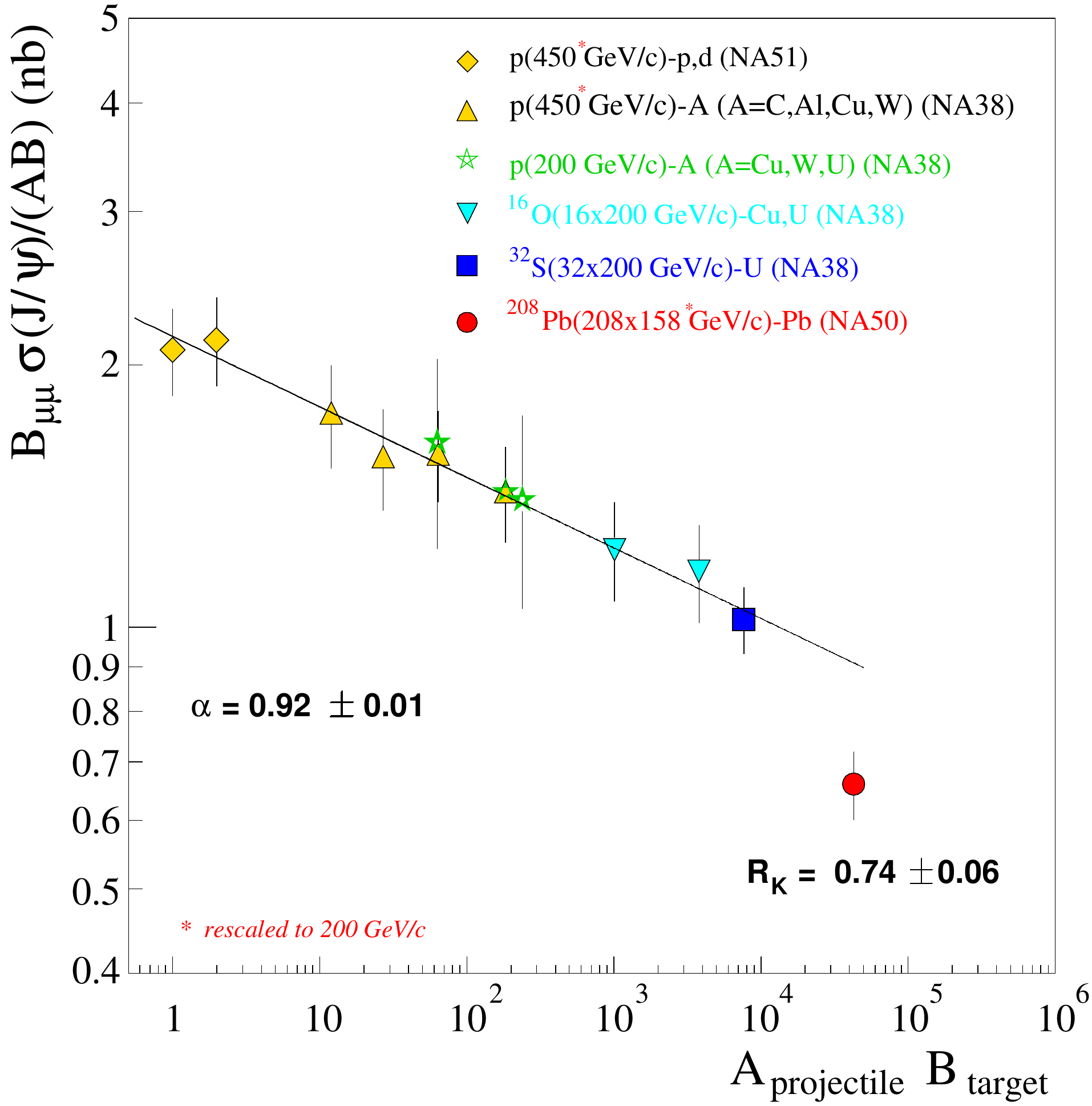}
\end{center}\vspace*{-1pc}
\caption[]{ The $J/\psi\rightarrow \mu^+ \mu^-$ cross section in A+B collisions divided by $A\times B$~\cite{Prino}. Line is $(AB)^{\alpha}$ with $\alpha=0.92\pm 0.01$. Only the Pb+Pb data fall below this line. However, note the {\Red *\Black}.}
\label{fig:JPsiAB}
\end{figure}

  In 1986, Matsui and Satz~\cite{MatsuiSatz} proposed that due to the Debye screening of the color potential in a {\QGP}, charmonium production would be suppressed since the $c-\bar{c}$ quarks couldn't bind. This paradigm drove the field of RHI physics for two decades; and the observation of ${J/\psi}$ suppression in A+A collisions~\cite{NA38firstsuppression,NA50PLB450} is the CERN fixed target Heavy Ion program's claim to fame. However, the interpretation is not straightforward because the ${J/\psi}$ is suppressed in p+A collisions (see Fig.~\ref{fig:JPsiAB}). The PHENIX experiment at RHIC was specifically designed to measure ${J/\psi}$ production at rest at mid-rapidity in p-p and Au+Au collisions~\cite{PXJpsiAuAu07} and found the same ${J/\psi}$ suppression at $\sqrt{s_{NN}}=200$ GeV as found at $\sqrt{s_{NN}}=17.2$ GeV (158A GeV bombarding energy) at CERN by NA50~\cite{NA50PLB450}, further complicating the $J/\psi$-suppression `paradigm'. Incredibly, the CERN fixed target RHI program is still making progress, having (after 21 years~\cite{NA38firstsuppression}) finally measured the p+A cold nuclear matter effect for $J/\psi$ at the same incident energy, 158 GeV, as for Pb+Pb~\cite{NA60-2010} (compare Fig.~\ref{fig:JPsiAB}), which materially changes the result. At best, the the $J/\psi$-suppression paradigm is presently inconclusive. Hopefully, results from the LHC may settle  the issue~\cite{egMJT09}.   
\section{A second paradigm for detecting the {\QGP}, BDMPSZ~\cite{BDMPS,Z} 1997--1998}
  In 1998, at the {\QCD} workshop in Paris, Rolf Baier asked me whether jets could be measured in Au+Au collisions because he had a prediction of a {\QCD} medium-effect (energy loss via gluon radiation induced by multiple scattering~\cite{BDPS}) on color-charged partons traversing a hot-dense-medium composed of screened color-charges~\cite{BaierQCD98}. I told him~\cite{MJTQCD98} that there was a general consensus~\cite{Strasbourg} that for Au+Au central collisions at $\sqrt{s_{NN}}=200$ GeV, leading particles are the only way to find jets, because in one unit of the nominal jet-finding cone,  $R=\sqrt{(\Delta\eta)^2 + (\Delta\phi)^2}$, there is an estimated $\pi R^2\times{1\over {2\pi}} {dE_T\over{d\eta}}\sim 375$ GeV of energy !(!) 
  
  The good news was that hard-scattering in p-p collisions had been discovered at the CERN ISR~\cite{CCR,SS,BS} by the method of leading particles, before the advent of {\QCD}, and it was proved by single inclusive and two-particle correlation measurements in the period 1972--1978 that high $p_T$ particles are produced from states with two roughly back-to-back jets which are the result of scattering of constituents of the nucleons as described by {\QCD}, which was developed during this period. The other good news was that these techniques could be used to study hard-scattering and jets in Au+Au collisions and that the PHENIX detector had been designed to make such measurements and could trigger, measure and separate $\gamma$ and $\pi^0$ out to $p_T > 25$ GeV in p-p and Au+Au collisions. In fact, in many talks, dating from as long ago as 1979~\cite{MJTDPF79}, I have been on record describing ``How everything you want to know about jets can be found using 2-particle correlations''.~\footnote{Recently, I had to amend this statement to `almost everything' because we found that the fragmentation function can not be measured in di-hadron correlations because the two particles, e.g. $\pi^0-h$, are both fragments of jets~\cite{ppg029,MJTCFRNC06} (see Secs.~\ref{sec:almost},~\ref{sec:nofrag} below).} 
  
\begin{figure}[!thb]
\begin{center}
\begin{tabular}{cc}
\includegraphics[width=0.50\linewidth]{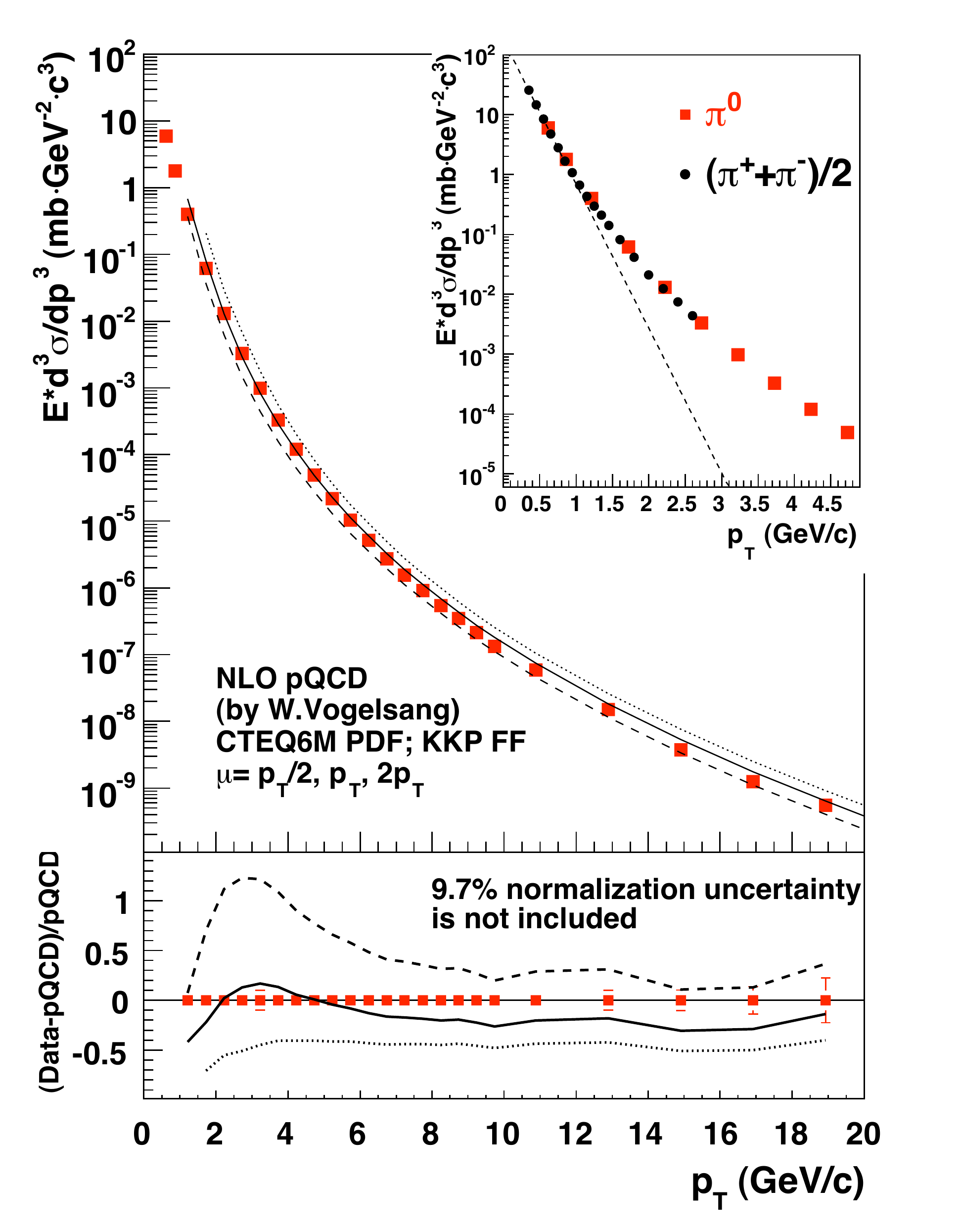}&
\hspace*{-0.02\linewidth}\includegraphics[width=0.56\linewidth]{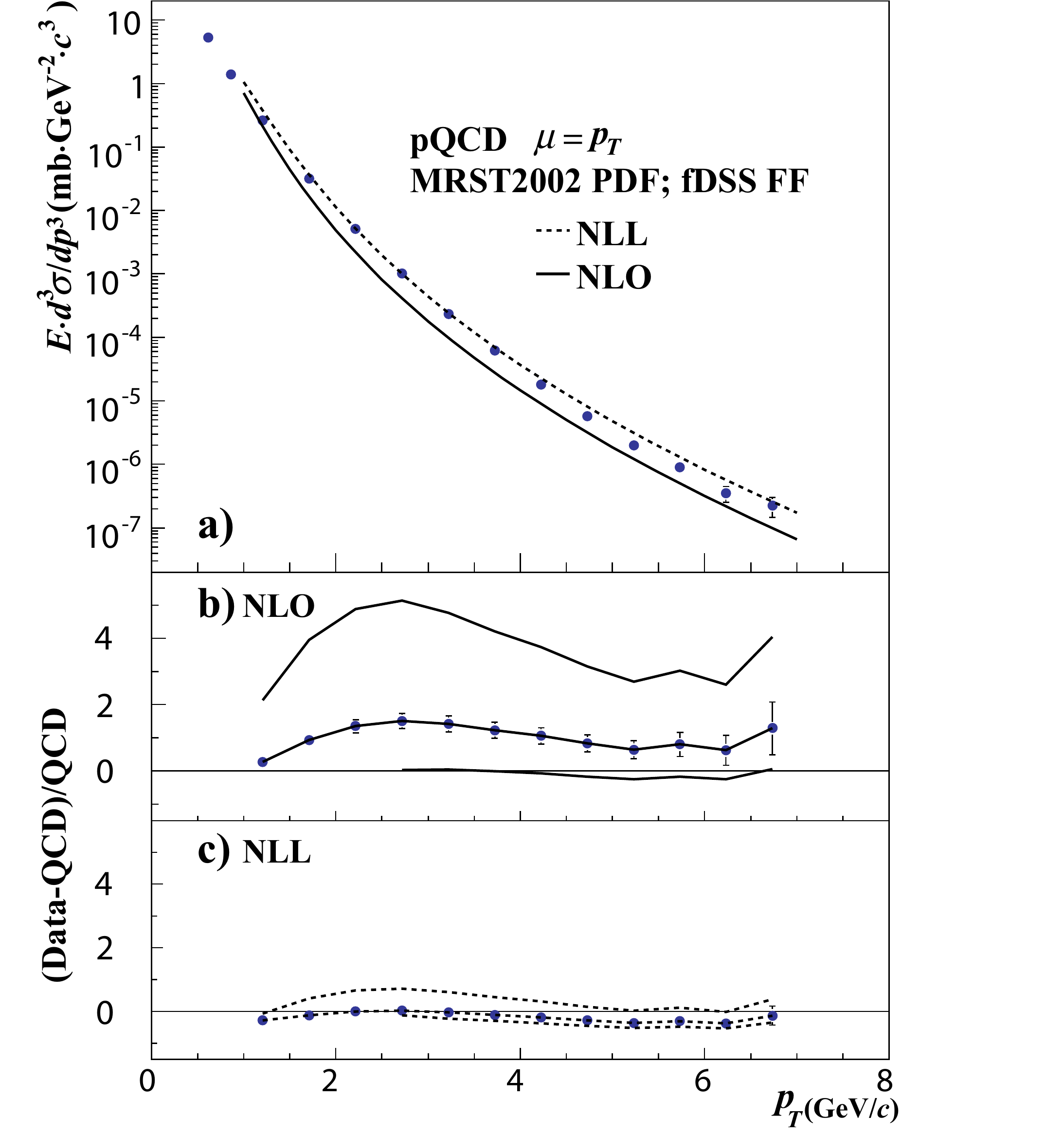}
\end{tabular}
\end{center}\vspace*{-0.25in}
\caption[]{(left) PHENIX measurement of invariant cross section, $E {d^3\sigma}/{d^3p}$, as a function of transverse momentum $p_T$ for $\pi^0$ production at mid-rapidity in p-p collisions at c.m. energy $\sqrt{s}=200$~GeV~\cite{ppg063}. (right) PHENIX measurement of $\pi^0$ in p-p collisions at $\sqrt{s}=62.4$~GeV~\cite{ppg087}.  }
\label{fig:PXpi0pp2}
\end{figure}

  One lesson from this new paradigm is that a good probe of {\QCD} in a fundamental system such as p-p collisions (Fig.~\ref{fig:PXpi0pp2})  provides a well calibrated probe of {\QCD} in more complicated collisions such as Au+Au.    In the 1980's when RHIC was proposed, hard processes were not expected to play a major role in A+A collisions. However, starting in 1997~\cite{MJTRHIC97}, inspired by Rolf and collaborators, and before them by the work of Gyulassy~\cite{MGyulassy} and Wang~\cite{XNWang}, I indicated that my best bet on discovering the {\QGP} was to utilize semi-inclusive $\pi^0$ or $\pi^{\pm}$ production in search for ``high $p_T$ suppression''.  
  
\subsection{Soft vs. hard-physics in p-p collisions}
Fig.~\ref{fig:PXpi0pp2}a nicely illustrates two important points for $\pi^0$ production in p-p collisions. Apart from the beautiful data, the main emphasis is the agreement of NLO pQCD with the measurement for $p_T\gsim 3$ GeV/c. 
Actually this surprised some of my colleagues from the lepton-scattering community, but was no surprise to me because such data in the 1970's is what made people believe in {\QCD} (see Sec.~\ref{sec:hardtime} below). The other important issue is the transition from the exponential ``soft physics'' spectrum for $p_T<1.5$ GeV/c to the power law spectrum of hard scattering at larger $p_T$. The $e^{-5.5p_T}$ exponential at $\sqrt{s}=200$ GeV, is very close to the original 
cosmic-ray ``Cocconi Formula''~\cite{Cocconi1961}, $e^{-6p_T}$, used by Lederman (Fig.~\ref{fig:E70}b) in 1970, but the power-law hard-scattering spectrum for $p_T>3$ GeV/c is much larger than the ``worst imaginable background''. This is the hard-scattering component which I shall discuss again below but I first wish to emphasize some important issues in Relativistic Heavy Ion physics which are dominated by the soft physics. 
\section{The important of soft-physics in Relativistic Heavy Ion Collisions}
 In dealing with Relativistic Heavy Ions, there are three dramatic differences from p-p physics. 
\begin{figure}[h]
\begin{center}
\begin{tabular}{cc}
\includegraphics[width=0.64\linewidth]{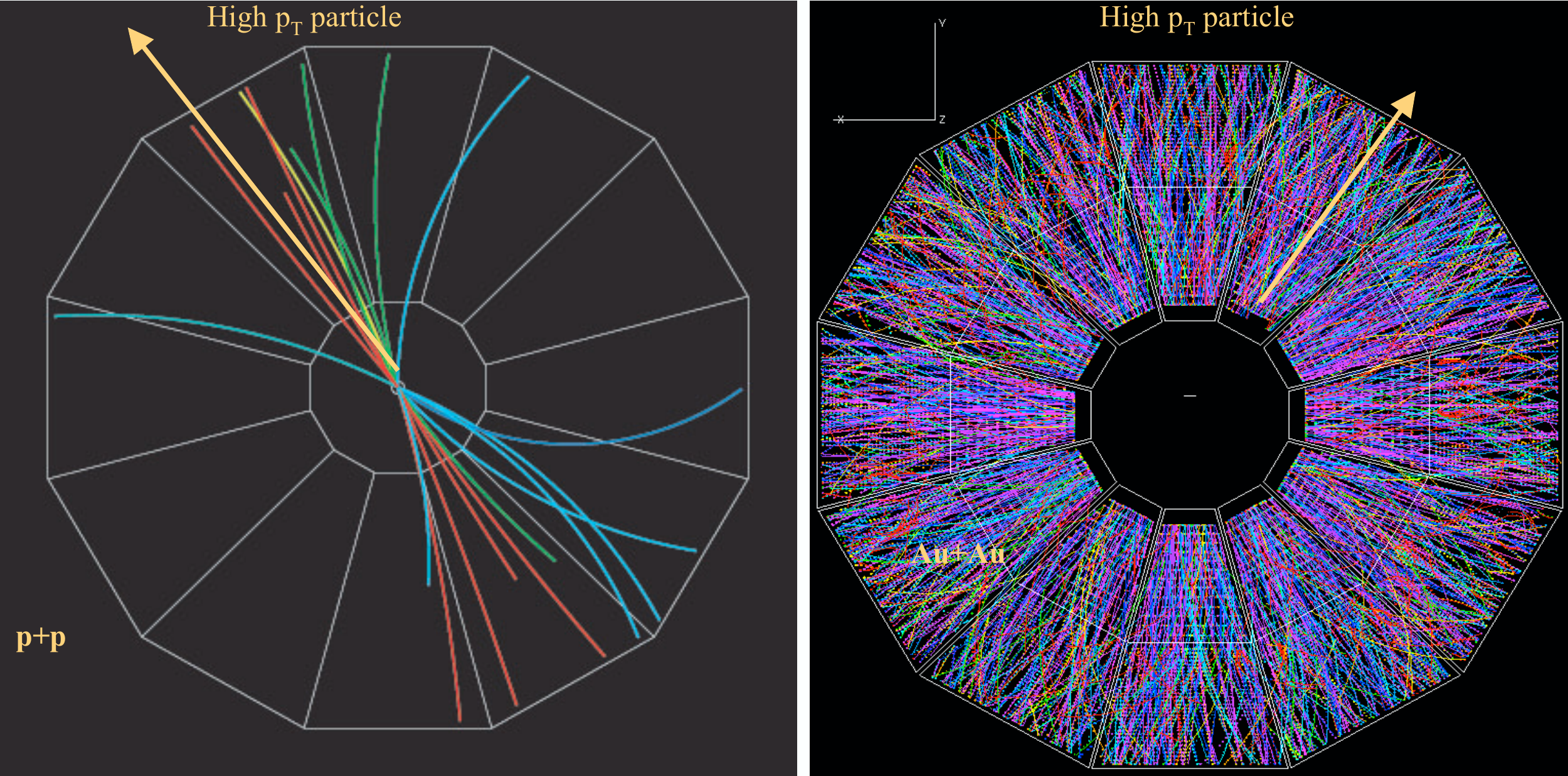}&\hspace*{-0.025\linewidth}
\includegraphics[width=0.315\linewidth,height=0.315\linewidth]{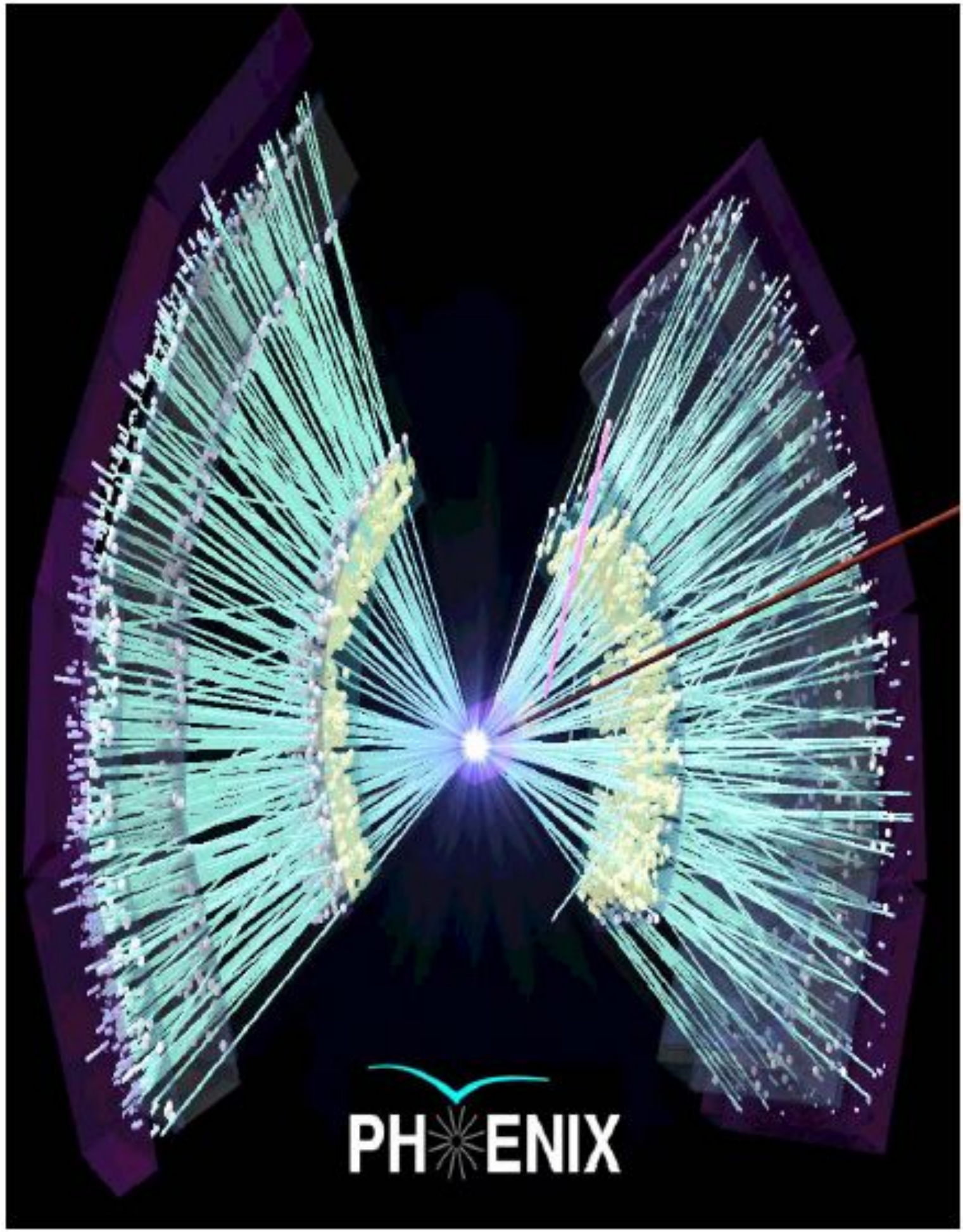}
\end{tabular}
\end{center}\vspace*{-0.15in}
\caption[]{a) (left) A p-p collision in the STAR detector viewed along the collision axis; b) (center) Au+Au central collision at $\sqrt{s_{NN}}=200$ GeV in the STAR detector;  c) (right) Au+Au central collision at $\sqrt{s_{NN}}=200$ GeV in the PHENIX detector.  
\label{fig:collstar}}
\end{figure}\vspace*{-1pc}
        \begin{itemize}
\item The particle multiplicity is $\sim$ A times larger in A+A central collisions than in p-p collisions (Fig.~\ref{fig:collstar}). This is why there is such a huge energy, $\sim 375\ R^2$ GeV, in a jet cone of radius $R$ at $\sqrt{s_{NN}}=200$ GeV.    
\item There are huge azimuthal anisotropies in A+A which don't exist in p-p which are interesting in themselves, and are useful, but sometimes can be troublesome.
\item Space-time issues, both in momentum space and coordinate space are important in RHI. For example:  
\begin{enumerate}
\item When in time and space does a parton fragment? Is this different for light and heavy quarks? When are particles formed?
\item The Dokshitzer textbook formula, $\tau_F=ER^2$, should, I think, be corrected to $\tau_F=\gamma R=R\times E/M=ER\lambda_C$, where $R$, $M$, are the radius and mass of the particle and $\lambda_C$ its Compton wavelength.
\item Would a proton embedded in a {\QGP} dissolve? How long does this take? How is this related to $J/\psi$ suppression?
\end{enumerate}
\end{itemize}
\subsection{Soft physics dominates multi-particle production in both p-p and A+A collisions}
\label{section:soft}
	A quantity related to multiplicity, but more relevant for the study of jets, is the distribution of transverse energy, $E_T=\sum_i E_i \sin \theta_i$, where the sum is taken over all particles emitted on an event into a fixed but large solid angle, typically measured in a calorimeter, corrected for the calorimeter response and then corrected to the solid angle $\Delta\eta=1$, $\Delta\phi=2\pi$~\cite{egseeMJT89}. $E_T$ is also related to $M_{p_T}$, the event-by-event average $p_T$ (Fig.~\ref{fig:NA49MJT}a), except that it is the sum of $p_T$ of all particles on every event   rather than the average. $dE_T/dy$ is thought to be related to the co-moving energy density in
a longitudinal expansion~\cite{BjorkenPRD27,PXWP}, and taken by all experiments as a measure of 
the energy density in space, ``the Bjorken energy density'', $\epsilon_{Bj}$:
\begin{equation}
\epsilon_{Bj}={d\mean{E_T}\over dy} {1\over \tau_F\pi R^2}
 \label{eq:eBj}
 \end{equation}
where $\tau_F$, the formation time, is usually taken as 1 fm/c,
$\pi R^2=A_{\perp}$ is the transverse overlap area of the collision, and $d\mean{E_T}/dy$ is the 
co-moving energy density. 

\begin{figure}[ht]
\begin{center}
a)\includegraphics[width=0.47\linewidth]{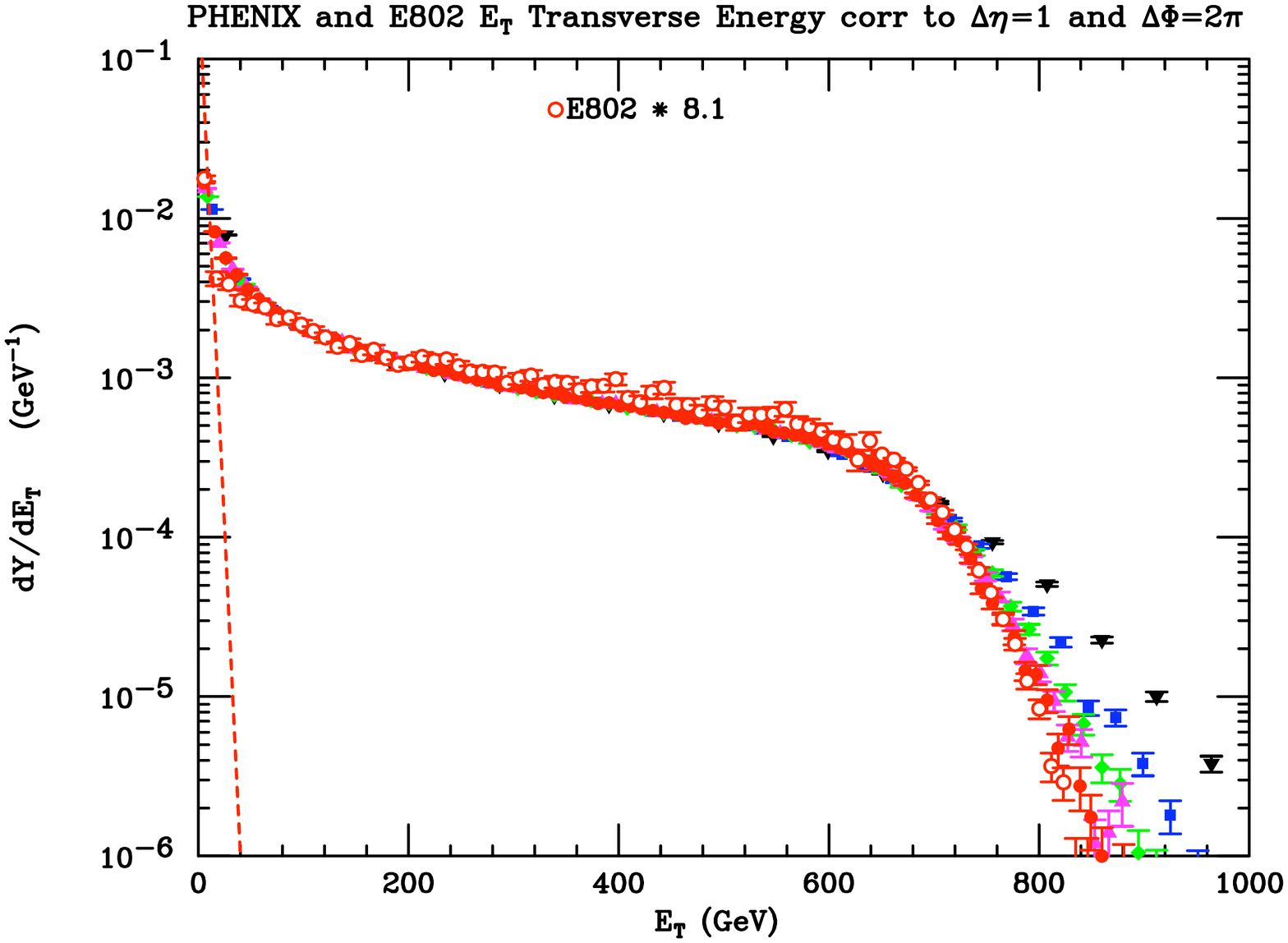} 
b)\includegraphics[width=0.47\linewidth]{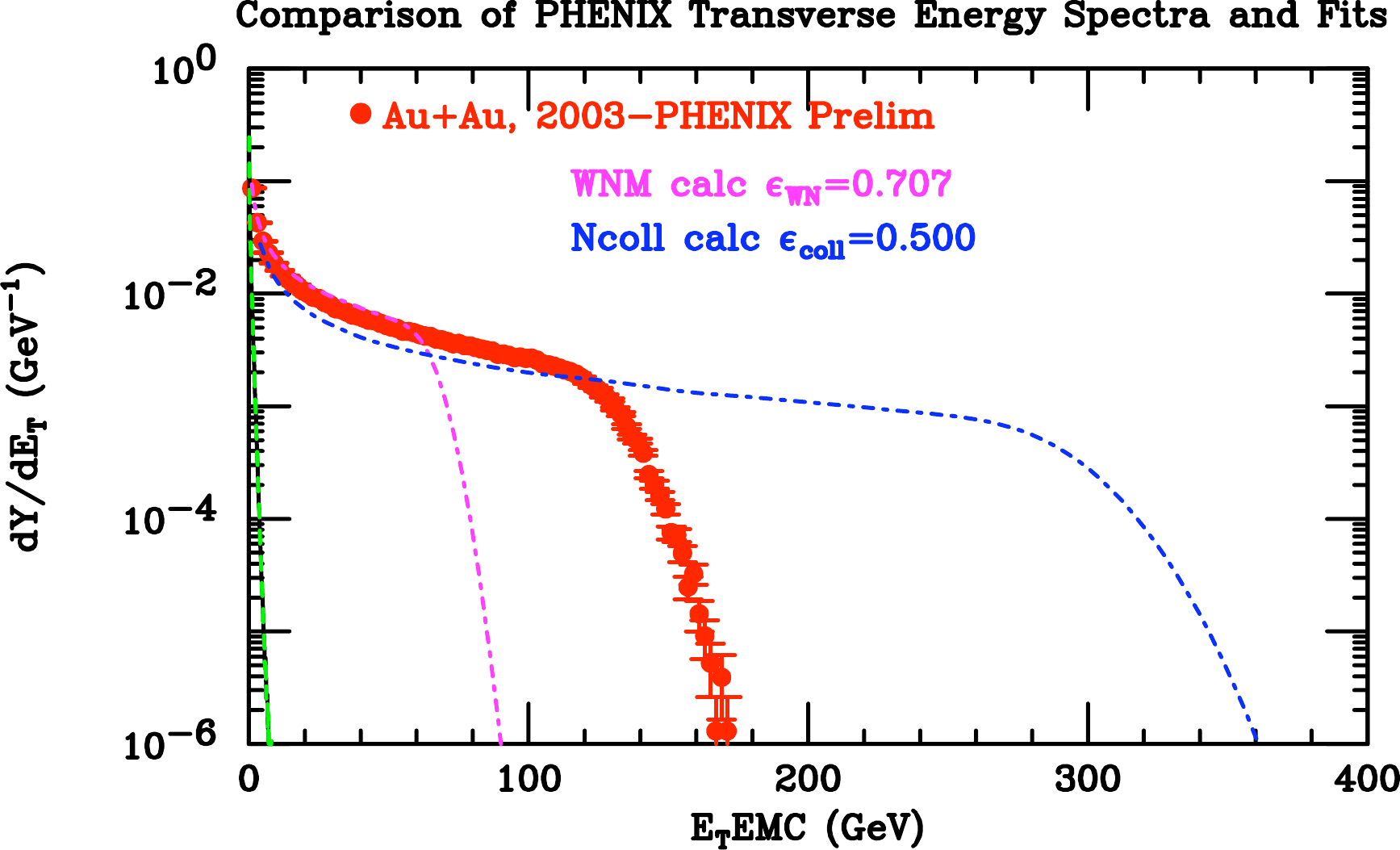}
\end{center}
\caption[]{a) $E_T$ distributions in Au+Au collisions: PHENIX at $\sqrt{s_{NN}}=200$ GeV~\cite{PXET} compared to the E802 (non-response-corrected) $E_T$ at $\sqrt{s_{NN}}=5.4$ GeV~\cite{E802ET,MJT04} scaled up by 8.1. The quantity plotted is the normalized yield, $dY/dE_T$  (GeV$^{-1}$), which integrates to 1, i.e $\int dE_T\;dY/dE_T=1$. Different filled symbols represent PHENIX measurements in solid angles $\Delta\eta=0.76$, $\Delta\phi=m\times \pi/8$ at mid rapidity with $m=1,...5$. Dashed line is an estimate of $E_T$ distribution in p-p collisions at $\sqrt{s}=200$ GeV for $m=5$. b) PHENIX $m=5$ raw $E_T$ distribution, ${E_{T}EMC}$, at $\sqrt{s_{NN}}=200$ GeV together with calculated distributions for participant scaling, WNM ($E_T\propto N_{\rm part}/2$), and Ncoll scaling ($E_T\propto N_{\rm coll}$) based on convolutions of the p-p distribution. }
\label{fig:E802PXpp} 
\end{figure}

	The main importance of $E_T$ distributions in RHI collisions (see Fig.~\ref{fig:E802PXpp}a) is that they are sensitive primarily to the nuclear geometry of the reaction, and hence can be used to measure the centrality of individual interactions on an event-by-event basis.\footnote{However, in PHENIX, centrality is determined far from mid-rapidity to avoid possibly biasing any of the mid-rapidity measurements.} Thus, the transverse energy or multiplicity distribution in A+A collisions has a characteristic shape which is quite different from that in p-p collisions since it is sensitive to the number of participants or centrality of an A+A collision and integrates over all impact parameters. This is shown in Fig.~\ref{fig:E802PXpp}a for Au+Au collisions at AGS ($\sqrt{s_{NN}}=5.4$ GeV) and RHIC ($\sqrt{s_{NN}}=200$ GeV), 
together with an estimate at RHIC for p-p collisions. The fact that the scaled AGS and RHIC $E_T$ distributions lie one on top of each other shows that the shape of $E_T$ distributions is essentially entirely dominated by the nuclear-geometry at both the AGS and RHIC; which implies that there are no critical fluctuations at either the AGS or RHIC.  Above the upper 0.5 percentile of the distribution ($E_T\approx 710$ GeV on Fig.~\ref{fig:E802PXpp}),  measurements in smaller solid angles show increasingly larger fluctuations  represented by  flatter slopes above the knee of the distribution; but this is a random fluctuation (fewer particles in the smaller solid angle fluctuate more). This is visible only for the most central collisions when the nuclear-geometrical fluctuations have saturated (all nucleons participating). The Bjorken energy density can be straightforwardly computed from from Fig.~\ref{fig:E802PXpp}a at RHIC ($\sqrt{s_{NN}}=200$ GeV), with  $\epsilon_{Bj}\times \tau_F =5.4\pm0.6$ GeV/fm$^2$~\cite{PXET} for the 5\% most central collisions ($E_T\geq 580$ GeV). For the AGS data, the ratio of 8.1 shown in Fig.~\ref{fig:E802PXpp} can not be used to get $\epsilon_{Bj}\times \tau_F$ at $\sqrt{s_{NN}}=5.4$ GeV because the E802 data have not been corrected for the calorimeter response, which is much more difficult in the fixed target geometry. However, Ref.~\cite{PXET} calculated the fully corrected E802 $E_T$ from charged particle measurements and obtained  $\epsilon_{Bj}\times \tau_F =1.0$ GeV/fm$^2$ at $\sqrt{s_{NN}}=4.8$ GeV.
\subsection{Another paradigm we should do away with}
   It has been popular to characterize the average values of $dE_T/d\eta$ (or the charged multiplicity density $dn^{ch}/d\eta$) in A+A collisions as a function of centrality by the equation~\cite{ppg001,KN01} 
	\begin{equation}
	{dn^{ch}_{AA}/d\eta}=(1-x)\ dn^{ch}_{pp}/d\eta\times {N_{\rm part}}/2 +x\ dn^{ch}_{pp}/d\eta\times {N_{\rm coll}}\qquad .
	\label{eq:crazy}
	\end{equation}
	 While this may seem reasonable for the average value $\mean{dn^{ch}_{AA}/d\eta}$ at a given centrality, is nonsense for the distribution, which if such a representation were true would be the weighted sum of $(1-x) \times$ the WNM curve + $x\  \times$ the Ncoll curve in Fig.~\ref{fig:E802PXpp}b, which obviously looks nothing like the actual $E_{T}${\rm EMC} measurement. (A more reasonable representation is ${dn^{ch}_{AA}/d\eta}\propto {N_{\rm part}}^\alpha$.) Also the claim that the Ncoll scaling of the {\em soft} p-p $E_T$ or multiplicity $n^{ch}_{pp}$ distributions (or average values) in Eq.~\ref{eq:crazy} has anything to do with hard-scattering~\cite{KN01} is equally nonsensical. This could be further elaborated by looking for jets in the p-p data; but such  measurements (not finding jet activity until 4-5 orders of magnitude down in $E_T$ cross section) were already done by COR~\cite{COR-ETpp} and UA2~\cite{UA2ET} 25 years ago (see Fig.~\ref{fig:ETdists} below).  
	 \subsection{Another soft-physics paradigm to test at the LHC---Hydrodynamics}
\begin{figure}[!h] 
\begin{center}
\begin{tabular}{ccc}
\hspace*{-0.014\linewidth}\includegraphics[width=0.35\linewidth]{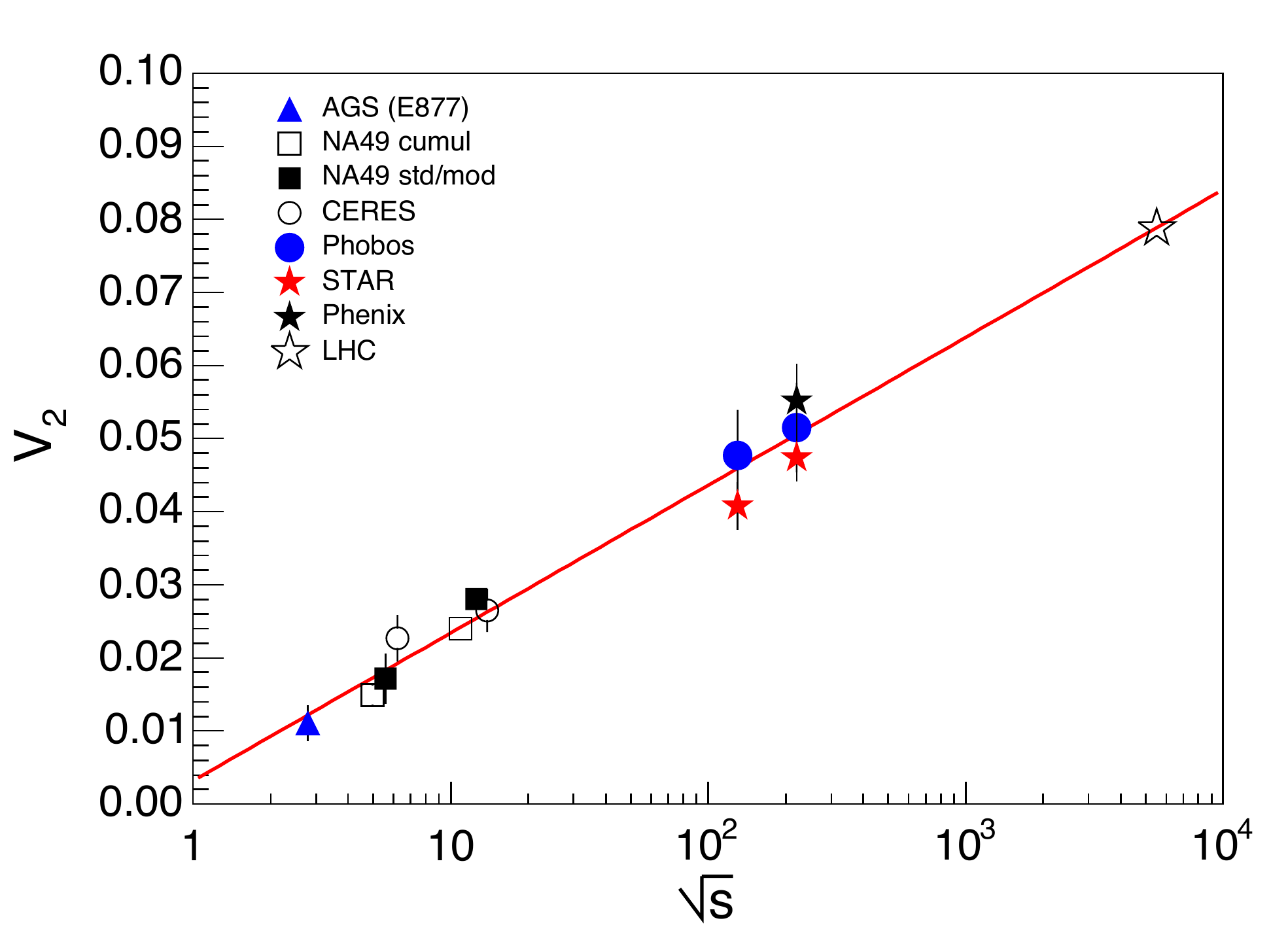} &
\hspace*{-0.027\linewidth}\includegraphics[width=0.33\linewidth]{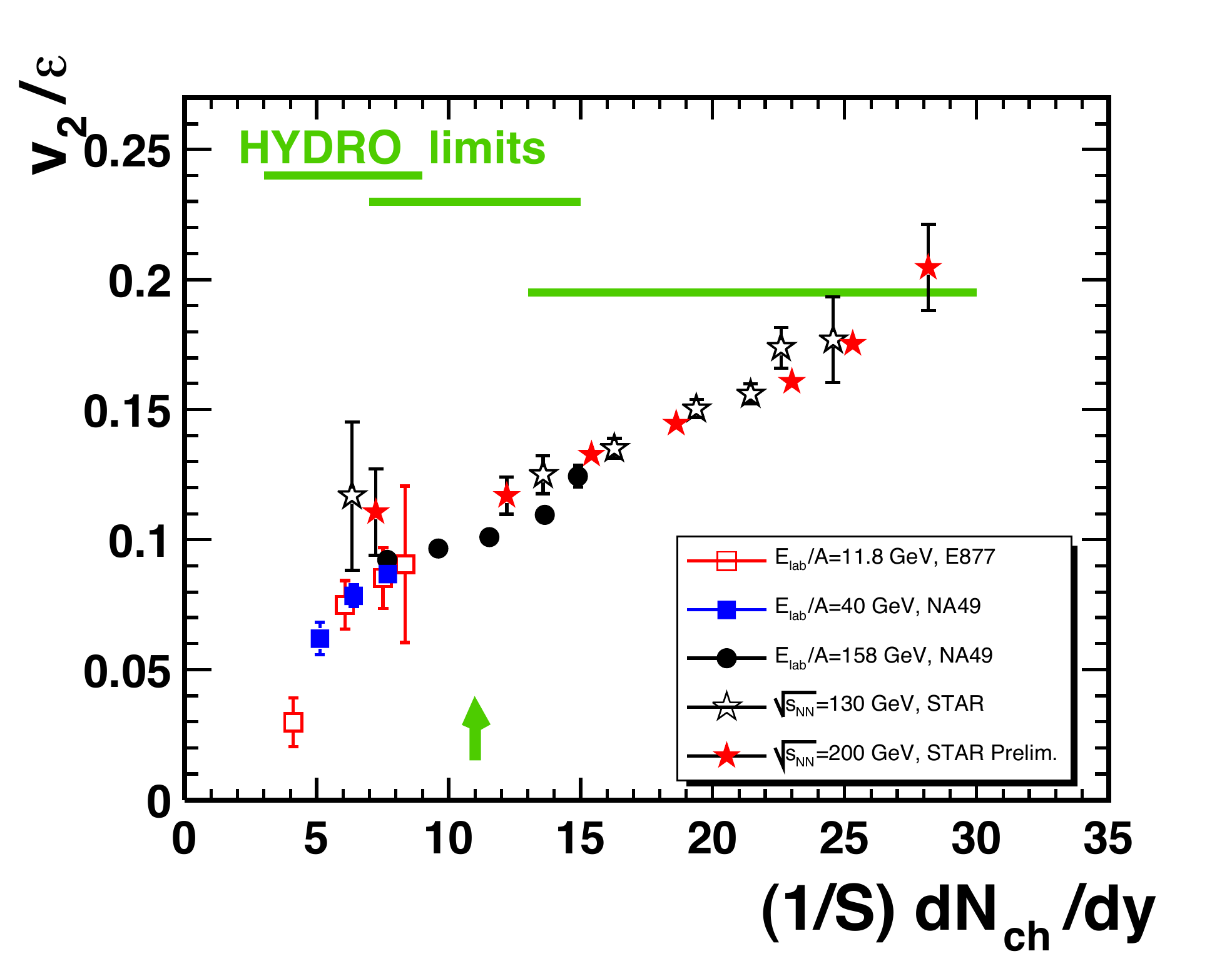}
\hspace*{-0.022\linewidth}\includegraphics[width=0.33\linewidth]{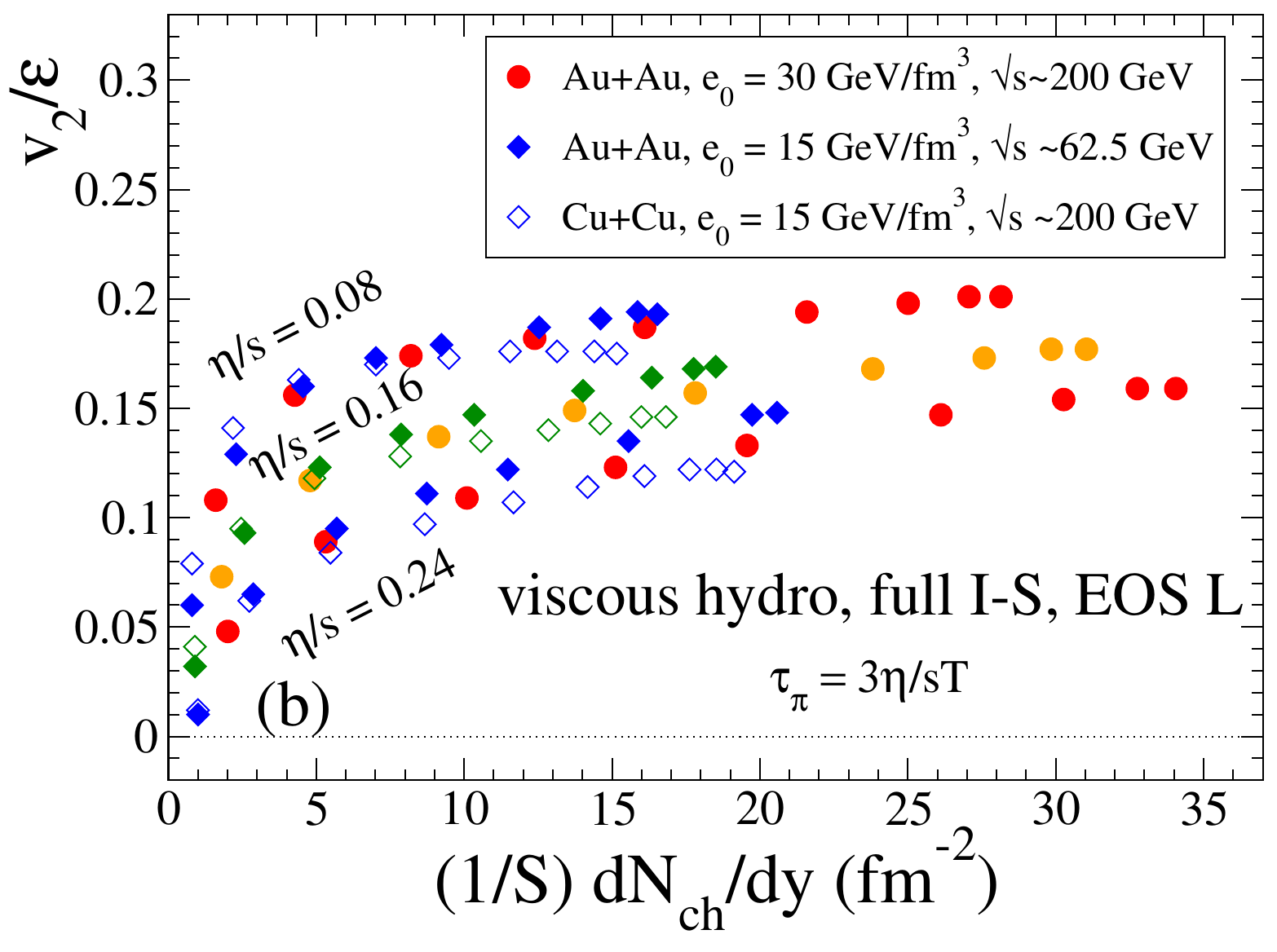}
\end{tabular}
\end{center}
\caption[]
{a) (left) Busza's extrapolation of $v_2$ to LHC~\cite{LastCall}. b) (center) $v_2/\varepsilon$ vs `Bjorken multiplicity density', $(1/S) dN_{\rm ch}/dy$~\cite{Alt}. 
c) (right) `Hydro Limit' calculated in viscous Hydrodynamics for several values of the initial energy density $e_0$~\cite{SongHeinz}.      
\label{fig:v2limit} }
\end{figure}

	    In my opinion, the most interesting soft physics issue for the LHC concerns the possible increase of the anisotropic flow $v_2$ beyond the `hydrodynamic limit'. Wit Busza's extrapolation~\cite{LastCall} of $v_2$ to the LHC energy is shown in Fig.~\ref{fig:v2limit}a, a factor of 1.6 increase from RHIC.  
A previous paper by NA49~\cite{Alt} which compared $v_2$ measurements from AGS and CERN fixed target experiments to RHIC as a function of the `Bjorken multiplicity density', $dn_{\rm ch}/d\eta /S$, where $S=$ is the overlap area of the collision zone, showed an increase in $v_2/\varepsilon$ from fixed target energies to RHIC leading to a ``hydro limit'', where $\varepsilon$ is the eccentricity of the collision zone (Fig.~\ref{fig:v2limit}b). This limit was confirmed in a recent calculation by Uli Heinz using viscous relativistic hydrodynamics~\cite{SongHeinz} which showed a clear hydro-limit of $v_2/\varepsilon=0.20$ (Fig.~\ref{fig:v2limit}c). This limit is sensitive to the ratio of the viscosity/entropy density, the now famous $\eta/s$, but negligibly sensitive to the maximum energy density of the collision, so I assume that this calculation would give a hydro-limit at the LHC not too different from RHIC, $v_2/\varepsilon\approx 0.20$. Busza's extrapolation of a factor of 1.6 increase in $v_2$ from RHIC to LHC combined with $v_2/\varepsilon$ from Fig.~\ref{fig:v2limit}b gives $v_2/\varepsilon=0.32$ at LHC. In my opinion this is a measurement which can be done to high precision on the first day of Pb+Pb collisions at the LHC, since it is high rate and needs no p-p comparison data. Personally, I wonder what the hydro aficionados would say if both Heinz' and Busza's predictions were correct? 

\section{Hard-Scattering in p-p collisions}
\label{sec:hardtime}
   We now go back to where we were in the search for high $p_T$ leptons and di-leptons before we got diverted by the $J/\psi$ discovery (and {\QGP} paradigm) and soft-physics issues. 
    \subsection{Bjorken scaling and the parton model}
\begin{figure}[!h] 
\begin{center}
\begin{tabular}{ccc}
a)\hspace*{-0.00\linewidth}\includegraphics[width=0.29\linewidth]{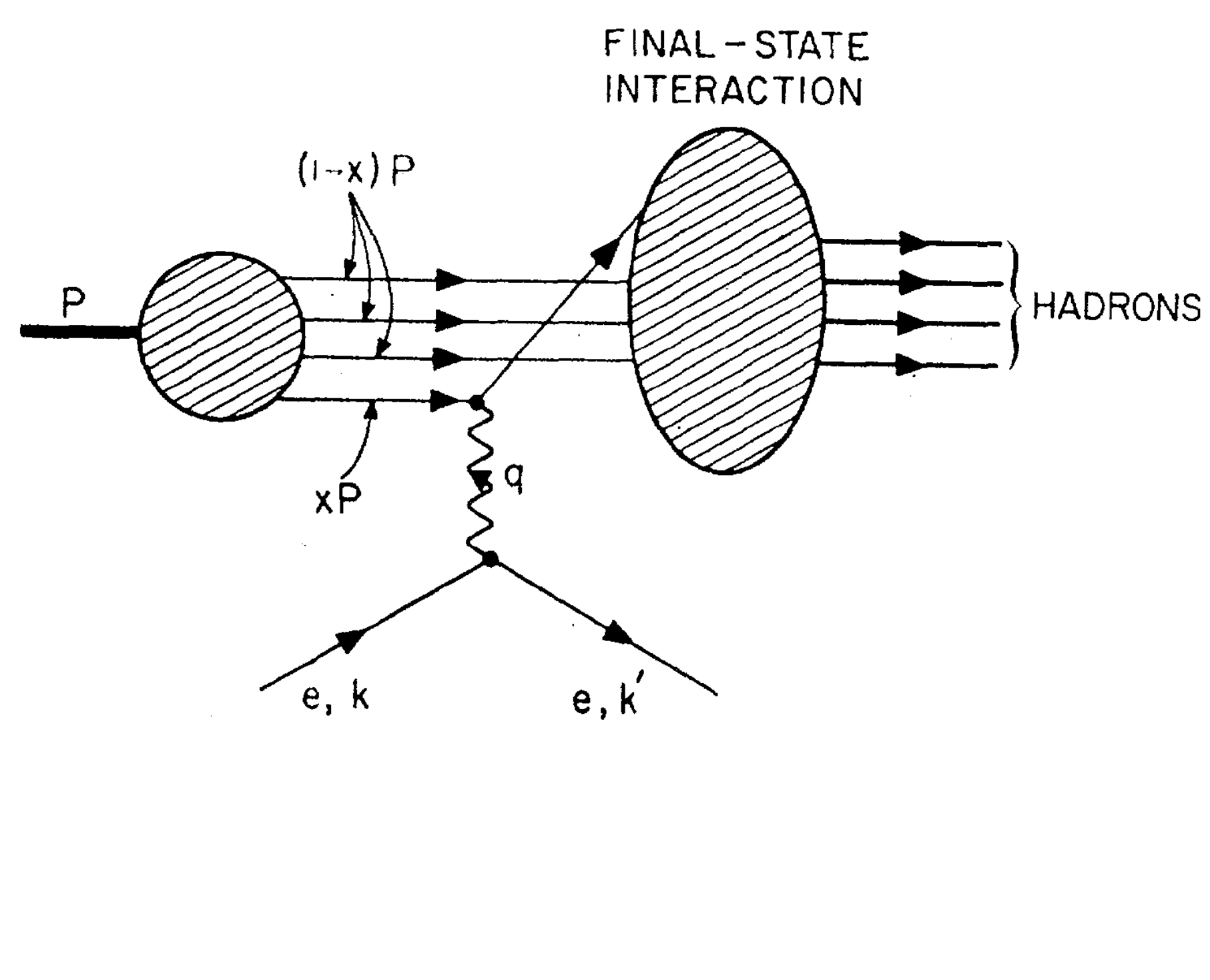} 
b)\hspace*{-0.00\linewidth}\includegraphics[width=0.29\linewidth]{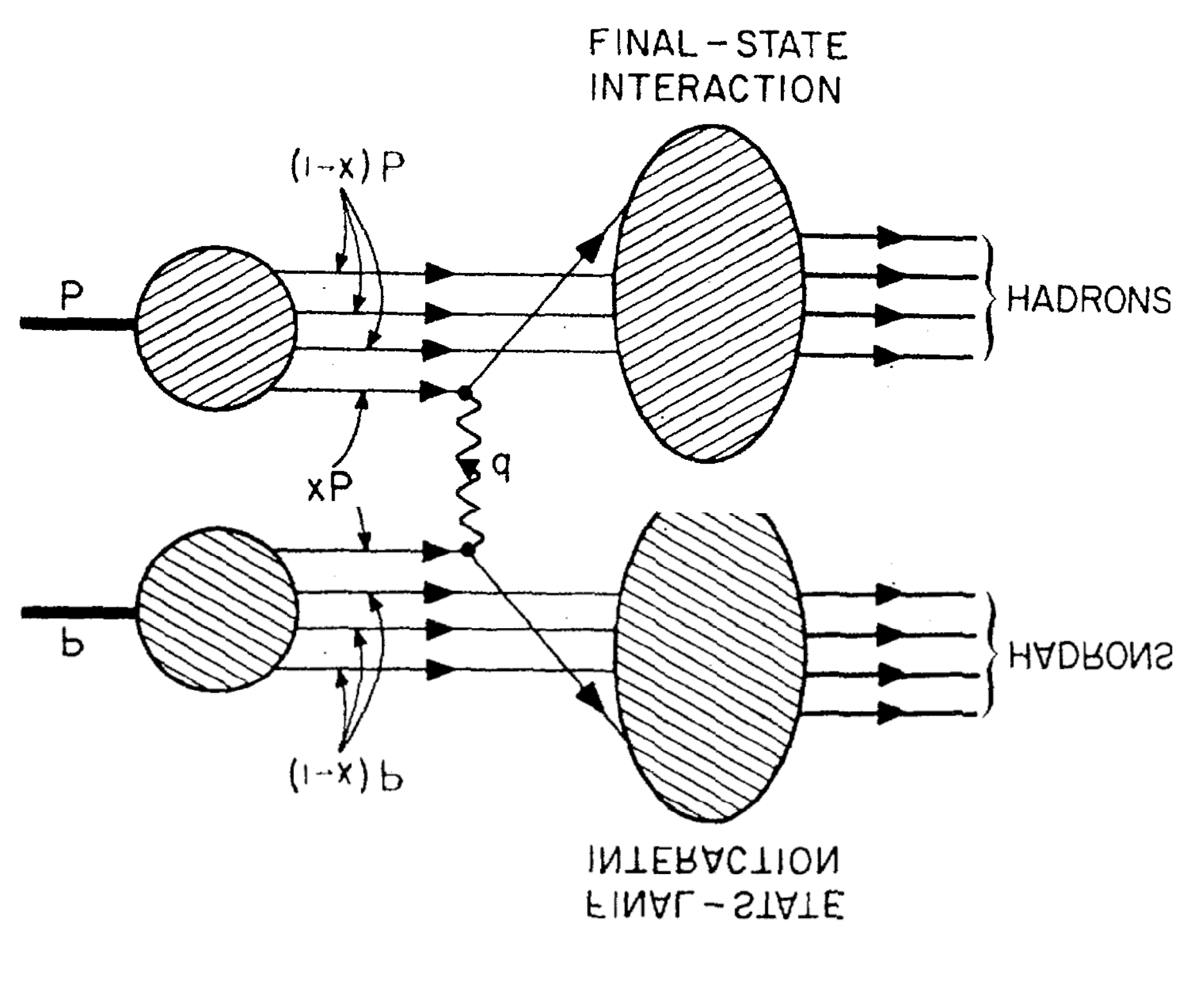}\hspace*{-0.01\linewidth} 
c)\hspace*{-0.00\linewidth}\includegraphics[width=0.33\linewidth]{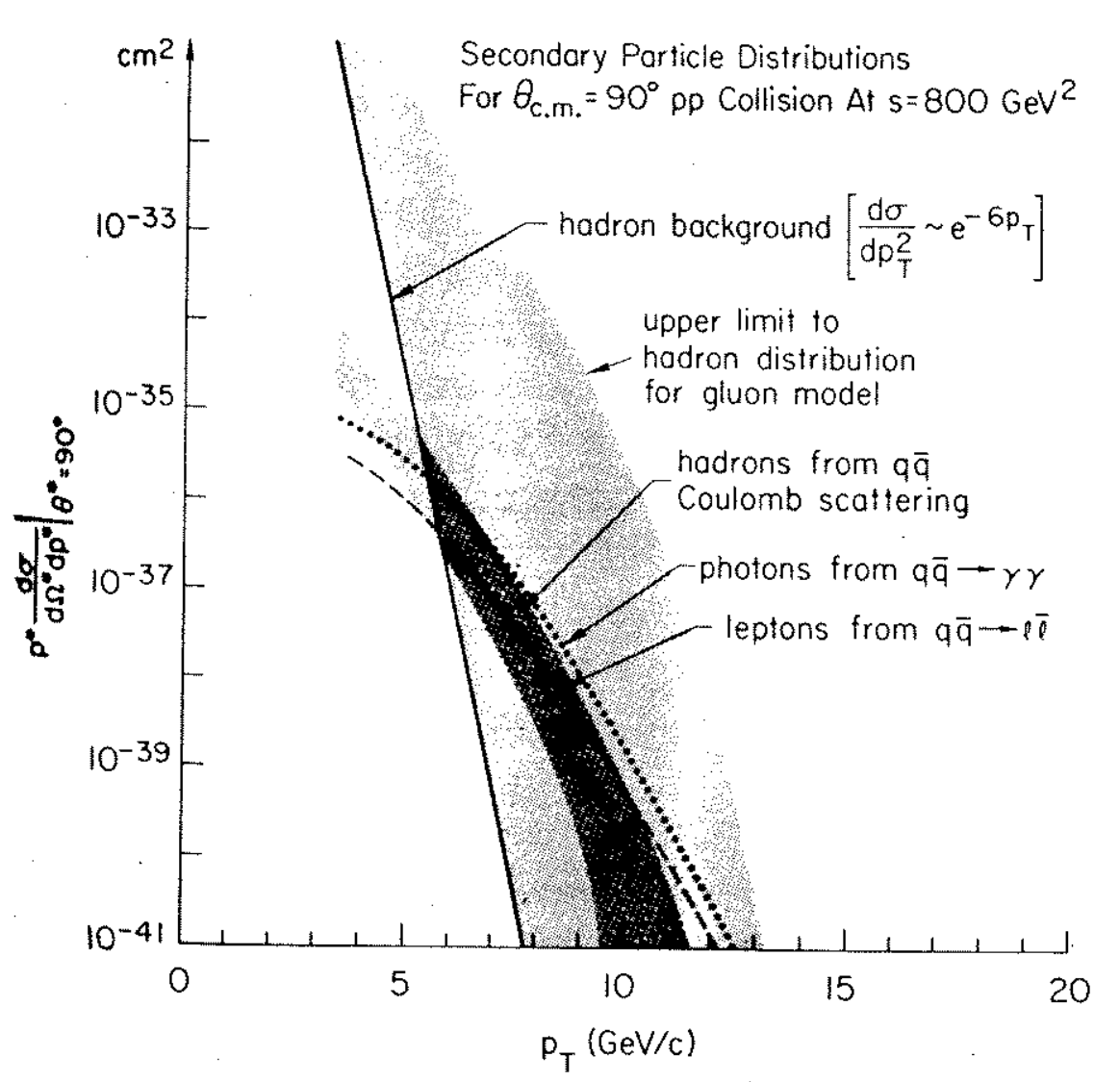}&
\end{tabular}
\end{center}
\caption[]
{a) Schematic e-p DIS via e-parton scattering~\cite{BjPaschos}  b) Schematic parton-parton scattering in p-p c) Predicted cross section according to reaction in (b).  
\label{fig:BBKpred} }
\end{figure}

 	   The idea of hard-scattering in p-p collisions dates from the first indication of pointlike structure inside the proton, in 1968, found at SLAC by deeply inelastic electron-proton scattering~\cite{DIS}, i.e. scattering with large values of 4-momentum transfer squared, $Q^2$, and energy loss, $\nu$. The discovery that the Deeply Inelastic Scattering (DIS) structure function 
\begin{equation}
F_2(Q^2, \nu)=F_2({Q^2 \over \nu})
\label{eq:F2scales}
\end{equation}
 ``scaled'' i.e just depended on the ratio 
\begin{equation}
x=\frac{Q^2}{2M\nu}
\label{eq:x_def}
\end{equation} 
independently of $Q^2$, as originally suggested by 
Bjorken~\cite{BjScaling}, led to the concept of a proton 
composed of point-like `partons'~\cite{BjPaschos}. The deeply inelastic scattering of an electron from a proton is simply quasi-elastic scattering of the electron from point-like partons of effective mass $Mx$, with quasi-elastic energy loss, $\nu=Q^2/2Mx$ (Fig.~\ref{fig:BBKpred}a). The probability for a parton to carry 
a fraction $x$ of the proton's momentum is measured by $F_2(x)/x$. 

	Since the partons of DIS are charged, and hence must scatter electromagnetically from each other in a p-p collision (Fig.~\ref{fig:BBKpred}b), Berman, Bjorken and Kogut (BBK)~\cite{BBK} and subsequent authors~\cite{CIM,CGKS}, derived a general formula for the cross section of the inclusive reaction  
\begin{equation}
 p + p\rightarrow C +X 
\label{eq:bbk1}
\end{equation} 
using the principle of factorization of the reaction into parton distribution functions for the protons, fragmentation functions to particle $C$ for the jet of particles from the scattered parton and a short-distance parton-parton hard scattering cross section. The invariant cross section for the inclusive reaction (Eq.~\ref{eq:bbk1}), where particle $C$ has transverse momentum $p_T$ near mid-rapidity was given by the general `scaling' form:~\cite{CIM}
\begin{equation}
E \frac{d^3\sigma}{dp^3}=\frac{1}{p_T^{n_{\rm eff}}} F({2 p_T \over \sqrt{s}})
= \frac{1}{\sqrt{s}^{\,n_{\rm eff}}} G({x_T})\quad \mbox{where}\quad x_T=2p_T/\sqrt{s} \qquad .
\label{eq:bbg}
\end{equation}
The cross section has 2 factors, a function $F$ $(G)$ which `scales', i.e. depends only on the ratio of momenta; and a dimensioned factor, ${p_T^{-n_{\rm eff}}}$   $(\sqrt{s}^{\,-n_{\rm eff}})$,   
where $n_{\rm eff}$ gives the form of the force-law 
between constituents. For QED or Vector Gluon exchange~\cite{BBK}, $n_{\rm eff}=4$, and for the case of quark-meson scattering by the exchange of a quark~\cite{CIM}, $n_{\rm eff}$=8.  Using this formalism, BBK predicted (Fig.~\ref{fig:BBKpred}c) hard-scattering must exist in p-p collisions since the charged partons of DIS must scatter electromagnetically, this ``electromagnetic contributions may be viewed as a lower bound on the real cross section at large $p_T$''.
\subsection{Discovery of high $p_T$ $\pi^0$ production in p-p collisions}
\begin{figure}[h]
\begin{center}
\includegraphics[width=0.45\linewidth]{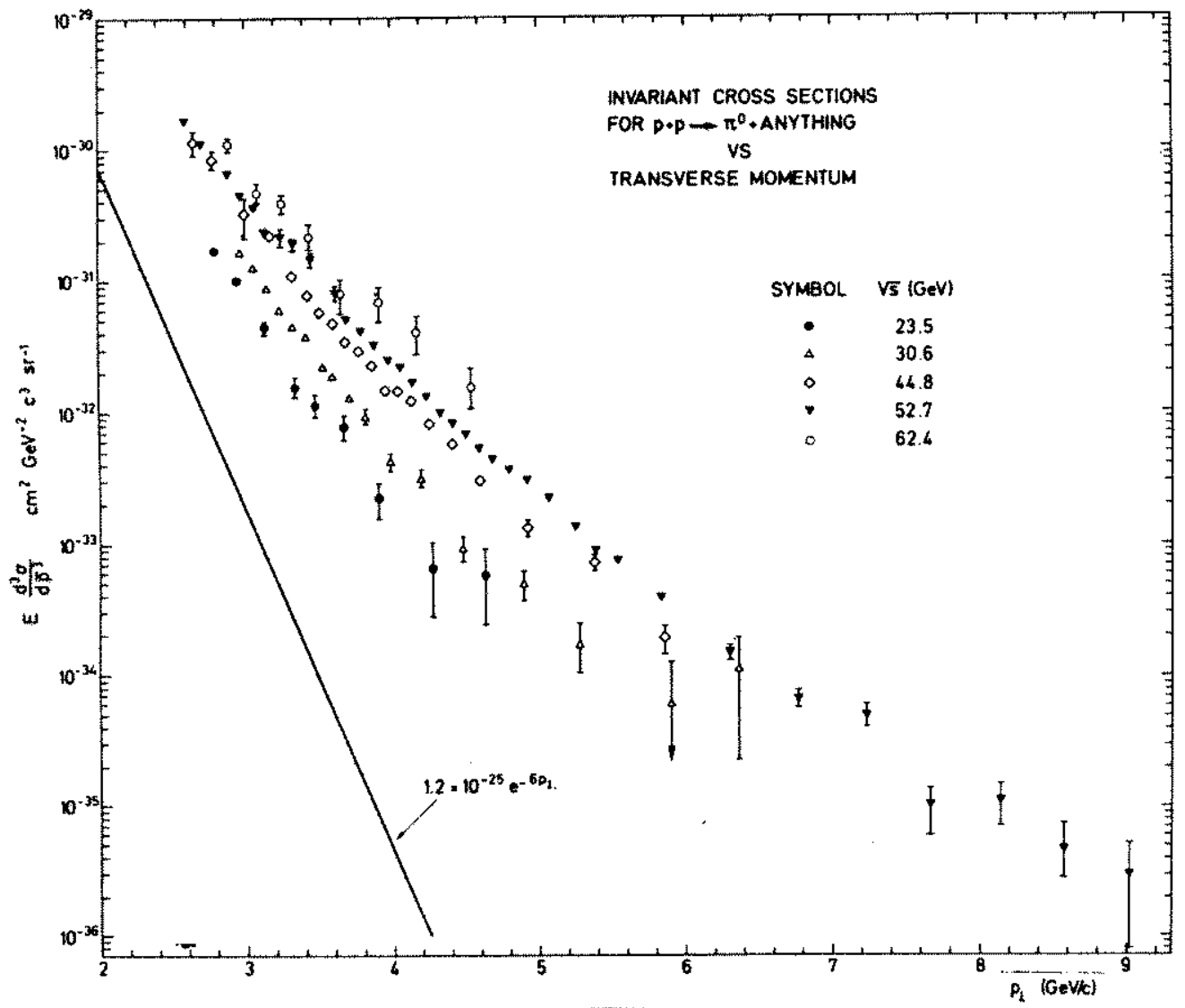}
\includegraphics[width=0.52\linewidth]{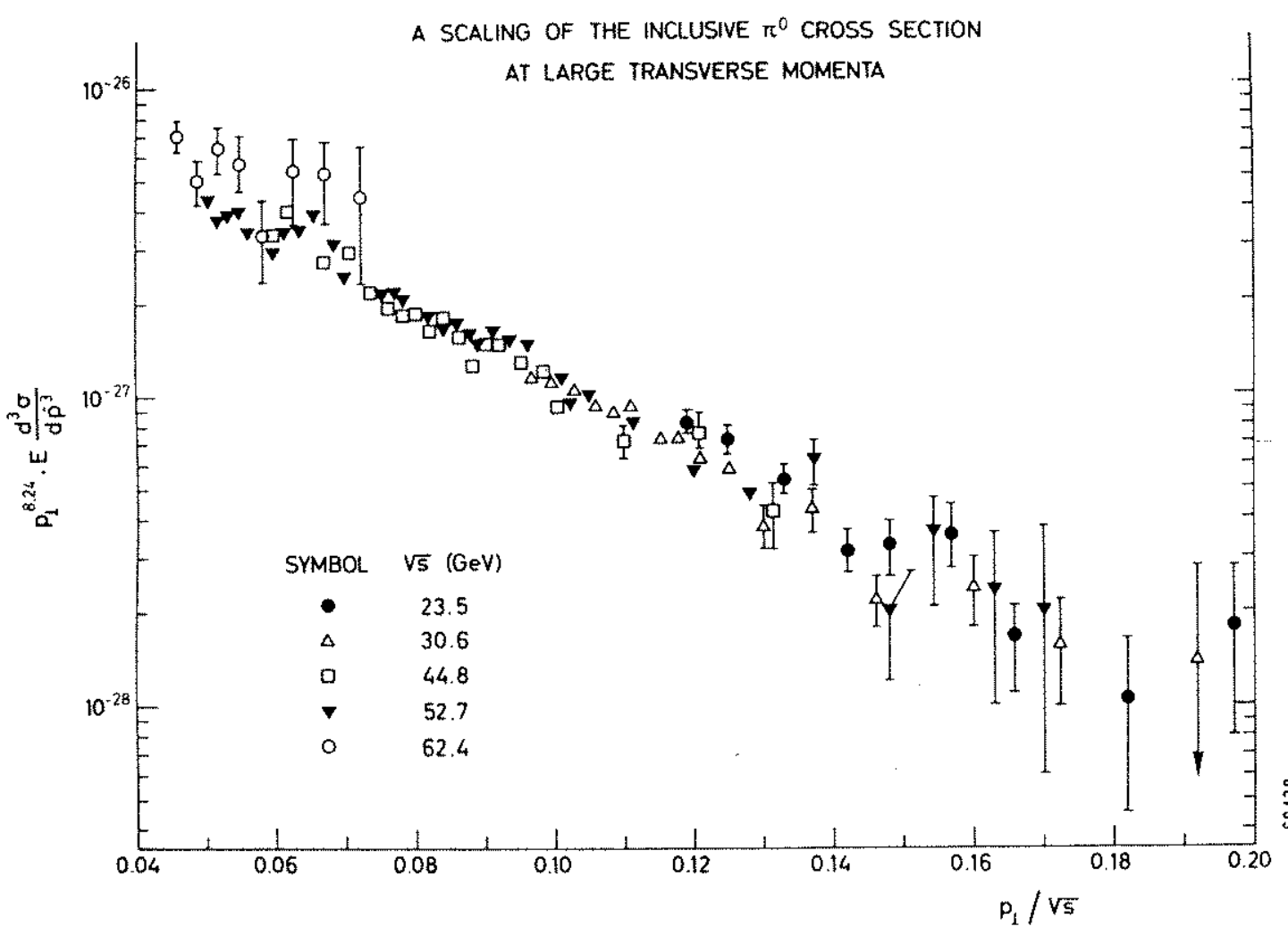}
\end{center}
\caption[]{a) (left) CCR~\cite{CCR} transverse momentum dependence of the invariant cross section at five center of mass energies. b) (right) The same data multiplied by $p_{\perp}^n$, using the best fit 
value of $n=8.24\pm0.05$, with $F=Ae^{-b x_{\perp}}$, plotted vs 
$p_{\perp}/\sqrt{s}$.   }
\label{fig:ccrpt}
\end{figure}

The CERN Columbia Rockefeller (CCR) Collaboration~\cite{CCR} 
(and also the Saclay Strasbourg~\cite{SS} and British Scandinavian~\cite{BS} 
collaborations) at the CERN-ISR measured 
pion production over a large range of transverse momenta, unavailable in cosmic ray studies or at lower $\sqrt{s}$. 
The $e^{-6p_T}$ soft-spectrum at low $p_T$ which depends very little on $\sqrt{s}$ breaks to a power law at high $p_T$ with a strong and characteristic $\sqrt{s}$ dependence (Fig.~\ref{fig:ccrpt}a). 
The large rate at high $p_T$, much larger than expected in Figs.~\ref{fig:E70}b and \ref{fig:BBKpred}c, indicates that {\em partons interact strongly ($\gg$ EM) with each other}.  
However, the QED or Vector Gluon form of BBK~\cite{BBK}, $p_{T}^{-4} F(p_{T}/\sqrt{s})$, was 
not observed in the experiment but scaling with $n_{\rm eff}\simeq 8$ was observed (Fig.~\ref{fig:ccrpt}b), which spawned a whole host of new theories.~\cite{CIM}.  The first application of {\QCD} to wide angle hadron collisions~\cite{CGKS}, showed that pure $x_T$ scaling breaks down: $n_{\rm eff}$ varies according to the $x_T$ and $\sqrt{s}$ regions used in the comparison, $n_{\rm eff}\rightarrow n_{\rm eff}(x_T, \sqrt{s})$; but this was not enough to resolve $n_{\rm eff}\simeq 8$ issue, which turned out to be due to ``$p_T$ broadening'' by initial state parton transverse momentum, the ``$k_T$ effect.''~\cite{egseeMJTCFRNC06}.
\subsection{Everything you want to know about jets can be found using 2-particle correlations.}
 \label{sec:almost}
   Following the discovery of hard-scattering in p-p collisions at the CERN-ISR by single particle inclusive measurements, the attention of experimenters turned to measuring the predicted di-jet structure of the hard-scattering events using two-particle correlations. 
At the CERN-ISR, from 1975--1982, two-particle correlations showed unambiguously that high $p_T$ particles come from di-jets.
   The outgoing jet-pair of hard-scattering obeys the kinematics of elastic-scattering (of partons) in a parton-parton c.m. frame which is longitudinally moving with rapidity $y=(1/2) \ln(x_1/x_2)$ in the p-p c.m. frame. Hence, the jet-pair formed from the scattered partons should be co-planar with the beam axis, with two jets of equal and opposite transverse momentum. Thus, the outgoing jet-pair should be back-to-back in azimuthal projection. It is not necessary to fully reconstruct the jets in order to measure their properties. In many cases two-particle correlations are sufficient to measure the desired properties. Many ISR experiments provided excellent 2-particle correlation measurements~\cite{Moriond79}. However, the CCOR experiment~\cite{Angelis79} was the first to provide charged particle measurement with full and uniform acceptance over the entire azimuth, with pseudorapidity coverage $-0.7\leq \eta\leq 0.7$, so that the jet structure of high $p_T$ scattering could be easily seen and measured.
        \begin{figure}[ht]
\begin{center}
\hspace*{-1pc}\includegraphics[width=0.38\linewidth]{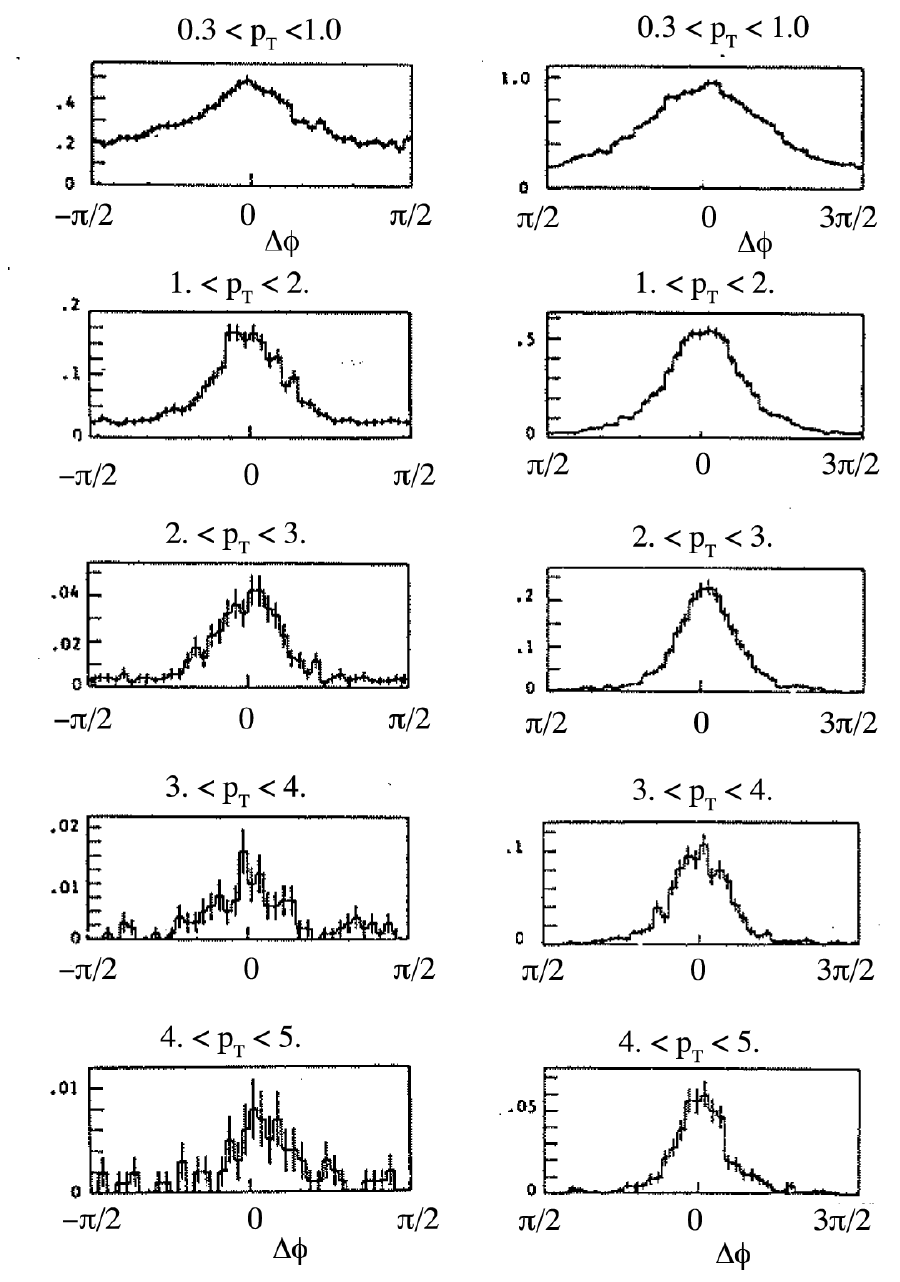}
\includegraphics[width=0.36\linewidth]{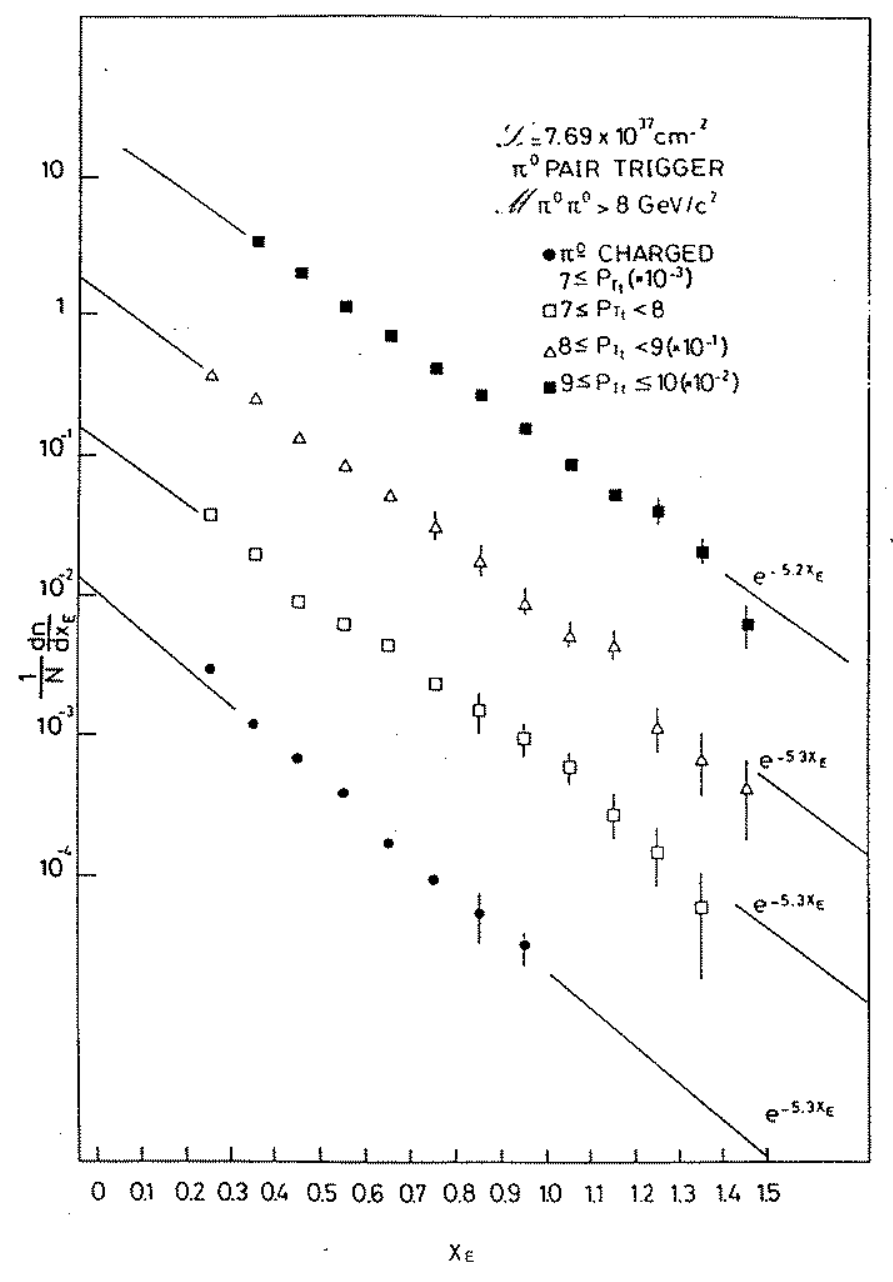}
\begin{tabular}[b]{c}
\includegraphics[width=0.24\linewidth]{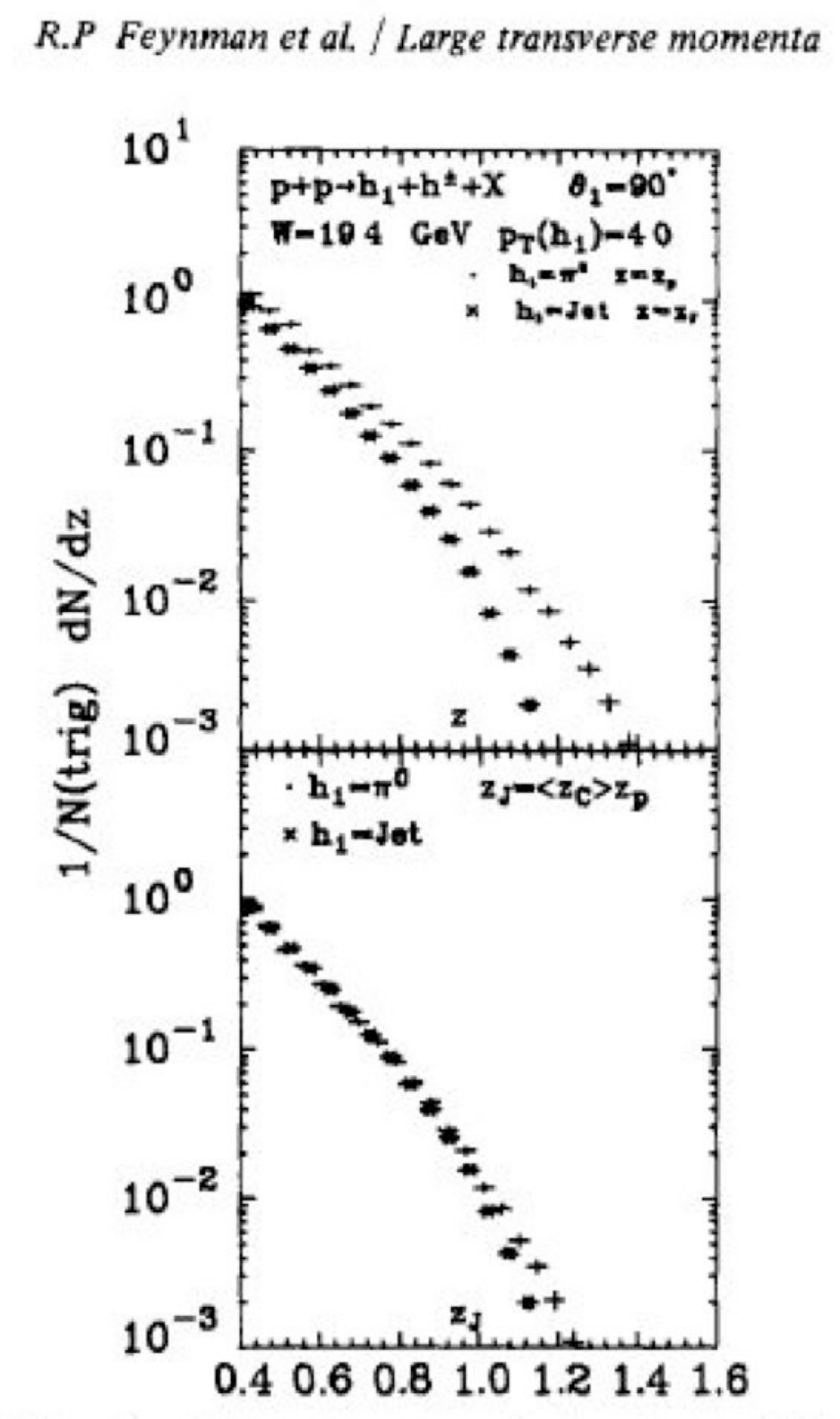}\cr
\hspace*{2pc}
\end{tabular}
\end{center}
\vspace*{-0.12in}
\caption[]{a) (left) trigger-side ($\Delta\phi=\pm \pi/2$) b) (left-center) away-side ($\Delta\phi=\pm \pi/2$ around $\pi$ rad.)  correlations of charged particles with indicated $p_T\equiv p_{T_a}$ for $\pi^0$ triggers with $p_{Tt}\geq 7$ GeV/c, for 5 intervals of $p_{T}$; c) (right-center) $x_E$ distributions from this data. d) right) [top] Comparison~\cite{FFF} of away side charged hadron distribution triggered by a $\pi^0$ or a jet, where $z_{\pi^0}=x_E$ and $z_j=p_{T_a}/\hat{p}_{T_a}$. [bottom] same distributions with $\pi^0$ plotted vs $z'_j=\mean{z_t} x_E$.    \label{fig:mjt-ccorazi}}\vspace*{-0.15in}
\end{figure}

The azimuthal distributions of associated charged particles 
relative to a $\pi^0$ trigger with transverse momentum $p_{Tt} > 7$ GeV/c are shown (Fig.~\ref{fig:mjt-ccorazi}a,b) for five intervals of associated particle transverse momentum $p_T$. In all cases, strong correlation peaks on flat backgrounds are clearly visible, indicating the di-jet structure which is contained in an interval $\Delta\phi=\pm 60^\circ$ about a direction towards and opposite to the trigger for all values of associated $p_T\, (>0.3$ GeV/c) shown.  The small variation of the widths of the away-side peaks for $p_{T}>1$ GeV/c (Fig.~\ref{fig:mjt-ccorazi}b) indicates out-of-plane activity of the di-jet system beyond simple jet fragmentation (the $k_T$ effect~\cite{egseeMJTCFRNC06}).  

Following the methods of previous CERN-ISR experiments~\cite{DarriulatNPB107,CCHK} and the best theoretical guidance~\cite{FFF}, the away jet azimuthal angular distributions  of Fig.~\ref{fig:mjt-ccorazi}b, which were thought to be unbiased, were analyzed in terms of the two variables: $p_{\rm out}=p_T \sin(\Delta\phi)$, the out-of-plane transverse momentum of a track,
 and $x_E$:\\ 
\hspace*{0.05\linewidth}\begin{minipage}[b]{0.45\linewidth}
\[ 
x_E=\frac{-\vec{p}_T\cdot \vec{p}_{T_t}}{|p_{T_t}|^2}=\frac{-p_T \cos(\Delta\phi)}{p_{T_t}}\simeq \frac {z}{z_{t}}  
\]
\vspace*{0.001in}
\end{minipage}
\hspace*{0.02\linewidth}
\begin{minipage}[b]{0.39\linewidth} 
\vspace*{0.003in}
\includegraphics[width=\linewidth]{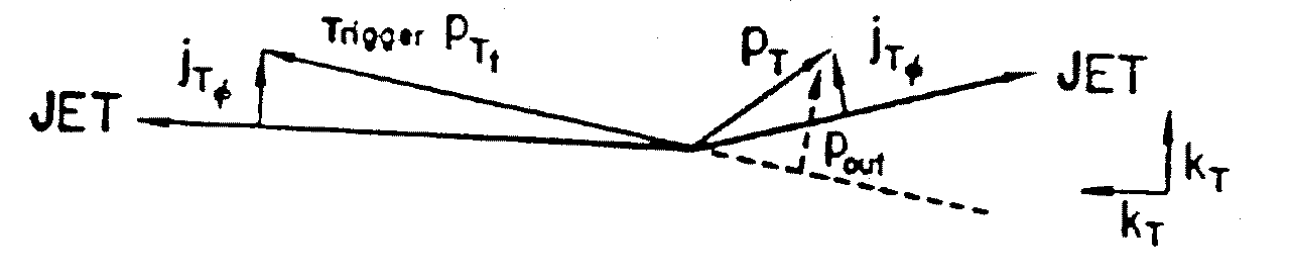}
\vspace*{-0.10in}
\label{fig:mjt-poutxe}
\end{minipage}
\vspace*{-0.12in}

\noindent where $z_{t}\simeq p_{T_t}/\hat{p}_{T_t}$ is the fragmentation variable of the trigger jet with $\hat{p}_{T_t}$,  and $z_a\simeq p_{T_a}/\hat{p}_{T_a}$ is the fragmentation variable of the away  jet (with $\hat{p}_{T_a}$).    
Note that $x_E$ would equal the fragmenation fraction $z$ of the away jet, for $z_{t}\rightarrow 1$, if the trigger and away jets balanced transverse momentum, i.e. if $\hat{x}_h\equiv\hat{p}_{T_a}/\hat{p}_{T_t}=1$. 
It was generally assumed, following the seminal article of Feynman, Field and Fox~\cite{FFF}, that the $p_{T_a}$ distribution of away side hadrons from a single particle trigger [with $p_{T_t}$], corrected for $\mean{z_t}$, would be the same as that from a jet-trigger and follow the same fragmentation function as observed in $e^+ e^-$  or DIS~\cite{Darriulat}. The $x_E$ distributions~\cite{Angelis79,JacobEPS79} for the data of Fig.~\ref{fig:mjt-ccorazi}b are shown in Fig.~\ref{fig:mjt-ccorazi}c and show the fragmentation behavior expected at the time, $e^{-6z}\sim e^{-6 x_E \langle z_{t}\rangle}$. Also, relevant to a recent claim~\cite{CDFPRD79}, Fig.~\ref{fig:mjt-ccorazi}c from 1979 showed that there were no di-jets each of a single particle, as claimed by another ISR experiment~\cite{JacobEPS79b}, since there is no peak at $x_E=1$. 

    As noted previously\footnotemark[1], the title of this section should start with ``Almost'' because we subsequently learned at RHIC~\cite{ppg029} that the $x_E$ distribution from di-hadrons (Fig~\ref{fig:mjt-ccorazi}c) does not measure the fragmentation function (see Sec.~\ref{sec:nofrag} below). 
\subsection{Direct searches for Jets 1975-1982: false claims, confusion, skepticism; finally success.}
\begin{figure}[ht]
\begin{center}
\includegraphics[width=0.30\linewidth]{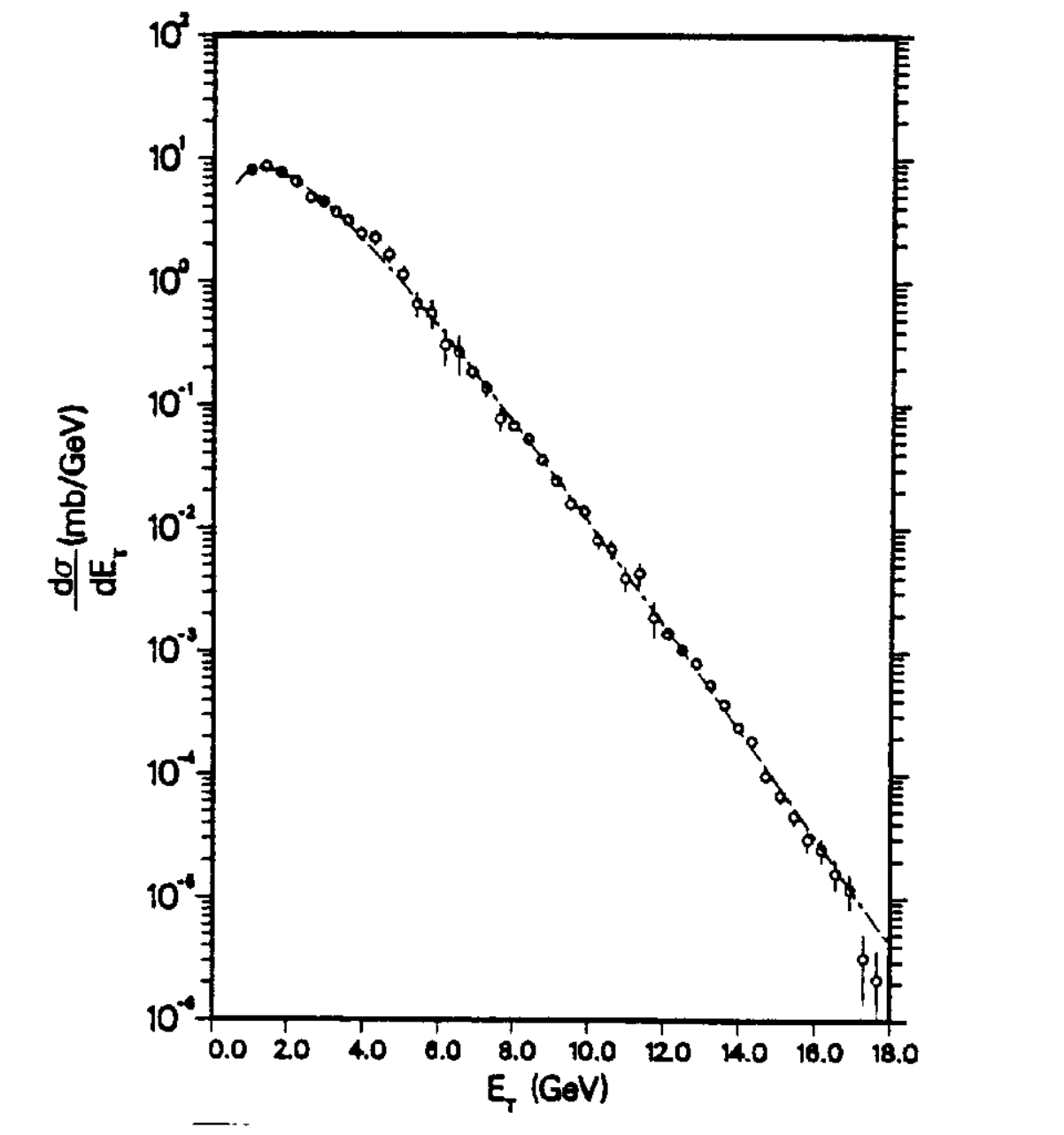}
\includegraphics[width=0.60\linewidth]{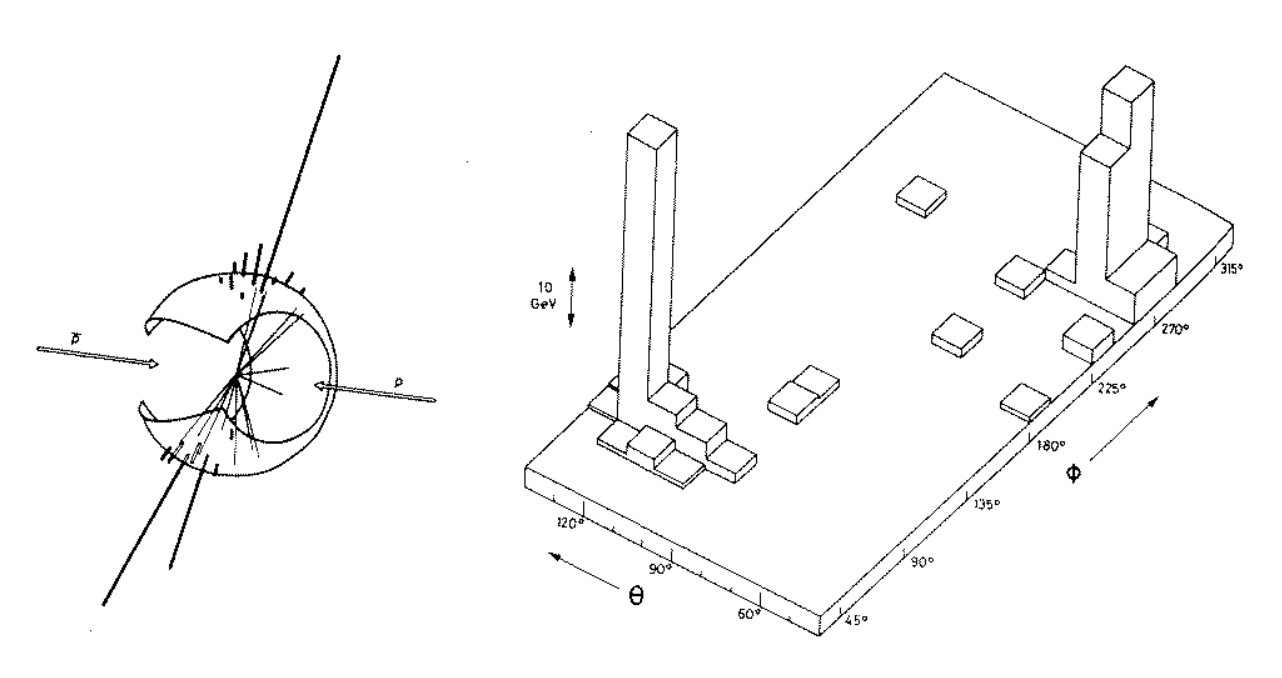} 
\end{center}
\caption[]
{ a)(left) NA5 structureless $E_T$ distribution~\cite{NA5}. b,c) UA2 jet event from 1982 ICHEP~\cite{Paris82}: b)(center) event shown in geometry of detector; c)(right) ``Lego" plot of energy in calorimeter cell as a function of angular position of cell in polar ($\theta$) and azimuthal ($\Phi$) angle space.  

\label{fig:UA2jet} }
\end{figure}

\begin{figure}[ht]
\begin{center}
\begin{tabular}{cc}
\hspace*{-0.1in}\includegraphics[width=0.75\linewidth]{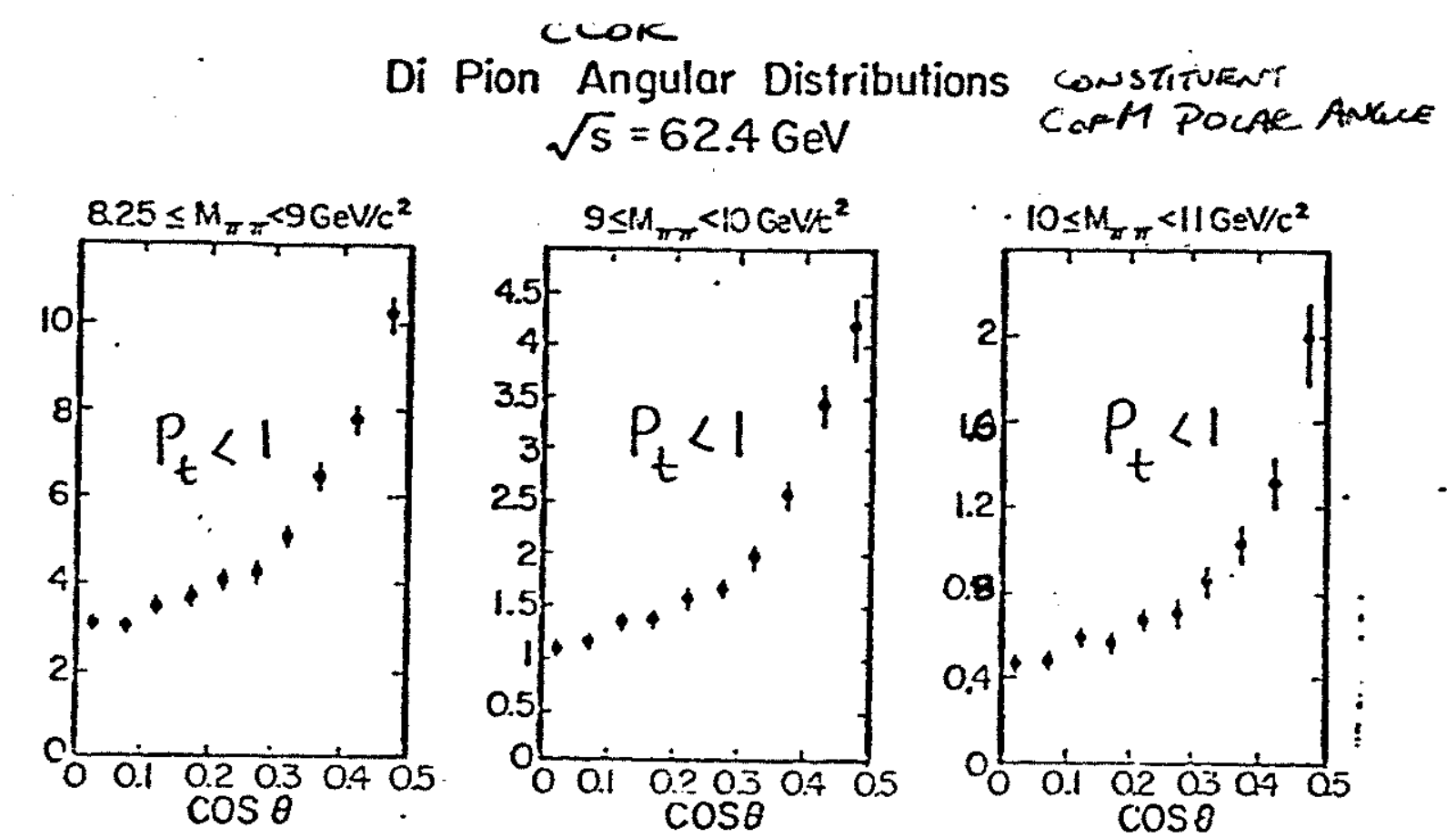} &
\hspace*{-0.35in}\includegraphics[width=0.288\linewidth]{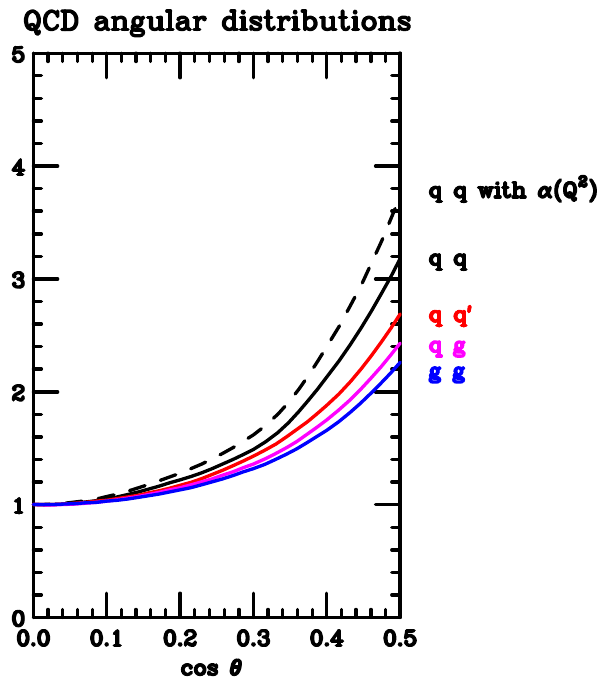}
\end{tabular}
\end{center}
\caption[]
{a) (left 3 panels) CCOR measurement~\cite{Paris82,CCOR82NPB} of polar angular distributions of $\pi^0$ pairs with net $p_T < 1$ GeV/c at mid-rapidity in p-p collisions with $\sqrt{s}=62.4$ GeV for 3 different values of $\pi\pi$ invariant mass $M_{\pi \pi}$. b) (rightmost panel) QCD predictions for $\Sigma^{ab}(\cos\theta^*)$ for the elastic scattering of $gg$, $qg$, $qq'$, $qq$, and $qq$ with $\alpha_s(Q^2)$ evolution.    
\label{fig:mjt-ccorqq} }
\end{figure}

 One of the original reasons for measuring jets in p-p collisions was that it was assumed that the jet cross section would measure the parton cross section without reference to the  fragmentation functions. 
 However, due to the fact (which was unknown in the 1970's) that jets in $4\pi$ calorimeters at ISR 
energies or lower are invisible below $\sqrt{\hat{s}}\sim E_T \leq 25$ 
GeV~\cite{Gordon}, there were many false claims of jet observation in the period 1977-1982~\cite{MJTIJMPA}.  The coup-de-gr\^ace to the concept of finding jet structures in large aperture calorimeters was provided by the first measurement of an $E_T$ distribution in the present usage of the terminology (see Sec.~\ref{section:soft} above) by the NA5 experiment at CERN (Fig.~\ref{fig:UA2jet}a)~\cite{NA5} who found that ``The events selected by the full calorimeter trigger show no
dominant jet structure. They appear to originate from processes other than two constituent scattering.'' 
This led to mass confusion in the hard-scattering community in the period 1980-1982 and 
to great skepticism about jets in hadron collisions, particularly in the USA.
A `phase change' in belief-in-jets was produced by one UA2 event 
at the 1982 ICHEP in Paris~\cite{Paris82}, which, together with the first direct measurement of the QCD constituent-scattering angular distribution, $\Sigma^{ab}(\cos\theta^*)$ (Eq.~\ref{eq:mjt-QCDabscat}), using two-particle correlations, presented at the same meeting (Fig.~\ref{fig:mjt-ccorqq})~\cite{Paris82,CCOR82NPB}, gave universal credibility to the pQCD description of high $p_T$ hadron physics~\cite{OwensRMP,Darriulat,DiLella}. Since that time, {\QCD} and jets have become the standard tools of High Energy Physics.

\begin{figure}[!bht]
\begin{center}
\begin{tabular}{cc}
\includegraphics[width=0.46\linewidth]{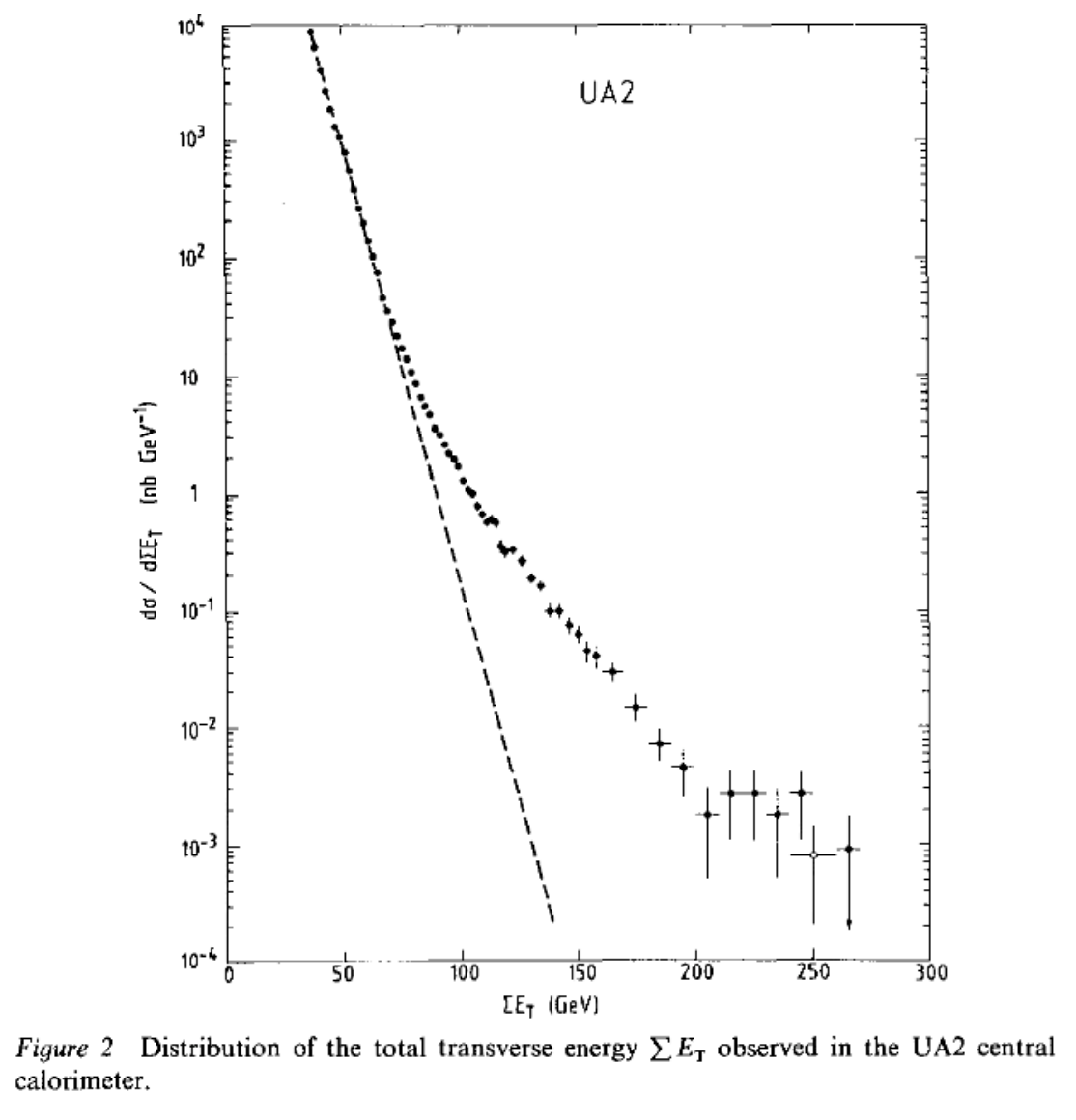}&
\includegraphics[width=0.39\linewidth,angle=+1]{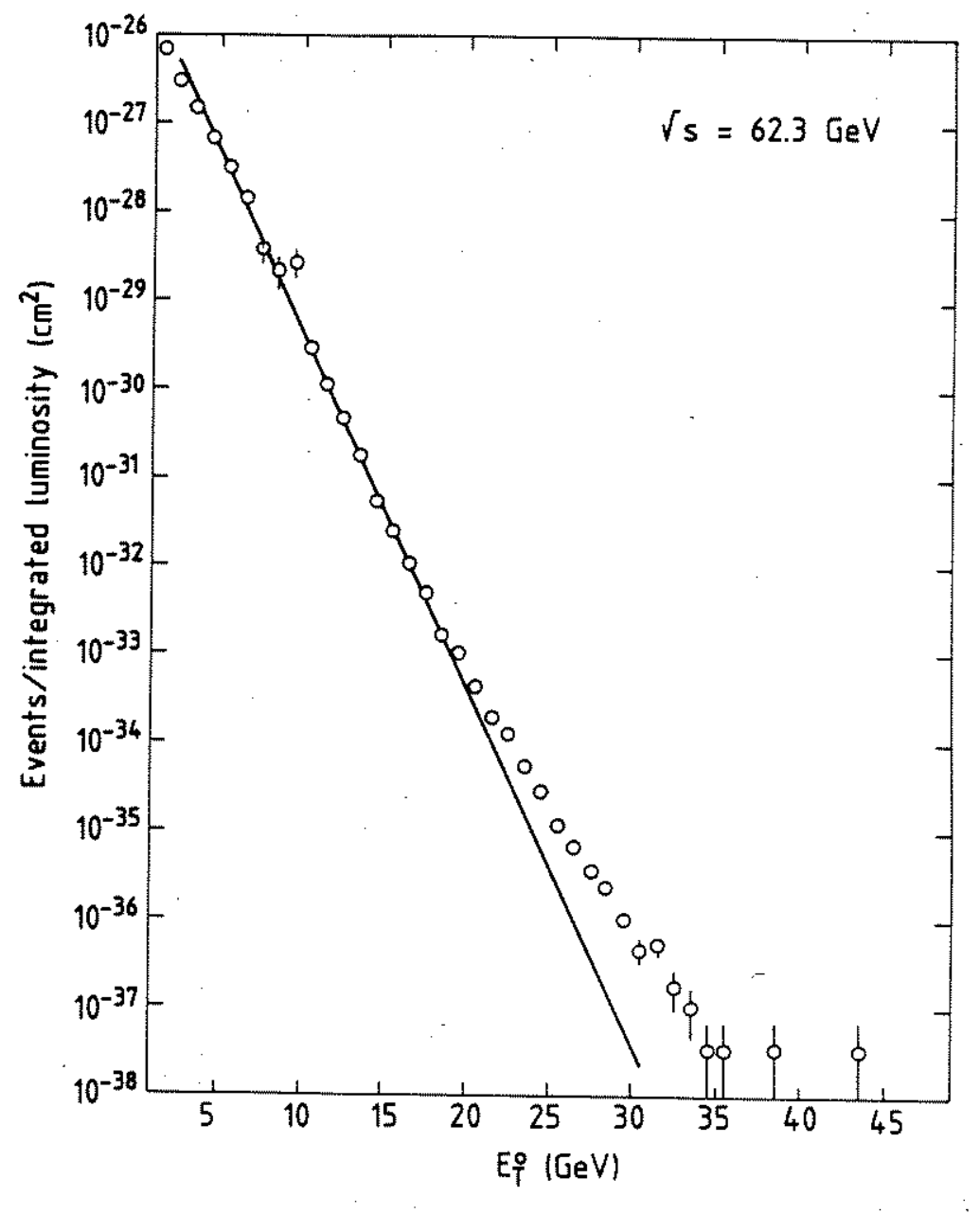}
\end{tabular}
\end{center}\vspace*{-0.25in}
\caption[]{$E_T$ measurements by: a)(left) UA2 in p-$\bar{\rm p}$ collisions at $\sqrt{s}=540$ GeV~\cite{UA2ET}; b)(right) CCOR in p-p collisions at $\sqrt{s}=62.3$ GeV~\cite{COR-ETpp}. The lines indicate extrapolation of the nearly exponential spectrum at lower $E_T$.}
\label{fig:ETdists}
\end{figure}

As a final note, recall that hard-scattering is visible by a break in the exponential single particle inclusive spectrum after roughly 3 orders of magnitude in cross section (Fig.~\ref{fig:PXpi0pp2}). However, finding jet structure in $E_T$ distributions measured in $4\pi$ calorimeters is much more difficult because the jets are only visible by a break in the $E_T$ spectrum after 5-7 orders of magnitude of cross section (Fig.~\ref{fig:ETdists})~\cite{UA2ET,COR-ETpp}, since the $E_T$ spectrum is even more dominated by soft physics than the single particle spectrum.   
\subsection{Status of theory and experiment, circa 1982}
 
Hard-scattering was visible both at ISR and FNAL (Fixed Target) energies 
via inclusive single particle production at large $p_T\geq$ 2-3 
GeV/c. Scaling and dimensional arguments for plotting 
data revealed the systematics and underlying physics. The theorists had the 
basic underlying physics correct; but many important details remained to 
be worked out, several by experiment. The transverse momentum 
imbalance of outgoing parton-pairs, the ``$k_T$-effect", was 
discovered by experiment~\cite{CCHK,MJT79}, and clarified by Feynman, Field and Fox~\cite{FFF}. The first modern QCD calculation and 
prediction for high $p_T$ single particle inclusive cross sections, including 
non-scaling and the $k_T$ effect was done in 1978, by Jeff 
Owens and collaborators~\cite{Owens78} closely followed and corroborated by Feynman and collaborators~\cite{FFF78}, under the assumption that high $p_T$ particles  
are produced from states with two roughly back-to-back jets
which are the result of scattering of constituents of the nucleons (partons). 
   The overall $p+p$ hard-scattering cross section in ``leading logarithm" pQCD   
is the sum over parton reactions $a+b\rightarrow c +d$ 
(e.g. $g+q\rightarrow g+q$) at parton-parton center-of-mass (c.m.) energy $\sqrt{\hat{s}}=\sqrt{x_1 x_2 s}$.  
\begin{equation}
\frac{d^3\sigma}{dx_1 dx_2 d\cos\theta^*}=
\frac{s d^3\sigma}{d\hat{s} d\hat{y} d\cos\theta^*}=
\frac{1}{s}\sum_{ab} f_a(x_1) f_b(x_2) 
\frac{\pi\alpha_s^2(Q^2)}{2x_1 x_2} \Sigma^{ab}(\cos\theta^*)
\label{eq:mjt-QCDabscat}
\end{equation} 
where $f_a(x_1)$, $f_b(x_2)$, are parton distribution functions, 
the differential probabilities for partons
$a$ and $b$ to carry momentum fractions $x_1$ and $x_2$ of their respective 
protons (e.g. $u(x_2)$), and where $\theta^*$ is the scattering angle in the parton-parton c.m. system. The parton-parton c.m. energy squared is $\hat{s}=x_1 x_2 s$,
where $\sqrt{s}$ is the c.m. energy of the p-p collision. The parton-parton 
c.m. system moves with rapidity $\hat{y}=(1/2) \ln (x_1/x_2)$ in the p-p c.m. system. Equation~\ref{eq:mjt-QCDabscat} gives the $p_T$ spectrum of outgoing parton $c$, which then
fragments into a jet of hadrons, including e.g. $\pi^0$.  The fragmentation function
$D^{\pi^0}_{c}(z)$ is the probability for a $\pi^0$ to carry a fraction
$z=p^{\pi^0}/p^{c}$ of the momentum of outgoing parton $c$. Equation~\ref{eq:mjt-QCDabscat}
must be summed over all subprocesses leading to a $\pi^0$ in the final state weighted by their respective fragmentation functions. In this formulation, $f_a(x_1)$, $f_b(x_2)$ and $D^{\pi^0}_c (z)$ 
represent the ``long-distance phenomena'' to be determined by experiment;
while the characteristic subprocess angular distributions,
{\bf $\Sigma^{ab}(\cos\theta^*)$},
and the coupling constant,
$\alpha_s(Q^2)=\frac{12\pi}{25} \ln(Q^2/\Lambda^2)$,
are fundamental predictions of QCD~\cite{CutlerSivers,Combridge:1977dm}.

\section{Hard-Scattering at RHIC}
      For the past decade the single and two-particle techniques developed in the period 1972--1982, largely at the CERN-ISR were used exclusively at RHIC to study hard-scattering in p-p A+A collisions, with outstanding results. 
      \subsection{Single particle (semi-) inclusive measurements}
\begin{figure}[h]
\begin{center}
a)\includegraphics[width=0.47\linewidth]{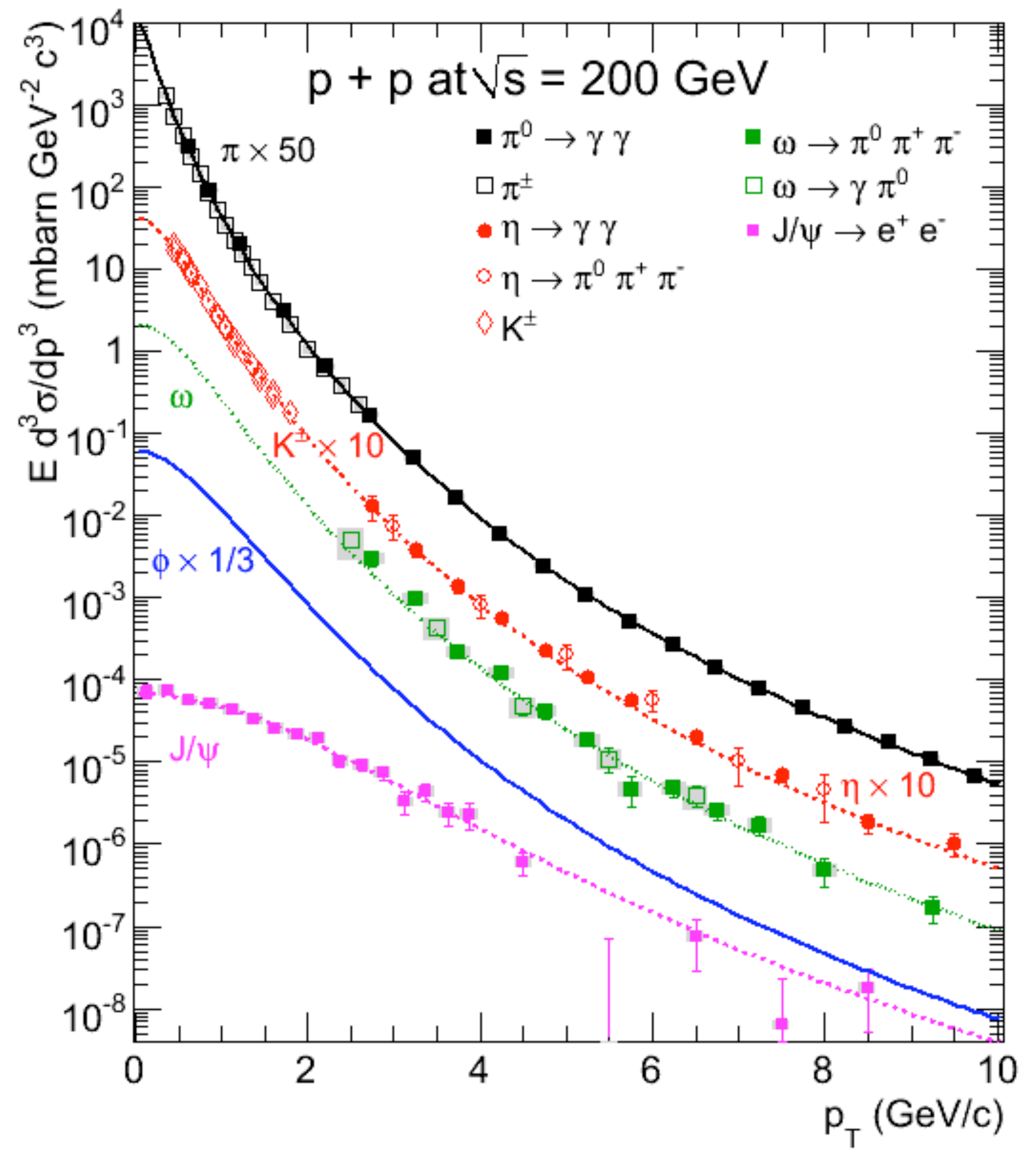}
b)\includegraphics[width=0.45\linewidth]{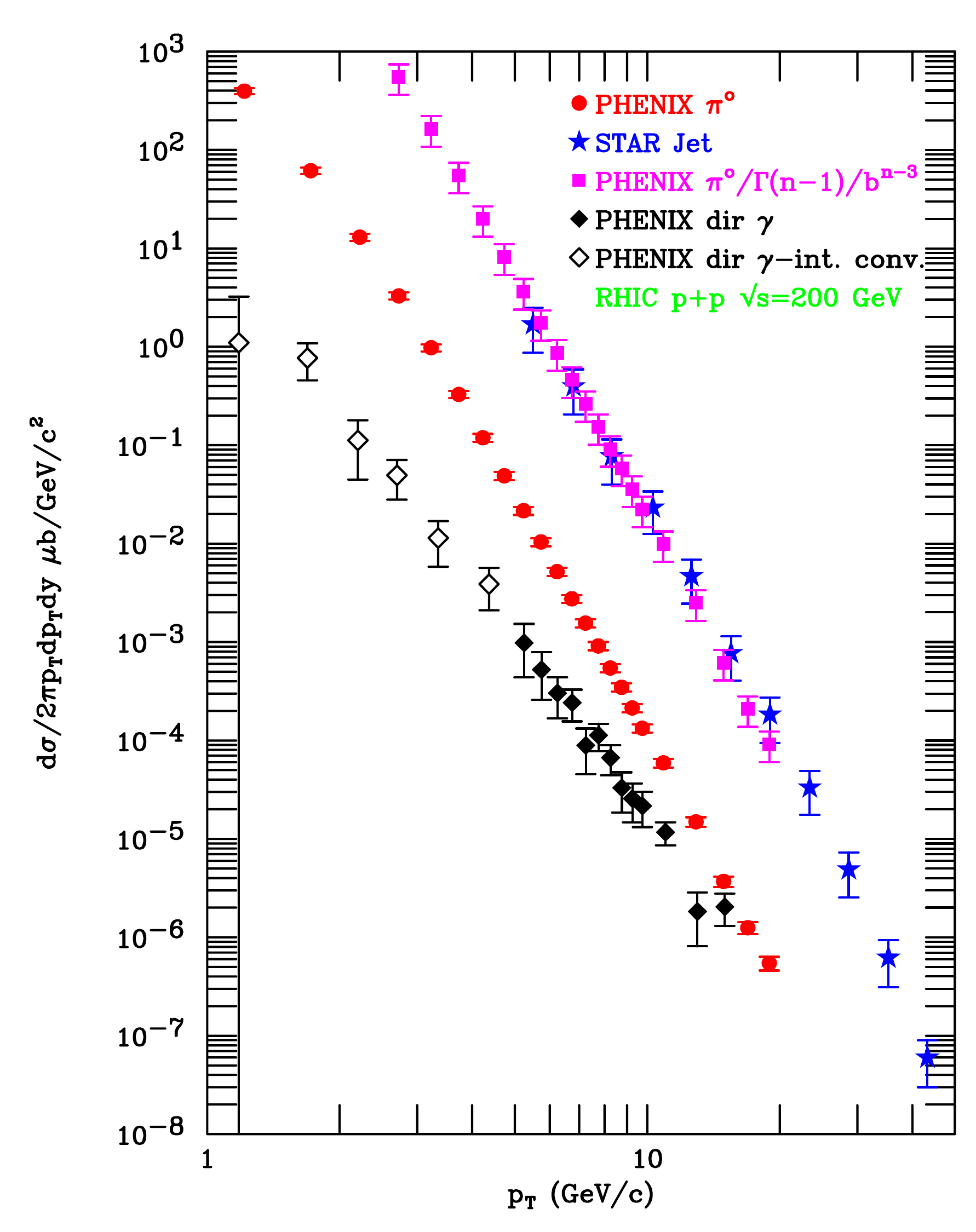}
\end{center}
\caption[]{PHENIX $p_T$ invariant cross section at mid-rapidity in p-p collisions at $\sqrt{s}=200$ GeV. a) Compilation of mesons needed for understanding leptonic background~\cite{PXPLB670} b) Direct-$\gamma$~\cite{PXintconv,PXdirg},$\pi^0$~\cite{ppg063} and STAR Jets~\cite{STARjetPRL97} for hard-scattering studies.}
\label{fig:PXpp}
\end{figure}

      The p-p single particle inclusive $p_T$ spectra measurements in Fig.~\ref{fig:PXpp} clearly illustrate the importance of particle identification since all the different particles have different systematic behavior. PHENIX concentrates on mesons and photons since they provide the main backgrounds to lepton and photon searches for which we built this specialized detector, while STAR concentrates on hyperons. Also, inclusive single  particle measurements have very precise $\sim 1\%$ $p_T$ scales in contrast to jets. As shown in Fig.~\ref{fig:PXpp}b for p-p collisions, jets are excellent if you can do it because they give the best rates. My favorite jet measurement is a search for parity violation in 100 GeV jets to look for quark substructure~\cite{MJTPSU}. Fig.~\ref{fig:PXpp}a, on a semi-log scale, emphasizes measurements at the lowest $p_T$ including the soft-physics range and the transition to hard-scattering, while the log-log scale on Fig.~\ref{fig:PXpp}b emphasizes the power-law behavior at high $p_T$. It is interesting to note that the $\pi^0$ and jet spectra are parallel to each other (the same $p_T^n$ dependence, see Eq.~\ref{eq:pioverjet} below), while the direct $\gamma$ are flatter, corresponding to an $n_{\rm eff}$ closer to 4 on an $x_T$ scaling plot. There should be a simple explanation, but I don't know it.  
\subsubsection{High $p_T$ suppression in Au+Au collisions}
\begin{figure}[h]
\begin{center}
\includegraphics[width=0.70\linewidth]{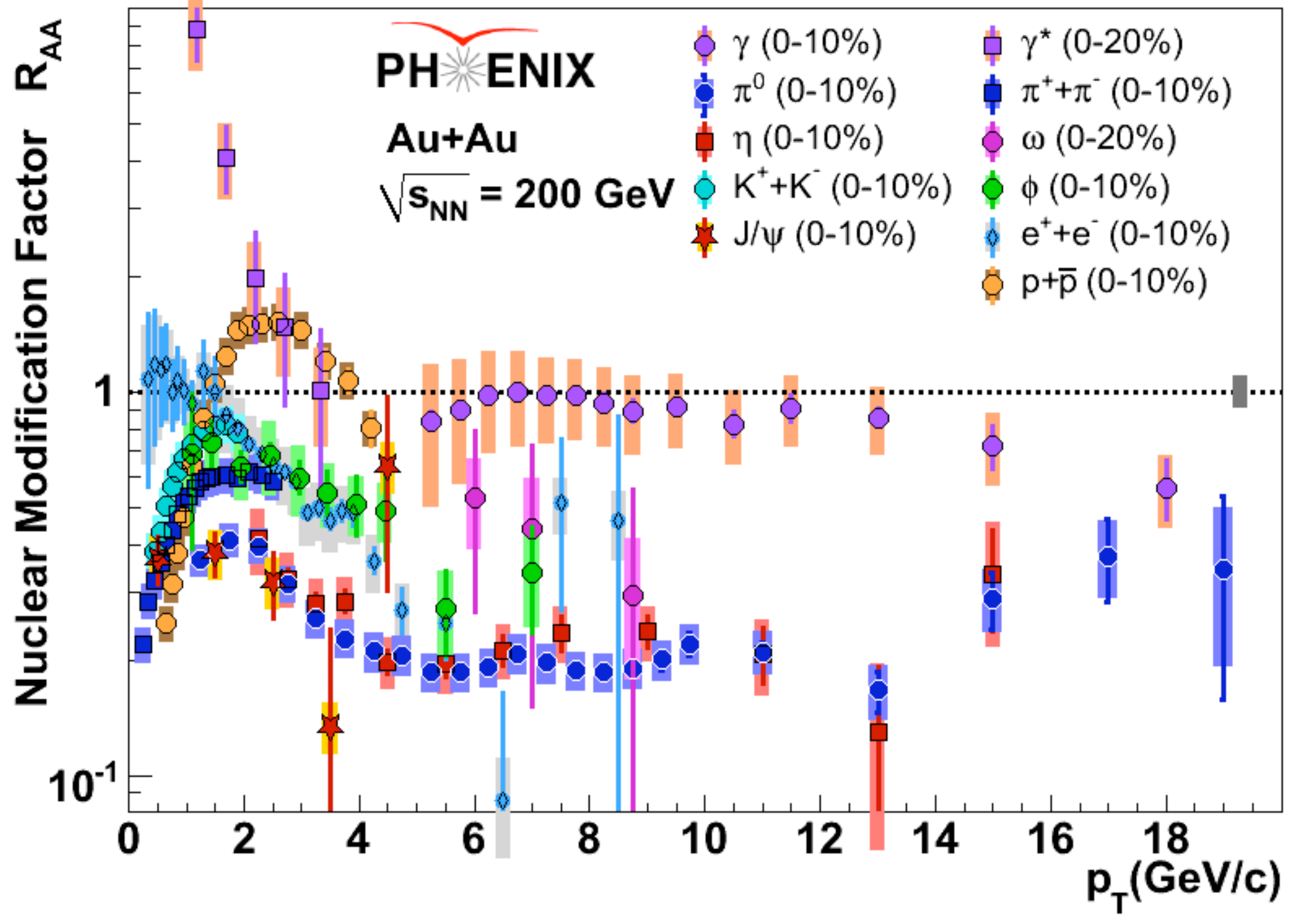}
\end{center}
\caption[]{$R_{AA}$ compilation from PHENIX for central Au+Au collisions. With the exception of the direct-$\gamma$ internal-conversion data ($\gamma*$) for $p_T<4$ GeV/c where a fit to the p-p data (which is not exponential) is used, all the other values of $R_{AA}$ are computed from the measured Au+Au and p-p $p_T$ spectra.}
\label{fig:Tshirt}
\end{figure}

   On of the major, if not the major discovery at RHIC is the huge suppression of particle production at large $p_T$ in A+A collisions. So far this appears to be consistent with the {\QCD} predictions of energy loss or absorption of the outgoing hard-scattered color-charged partons due to interactions with the presumably deconfined and thus color-charged medium (a liquid version of the {\QGP}) produced in central A+A collisions at RHIC~\cite{BSZ}. The suppression is represented by the nuclear modification factor, the ratio $R_{AA}$ of the measured semi-inclusive yield (e.g of $\pi$) for a given centrality class to  the p-p cross section scaled by $\mean{T_{AA}}$ the average overlap integral of the nuclear thickness functions for that centrality class (Eq.~\ref{eq:RAA}). For pure point-like hard-scattering, $R_{AA}=1$. 
\begin{equation}
R_{AA} = \frac{ d^2 N_{AA}^{\pi} / p_T dp_T dy\, N^{inel}_{AA} }{\langle T_{AB} \rangle  \times d^2 \sigma_{pp}^{\pi}/p_T dp_T dy}
       \label{eq:RAA}
\end{equation}
The striking differences of $R_{AA}(p_T)$ in central Au+Au collisions for the many particles measured by PHENIX at RHIC (Fig.~\ref{fig:Tshirt}) illustrates the importance of particle identification for understanding the physics of the medium produced at RHIC.    
For the entire region $p_T<20$ GeV/c so far measured at RHIC, all particles show suppression for $p_T>3$ GeV/c with the exception of $p+\bar{p}$, which are enhanced in the region $2\leq p_T\leq 4$ GeV/c (the ``baryon anomaly''~\cite{PXBanomaly}), and the direct $\gamma$, which are not suppressed, presumably because the outgoing $\gamma$'s do not interact with the color charged medium. New and truly notably this year is the behavior of $R_{AA}$ of direct-$\gamma$ for $p_T<2$ GeV/c~\cite{PXintconv}, which is totally and dramatically different from all other particles, exhibiting an order of magnitude exponential enhancement as $p_T\rightarrow 0$. This exponential enhancement is certainly suggestive of a new production mechanism in central Au+Au collisions different from the conventional soft and hard particle production processes in p-p collisions and its unique behavior is attributed to thermal photon production by many authors (e.g. see citations in reference~\cite{ppg086}).   

 \subsection{Two-particle correlations at RHIC}
 \subsubsection{A RHIC discovery--di-hadron correlations do not measure the fragmentation function}
 \label{sec:nofrag}
 The two-particle (di-hadron) correlations as measured at RHIC follow the systematics from previous measurements at the CERN-ISR as already shown in Sec.~\ref{sec:almost}. However, we learned something new and important at RHIC~\cite{ppg029}--the di-hadron correlation, e.g. $\pi^0$-h, where both hadrons are fragments of jets, does not measure the fragmentation function. 
 
\begin{figure}[h]
\begin{center}
a)\includegraphics[width=0.47\linewidth]{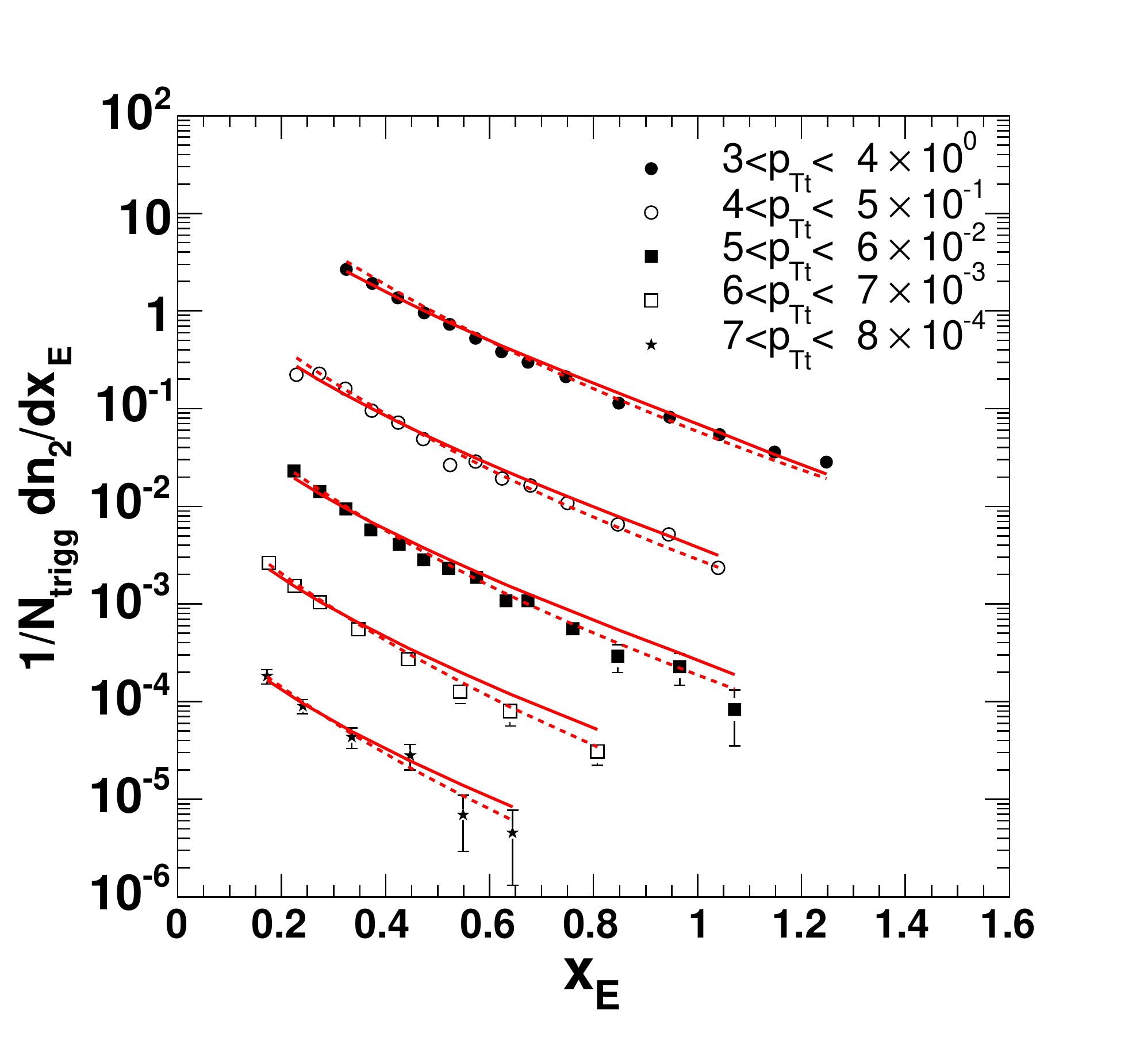}
b)\includegraphics[width=0.47\linewidth]{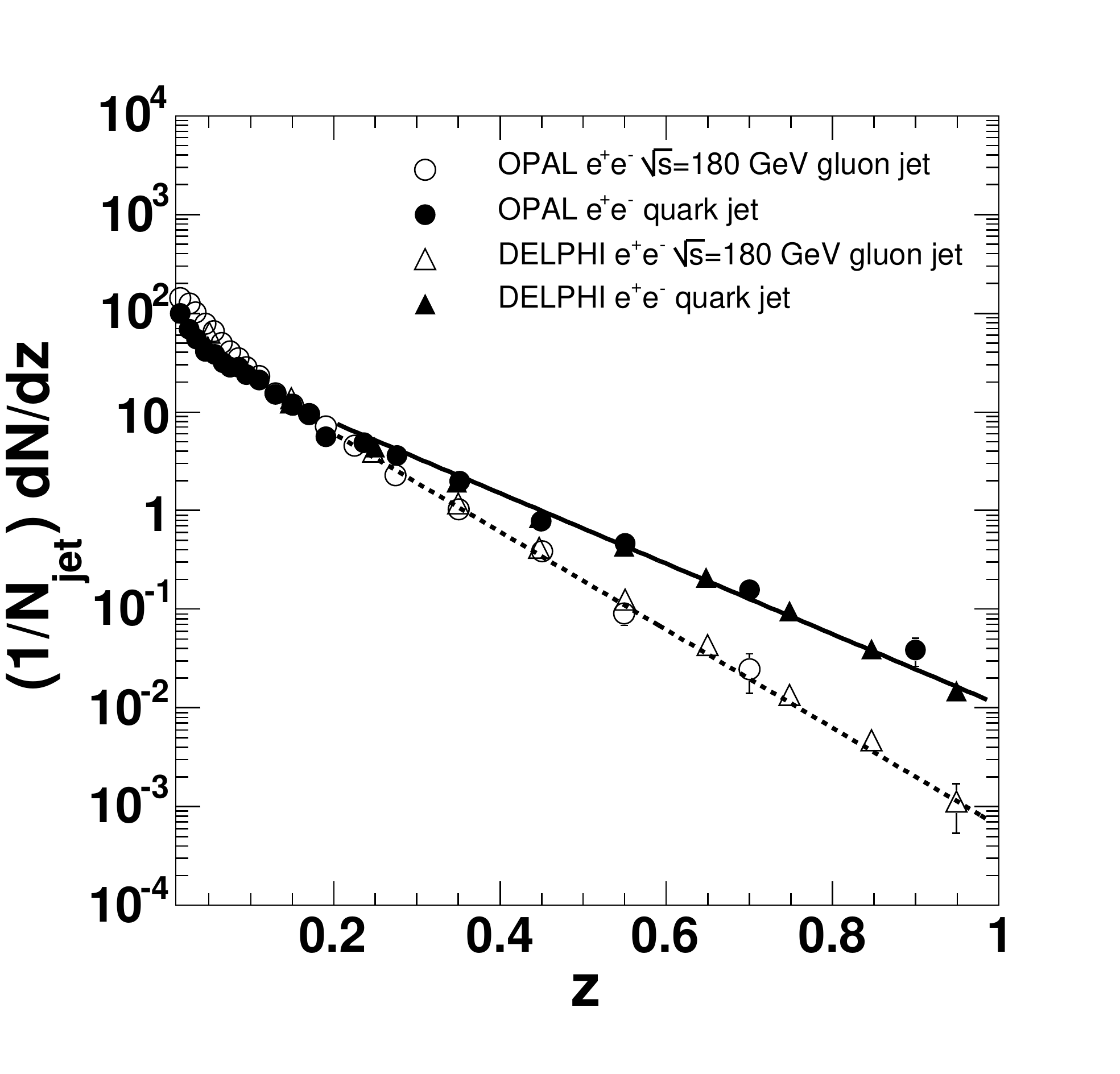}
\end{center}
\vspace*{-7mm}
\caption[]
{a) $x_E$ distributions from PHENIX~\cite{ppg029} for $\pi^0$-h correlations in p-p collisions at $\sqrt{s}=200$ GeV for several values of $p_{T_t}$. The solid and dashed lines represent calculations of the distribution for quark (solid lines) and gluon (dashed lines) fragmentation functions based on exponential fits to the LEP measurements~\cite{OPAL,DELPHI} shown in b). See Ref.~\cite{ppg029} for details
\label{fig:wow} }
\end{figure}

     PHENIX~\cite{ppg029} attempted to measure the mean net transverse momentum of the di-jet, $\mean{p_{T{\rm pair}}}=\sqrt{2}\mean{k_T}$, in p-p collisions at RHIC, where $k_T$ represents the out-of-plane activity of the hard-scattering~\cite{FFF,Darriulat}. This requires the knowledge of $\mean{z_t}$ of the trigger $\pi^0$, which PHENIX attempted to calculate using a fragmentation function derived from the measured $x_E$ distributions (Fig.~\ref{fig:wow}a).  
     
It didn't work. Finally, it was found that starting with either the quark $\approx \exp (-8.2 \cdot z)$ or the gluon $\approx \exp (-11.4 \cdot z)$ fragmentation functions from LEP (Fig.~\ref{fig:wow}b solid and dotted lines), which are quite different in shape, the results obtained for the $x_E$ distributions (solid and dotted lines on Fig.~\ref{fig:wow}a) do not differ significantly! Although nobody had noticed this for nearly 30 years, the reason turned out to be quite simple. The integration over $z_t$ of the trigger jet for fixed trigger particle $p_{T_t}$ is actually an integral over the trigger jet $\hat{p}_{T_t}$. However since the trigger and away-jets are always roughly equal and opposite in transverse momentum, integrating over $\hat{p}_{T_t}$ simultaneously integrates over $\hat{p}_{T_a}$ and thus also integrates over $z$ of the away-jet. 

	With no assumptions other than a power law for the jet $\hat{p}_{T_t}$ distribution (${{d\sigma_{q} }/{\hat{p}_{T_t} d\hat{p}_{T_t}}}= A \hat{p}_{T_t}^{-n}$) , an exponential fragmentation function ($D^{\pi}_q (z)=B e^{-bz}$), and constant $\hat{x}_h$, for fixed $p_{T_t}$ as a function of $p_{T_a}$, it was possible to derive the $x_E$ distribution in the collinear limit, where $p_{T_a}=x_E p_{T_t}$~\cite{ppg029}: 
	     \begin{equation}
\left.{dP_{\pi} \over dx_E}\right|_{p_{T_t}}\approx {N(n-1)}{1\over\hat{x}_h} {1\over
{(1+ {x_E \over{\hat{x}_h}})^{n}}} \, \qquad ,  
\label{eq:condxe2}
\end{equation}
and $N=\mean{m}$ is the multiplicity of the unbiased away-jet. The shape of the $x_E$ distribution (Fig.~\ref{fig:wow}a) is given by the power $n$ of the partonic and inclusive single particle invariant transverse momentum spectra and does not depend on the exponential slope of the fragmentation function (Fig.~\ref{fig:wow}b). Note that Eq.~\ref{eq:condxe2} provides a relationship between the ratio of the away and trigger particle's transverse momenta, $x_{E}\approx p_{T_a}/p_{T_t}$, which is measured, to the ratio of the transverse momenta of the away and trigger jets, $\hat{x}_h=\hat{p}_{T_a}/\hat{p}_{T_t}$, which can thus be deduced. In p-p collisions the imbalance of the away-jet and the trigger jet ($\hat{x}_h\sim 0.7-0.8$) is caused by $k_T$-smearing~\cite{FFF,ppg029}.  In A+A collisions, $\hat{x}_h$ is sensitive to the relative energy loss of the trigger and associated jets in the medium, which can thus be measured~\cite{egseeMJTCFRNC06}. 

     The same derivation gives a simple formula for Bjorken's parent-child ratio, the ratio of the number of $\pi$ at a given $p_{T_t}$ to the number of partons($q$) at the same $p_{T_t}$: 
\begin{equation}
\left . {\pi^0 \over q }\right|_{\pi^0} (p_{T_t})\approx {\Gamma(n-1) \over b^{n-3}} \qquad . 
\label{eq:pioverjet}
\end{equation}
  I used Eq.~\ref{eq:pioverjet} together with a half and half mixture of quark and gluon jets with $b_q=8.2$ 
  $b_g=11.7$ from Fig.~\ref{fig:wow}b to obtain the parton cross section from the $\pi^0$ cross section on Fig.~\ref{fig:PXpp}b~\cite{MJTPoS07}.
\subsubsection{Two-particle correlations in A+A collisions are much more complicated than for p-p}
\begin{figure}[!hbt]
\begin{center}
\begin{tabular}{cc}
\begin{tabular}[b]{c}
\hspace*{-0.10in}\includegraphics[width=0.50\linewidth]{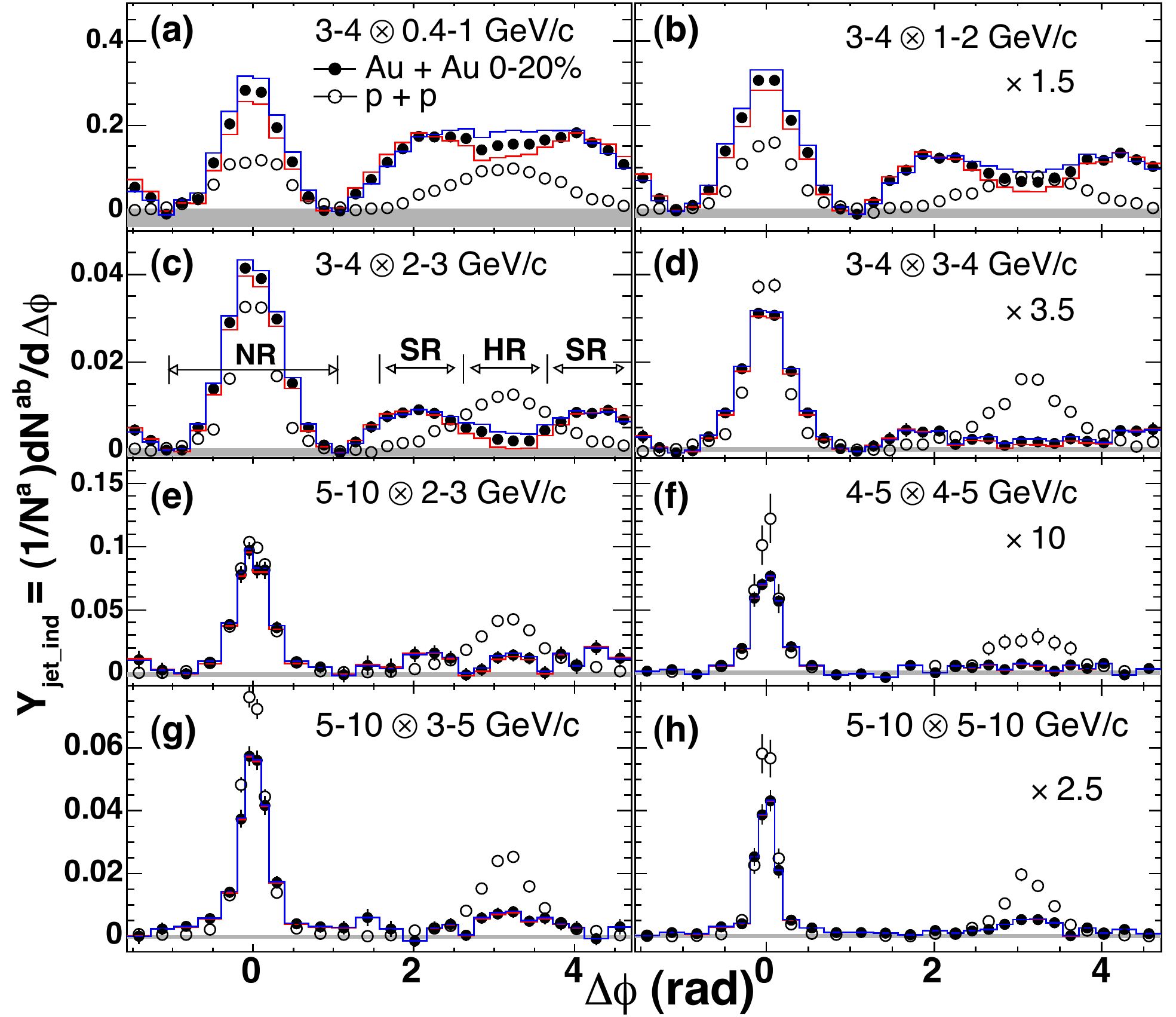}
\end{tabular}
\begin{tabular}[b]{c}
\includegraphics[width=0.45\linewidth]{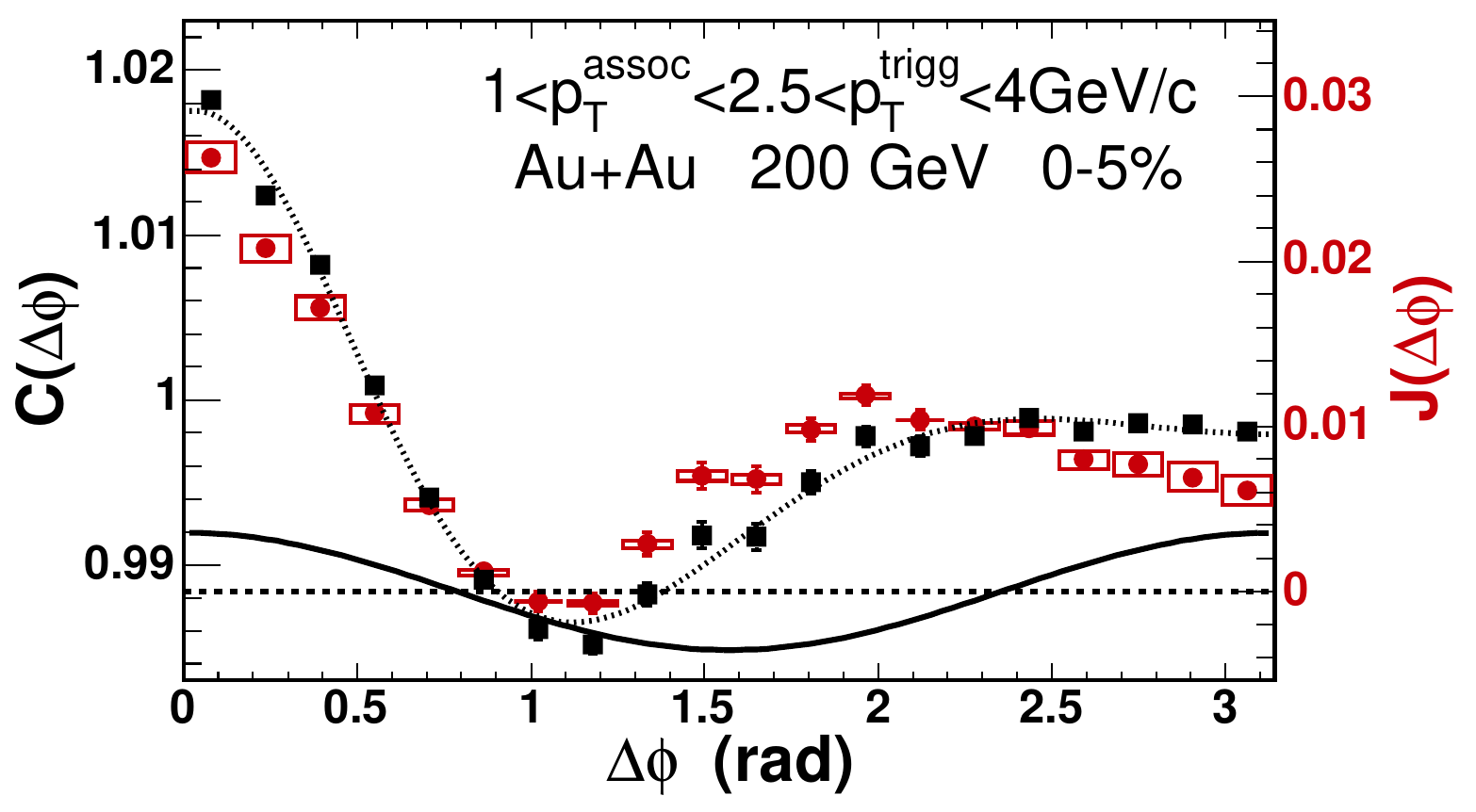}\cr
\includegraphics[width=0.42\linewidth]{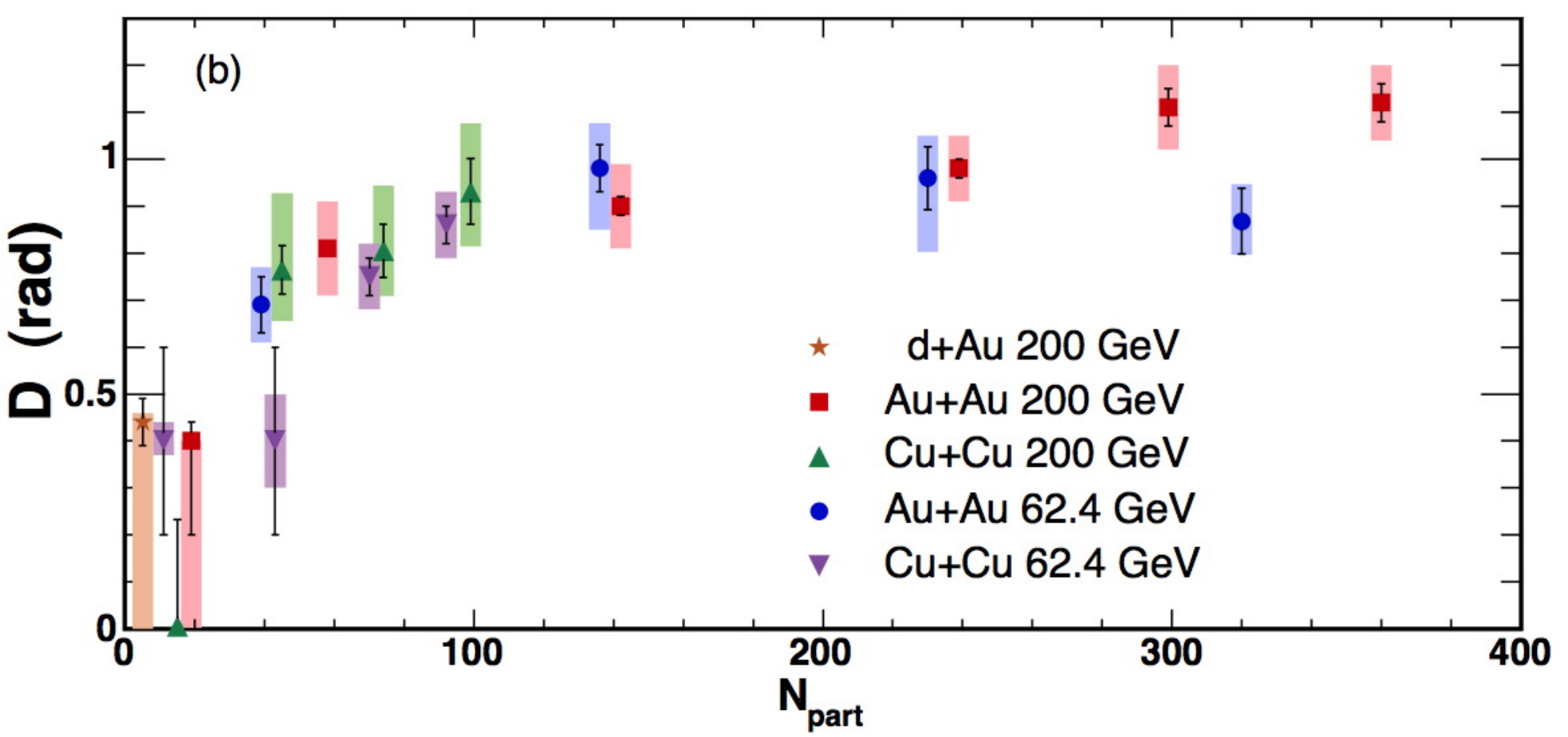}
\end{tabular}

\end{tabular}

\end{center}
\caption[]
{a-h) (left) PHENIX~\cite{ppg083} azimuthal correlation conditional yield of associated $h^{\pm}$ particles with $p_T^b$ for trigger $h^{\pm}$ with $p_T^a$ for the various $p_T^a \otimes p_T^b$ combinations shown. i) (right)-(top)  PHENIX~\cite{ppg067} azimuthal correlation function $C(\Delta\phi)$ of $h^{\pm}$ with $1\leq p_{T_a}\leq 2.5$ GeV/c with respect to a trigger $h^{\pm}$ with $2.5\leq p_{T_t} \leq 4$ GeV/c in Au+Au central collisions, where the line with data points indicates $C(\Delta\phi)$ before correction for the azimuthally modulated ($v_2$) background, and the other line is the $v_2$ correction which is subtracted to give the jet correlation function $J(\Delta\phi)$ (data points). j) (right)-(bottom) PHENIX $D$ parameters~\cite{ppg067}, the angular distance of the apparently displaced peak of the $J(\Delta\phi)$ distribution from the angle $\Delta\phi=\pi$ as a function of centrality, represented as the number of participants $N_{\rm part}$, for the systems and c.m. energies indicated. 
\label{fig:HSD} }
\end{figure}

    In analogy to Fig.~\ref{fig:mjt-ccorazi} (above), the di-hadron correlations in Au+Au collisions (Fig.~\ref{fig:HSD}) show a di-jet structure. 
However, one of the many interesting new features in Au+Au collisions is that the away side azimuthal jet-like correlations (Fig.~\ref{fig:HSD}c) are much wider than in p-p collisions and show a two-lobed structure (``the shoulder'' (SR)) at lower $p_{T_t}$ with a dip at 180$^\circ$,  reverting to the more conventional structure of a peak at 180$^\circ$ (``the head'' (HR)) for larger $p_{T_t}$. The wide away-side correlation in central Au+Au collisions is further complicated by the large multiparticle background which is modulated in azimuth by the $v_2$ collective flow of a comparable width to the jet correlation (Fig.~\ref{fig:HSD}i). After the $v_2$ correction, the double peak structure $\sim \pm 1$ radian from $\pi$, with a dip at $\pi$ radians, becomes evident. The double-peak structure may indicate a reaction of the medium to a passing parton in analogy to a ``sonic boom'' or the wake of a boat and is under active study both theoretically~\cite{egseeppg083} and experimentally. PHENIX characterizes this effect by the half-width $D$ ($\sim 1.1$ radian) of the Jet function, $J(\Delta\phi)$, the angular distance of the displaced peak of the distribution from the angle $\Delta\phi=\pi$. One of the striking features of the wide away side correlation is that the width $D$ (Fig.~\ref{fig:HSD}j)  does not depend on centrality, angle to the reaction plane,  $p_{T_a}$ and $\sqrt{s_{NN}}$, which seems problematic to me if the effect is due to a reaction to the medium.     

\begin{figure}[h]
\begin{center}
a)\includegraphics[width=0.47\linewidth]{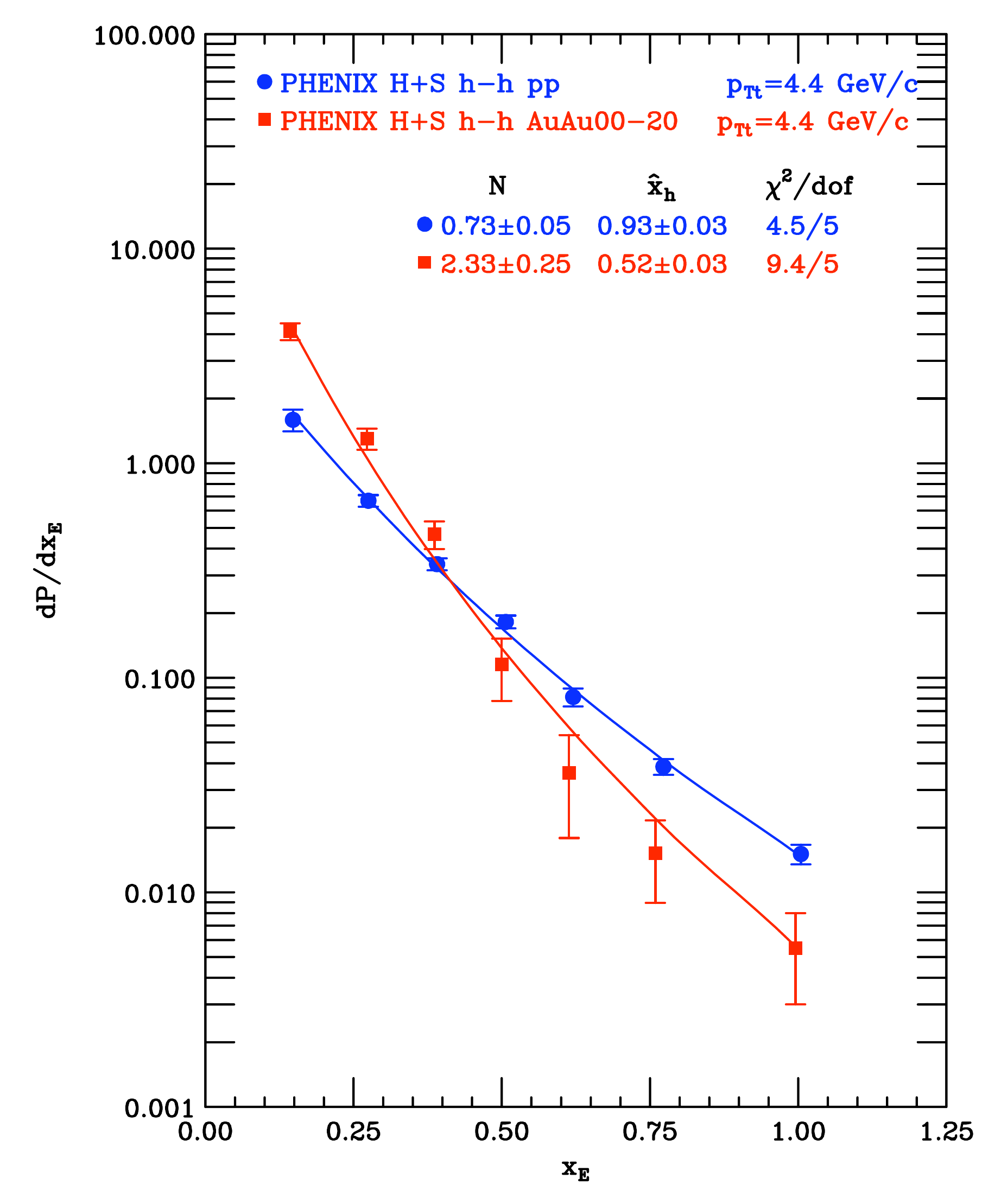}
b)\includegraphics[width=0.47\linewidth]{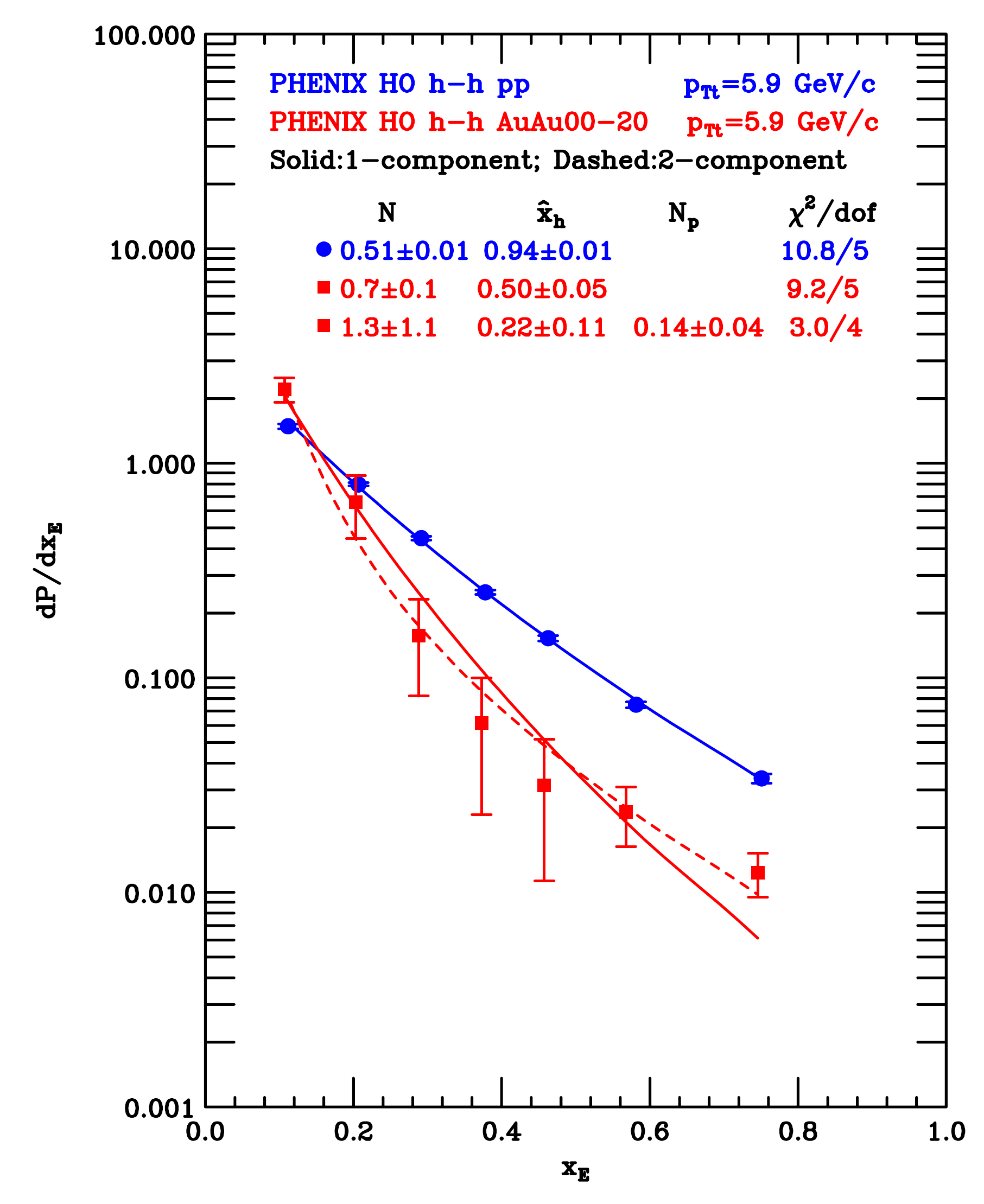}
\end{center}
\caption[]{a) $x_E$ distribution ~\cite{ppg083} $dP/dx_E$ of associated $h^{\pm}$ per trigger $h^{\pm}$ with $4\leq p_{T_t}\leq 5$ GeV/c for the full away side azimuth, $|\Delta\phi-\pi|<\pi/2$, in p-p (circles) and Au+Au central (0-20\%) collisions (squares). Fits to Eq.~\ref{eq:condxe2} are shown with best-fit parameters indicated. b) Same plot for $5\leq p_{T_t}\leq 10$ GeV/c data~\cite{ppg083}, for the Head region only (HO), $|\Delta\phi-\pi|<\pi/6$. }
\label{fig:xEAA}
\end{figure}

As discussed above, the $x_E$ (also denoted $z_T$) distribution from di-hadron collisions does not measure the fragmentation function. Instead, it is sensitive to the ratio of the away parton transverse momentum to the trigger parton transverse momentum, $\hat{x}_h=\hat{p}_{T_a}/\hat{p}_{T_t}$, which is a measure of the differential energy loss of the away parton relative to the trigger parton which is surface biased due to the steeply falling $\hat{p}_{T_t}$ spectrum~\cite{MagestroRef,Magestro}. 
    The energy loss of the away-parton is indicated by the fact that the $x_E$ distribution in Au+Au central collisions e.g. for $4\leq p_{T_t}\leq 5$ GeV/c (Fig.~\ref{fig:xEAA}a) from the data in Fig.~\ref{fig:HSD}~\cite{ppg083} is steeper than that from p-p collisions. 
Eq.~\ref{eq:condxe2} provides excellent fits to these distributions in both p-p and Au+Au collisions (Fig.~\ref{fig:xEAA}a),  where fits to the full away side azimuth Head + Shoulder (HS) region are shown.      
In p-p collisions, the imbalance of the away-parton and the trigger parton  indicated by the fitted value of $\hat{x}_h=0.93\pm 0.03$ in Fig.~\ref{fig:xEAA}a is caused by $k_T$-smearing.  In A+A collisions, the fitted value $\hat{x}_h=0.52\pm 0.03$ indicates that the away parton has lost energy relative to the trigger parton.

\begin{figure}[h]
\begin{center}
a)\includegraphics[width=0.43\linewidth]{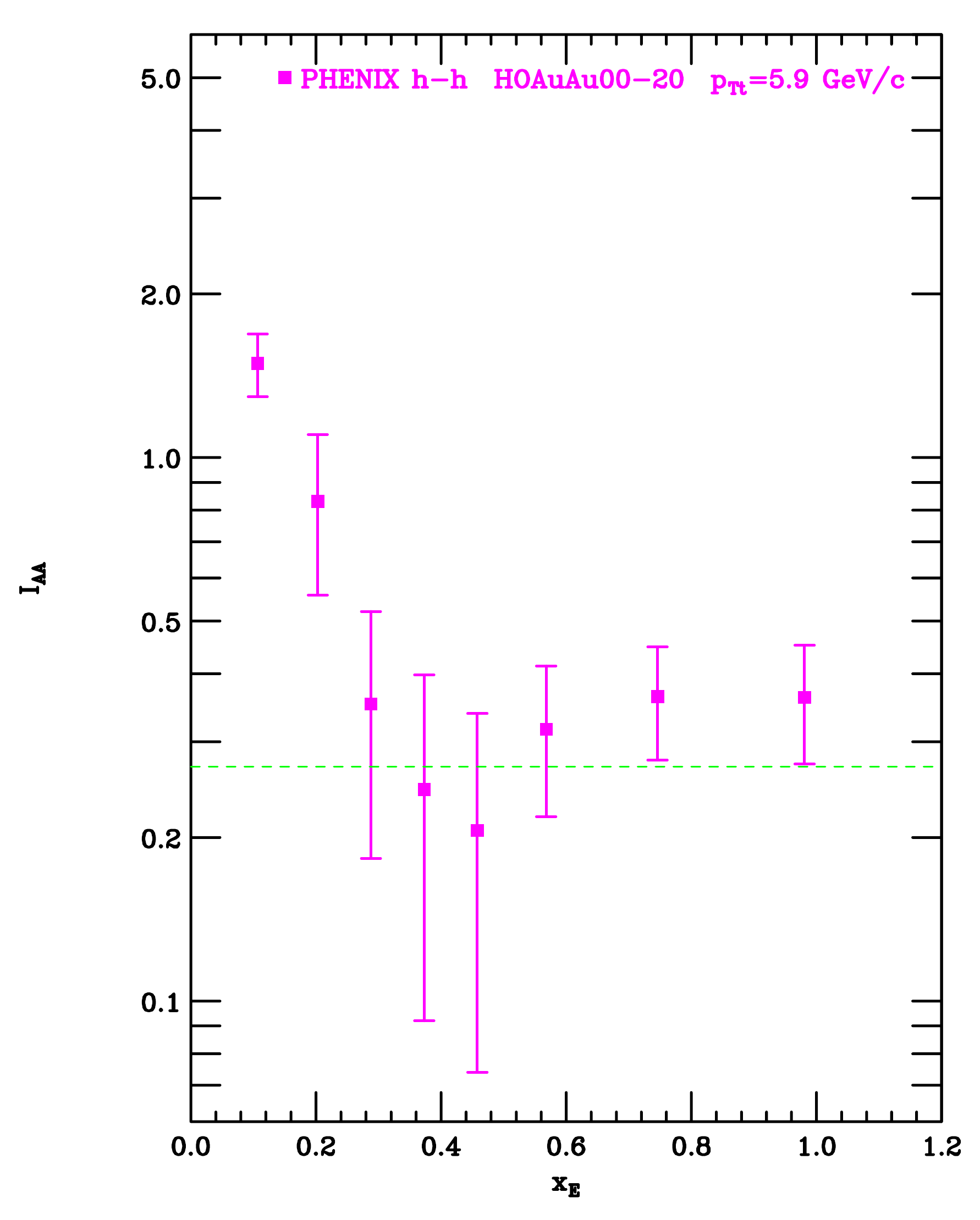}
b)\includegraphics[width=0.51\linewidth]{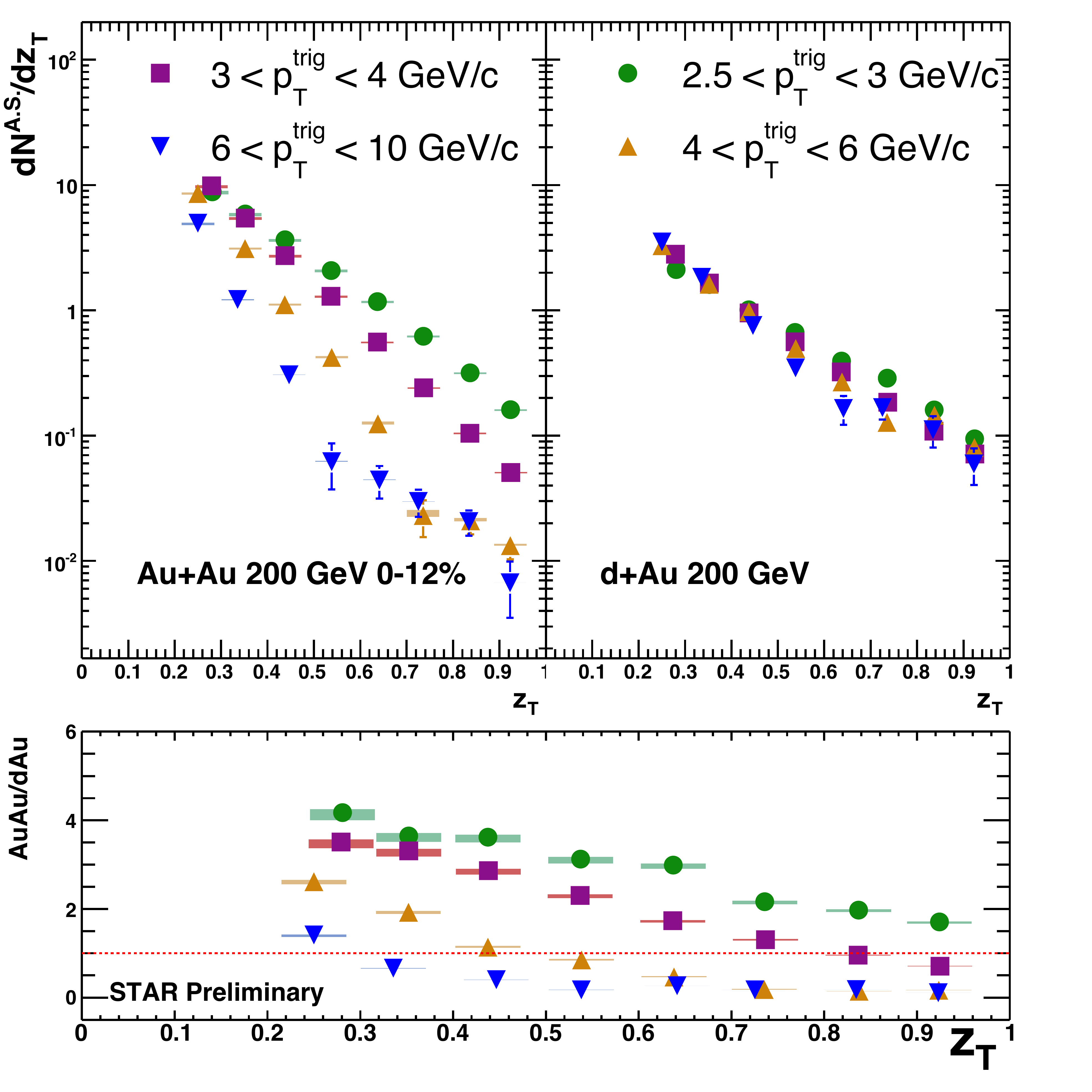}
\end{center}
\caption[]{a) $I_{AA}(x_{E})$ from the ratio of Au+Au/pp distributions in Fig.~\ref{fig:xEAA}b. Dashed line is 27\% punch-through from fit. b) STAR~\cite{Horner} away side $z_T$ (same as $x_E$) distributions for various $p_{T_t}$ in Au+Au, d+Au and their ratio in $\sqrt{s_{NN}}=200$ GeV collisions.}
\label{fig:IAA}
\end{figure}

	For larger $p_{T_t}$, another interesting effect takes place, punch-through of narrow away-side peaks consistent with roughly the same azimuthal width and $x_E$ distributions as in p-p collisions~\cite{Magestro}.
Fig.~\ref{fig:xEAA}b shows the $x_E$ distributions and fits at higher $5\leq p_{T_t}\leq 10$ GeV/c where the region of integration of the away-side azimuthal distribution has been restricted to the head-region only (HO) in both p-p and Au+Au collisions. The fitted values of $\hat{x}_h$ are nearly identical in Fig.~\ref{fig:xEAA}b ($\hat{x}_h=0.94\pm 0.01$ in p-p and $\hat{x}_h=0.50\pm 0.05$) and Fig.~\ref{fig:xEAA}a but the integrated away-side multiplicity $N$ is reduced in the HO region compared to the full-away side (HS). Of more importance to note is that the fit to the Au+Au HO data (Fig.~\ref{fig:xEAA}b) is greatly improved ($\Delta\chi^{2}=6.2/1$) if a second component with the same $\hat{x}_h$ as the p-p distribution is added (dashed-curve), statistically indicating a parton that has apparently punched through the medium without losing energy. The two-component fit shows 27\% punch-through and a much larger energy loss, 77\%. This is more clearly exhibited by taking the ratio, $I_{AA}$ of the $x_E$ distributions for Au+Au to p-p (Fig.~\ref{fig:IAA}a), and is evident directly in the STAR $z_T$(same as $x_E$) distribution~\cite{Horner} (Fig.~\ref{fig:IAA}b) by the sharp change in slope at $z_T~(x_E)=0.5$ in Au+Au for $6< p_{T_t}< 10$ GeV/c. 
  
   The punch-through and normal fragmentation of partons which have lost energy are standard features of the {\QCD} energy loss models. For instance the ZOWW~\cite{ZOWW} model breaks down the away-side fragmentation function to: 

\begin{center}
\fbox{   \begin{minipage}[b]{0.5\linewidth} 
\vspace*{0.003in}
\includegraphics[width=\linewidth]{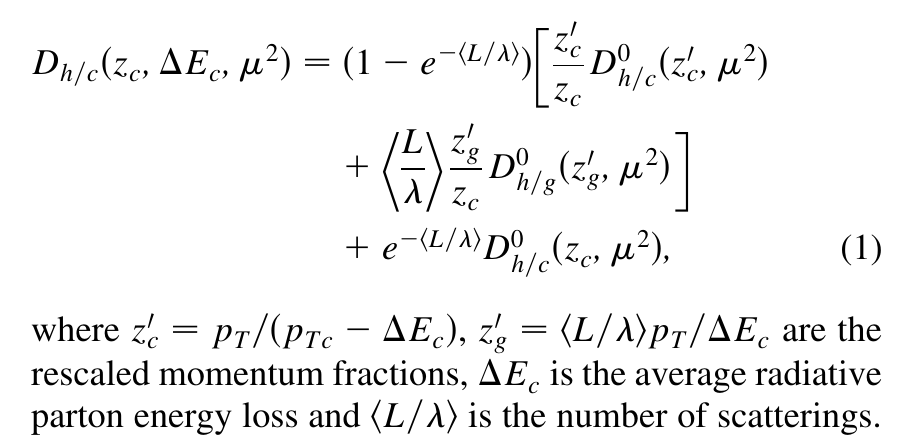}
\vspace*{-0.10in}
\end{minipage} }
\end{center} 
In words, the three terms in the above equation correspond to normal fragmentation of partons which have lost energy; the fragmentation of the lost energy, assumed to be a gluon; normal fragmentation with full original energy of punch-through partons. This `paradigm' is consistent, as far as I can tell, with all measured $I_{AA}(x_E)$ distributions in di-hadron correlations (Fig.~\ref{fig:IAA})--an exponential decrease at low $x_E$ from energy-loss and a flat distribution at large $x_E$ due to punch-through when no radiation occurs.  

	The discrete radiation of relatively large energy gluons is characteristic of the LPM effect~\cite{BDMPS,alsoRenk} in {\QCD}, where in distinction to QED, it is the radiated gluon that interacts coherently with the medium. This suppresses the gluon spectrum for large radiated energy $k$ (rather than at low $k$ as in QED, Fig.~\ref{fig:effects}a) but the coherence comes from gluons with a $k$ larger than some critical value ($\omega_{BH}$) which is relatively hard, 1--4 GeV. Thus if LPM radiation occurs there is a large radiation loss, but there is a Poisson probability $e^{-L/\lambda}$ that no radiation occurs leading to punch-through. However, as I understand it, multiple scattering ($p_T$ broadening), elastic and normal Bethe-Heitler radiative loss should take place even with punch-through, so we still have a long way to go on this issue. 

The energy-loss models all do quite well in describing the suppression, $R_{AA}(p_T)$, for single-inclusive $\pi^0$, although each model has different parameters~\cite{ppg079}. However, the suppression of direct-$e^{\pm}$ from heavy quark decays (recall Fig.~\ref{fig:Tshirt}) disfavors radiative models since, naively, heavy quarks should radiate much less than light quarks in the medium. This has attracted much theoretical attention and is still not explained~\cite{egMGPL,MJTWWND10}. 

	For two-particle correlations, no clear paradigm has emerged for the two-lobed wide away-jet structure. Some suggestions are:
\begin{itemize}
\item Mach or Cerenkov cone due to medium reaction. Since the effect vanishes for larger $p_{T_t}$, with no dependence on centrality or $p_{T_a}$ (recall Fig.~\ref{fig:HSD}j), I find this explanation unlikely. Further studies with particle identification might show whether the composition of the shoulder is the same as the medium or the jet for a given $p_{T_a}$--kind of like the difference of being hit by the wake (water) or the boat (wood).
\item Fluctuations of $v_3$~\cite{Sorensen0810,Alver10}: $\mean{v_3}=0$ at mid-rapidity but $\mean{v_3^2}\neq 0$. This gives the best fits to the azimuthal correlation functions $C(\Delta\phi)$ and the constant value of $D\approx 60^\circ$ (Fig.~\ref{fig:HSD}) and is consistent with vanishing at large $p_{T_t}$. Lots of ongoing work at the present time.
\item NLO 3-jet events less suppressed than di-jets due to smaller path-length in medium~\cite{Ayala10}.
\end{itemize}

\section{Direct-$\gamma$-h correlations do measure the fragmentation function} 

Direct single-$\gamma$ production via the inverse QCD-compton process~\cite{QCDCompton}:
\begin{equation} 
g+q \rightarrow \gamma+q
\label{eq:QCDcompton}
\end{equation}
 is an important probe in p-p collisions because it is sensitive the gluon structure function at leading order as well as higher orders~\cite{Aurenche}. It is possibly even more important in A+A collisions because the $\gamma$ is a direct participant in the hard-scattering (at the constituent level), which emerges from the medium without interacting and can be measured precisely. This means that, modulo $k_T$, the $p_T$ of the quark jet which is equal and opposite to that of the direct-$\gamma$ is known precisely. Furthermore, since the direct-$\gamma$ is not a fragment of a jet, but a direct participant in the hard-scattering, this means that the fragmentation function can be measured by the direct-$\gamma$-hadron $x_E$ ($z_T$) distribution, as originally suggested by Wang, Huang and Sarcevic in 1996~\cite{WHZ96}, `the golden channel'~\cite{Renk09}. 

    There are nice reviews of both the theory~\cite{OwensRMP} and the experimental discovery of direct-single-$\gamma$ production at the CERN-ISR in 1978--80 and other early measurements~\cite{FerbelMolzon}. Measurements of direct-$\gamma$ at RHIC in both p-p~\cite{PXgampp07} and Au+Au~\cite{PXgamAuAu05} collisions have shown the absence of suppression for this probe which does not interact with the medium (recall Fig.~\ref{fig:Tshirt}). Calculations of direct-$\gamma$ production in {\QCD} give impressive agreement with the data over a wide range of c.m. energies~\cite{Aurenche}. Other nice features of direct-$\gamma$ production from the inverse QCD-compton effect (Eq.~\ref{eq:QCDcompton}) is that the $\gamma$ are isolated and the away-jet is a quark-jet, 8 times more likely $u$ than $d$.

\begin{figure}[h]
\begin{center}
\includegraphics[width=0.33\linewidth]{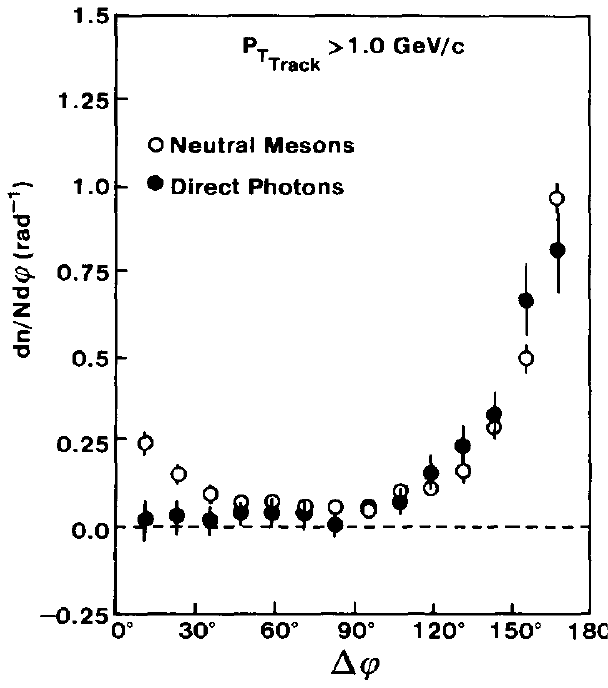}
\hspace*{0.1in}\includegraphics[width=0.64\linewidth]{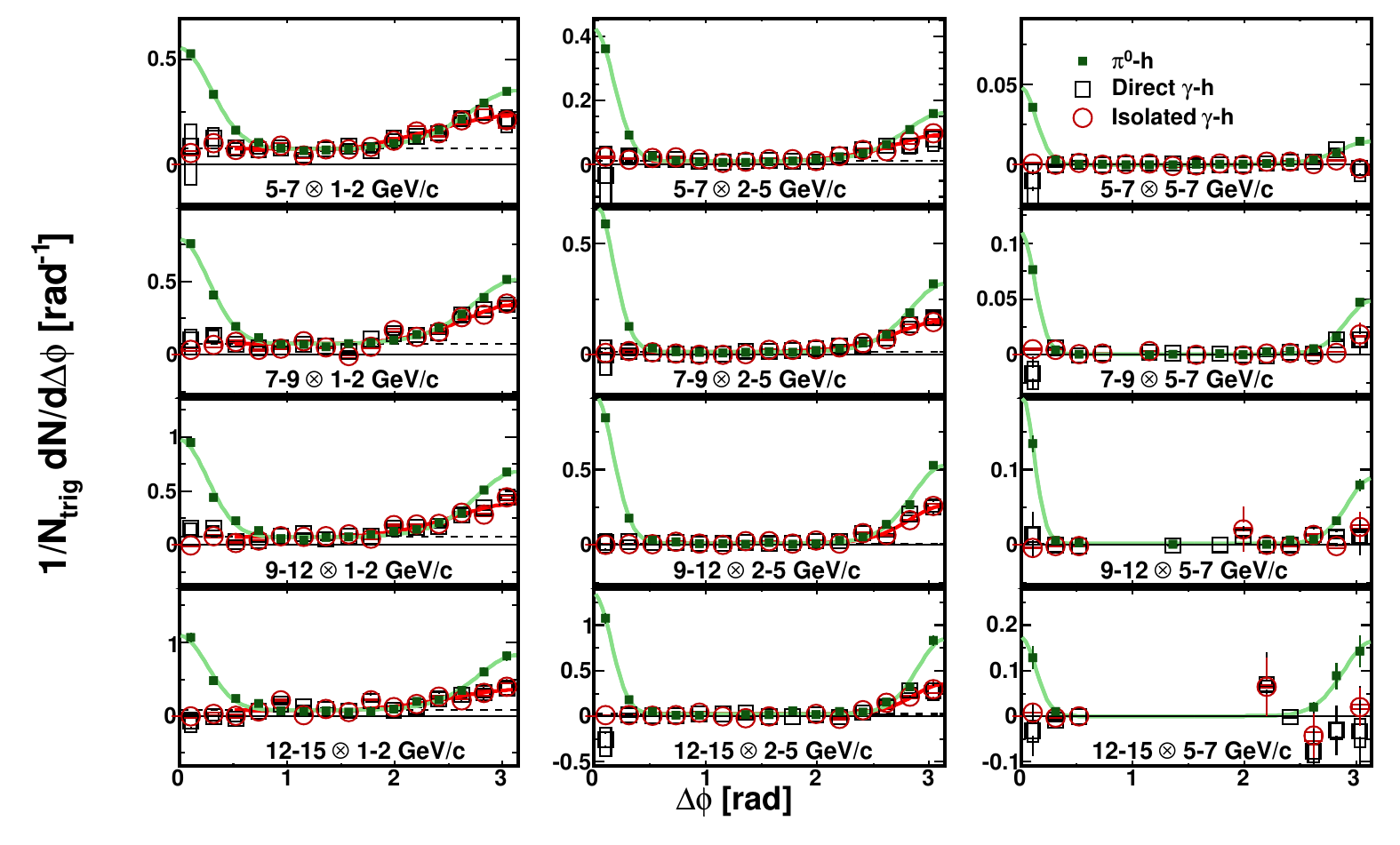}
\end{center}
\caption[]{Direct single-$\gamma$-h and $\pi^0$-h azimuthal correlations in p-p collisions: a)$\sqrt{s_{NN}}=62.4$ GeV, $p_{T_t}>6$ GeV/c~\cite{CMOR};  b) $\sqrt{s}=200$ GeV, $p_{T_t}$, $p_{T_a}$ as indicated~\cite{ppg095}.
\label{fig:pi0gam-azi}}
\end{figure}

    Beyond Leading Order, single-$\gamma$ may be produced in the fragmentation process~\cite{OwensRMP}; but these $\gamma$ are not isolated. However, measurements of $\gamma-h$ azimuthal correlations in p-p collisions at both the CERN-ISR~\cite{CMOR} and RHIC~\cite{ppg095} (Fig.~\ref{fig:pi0gam-azi}) show that direct-$\gamma$ are isolated, with very few, if any, accompanying same-side particles, while $\pi^0$ have accompanying particles since they are fragments of jets from high $p_T$ partons. This is strong evidence against a significant fragmentation component for direct-$\gamma$ production, easily $\lsim$10\%.   
    
\begin{figure}[h]
\begin{center}
a)\includegraphics[width=0.47\linewidth]{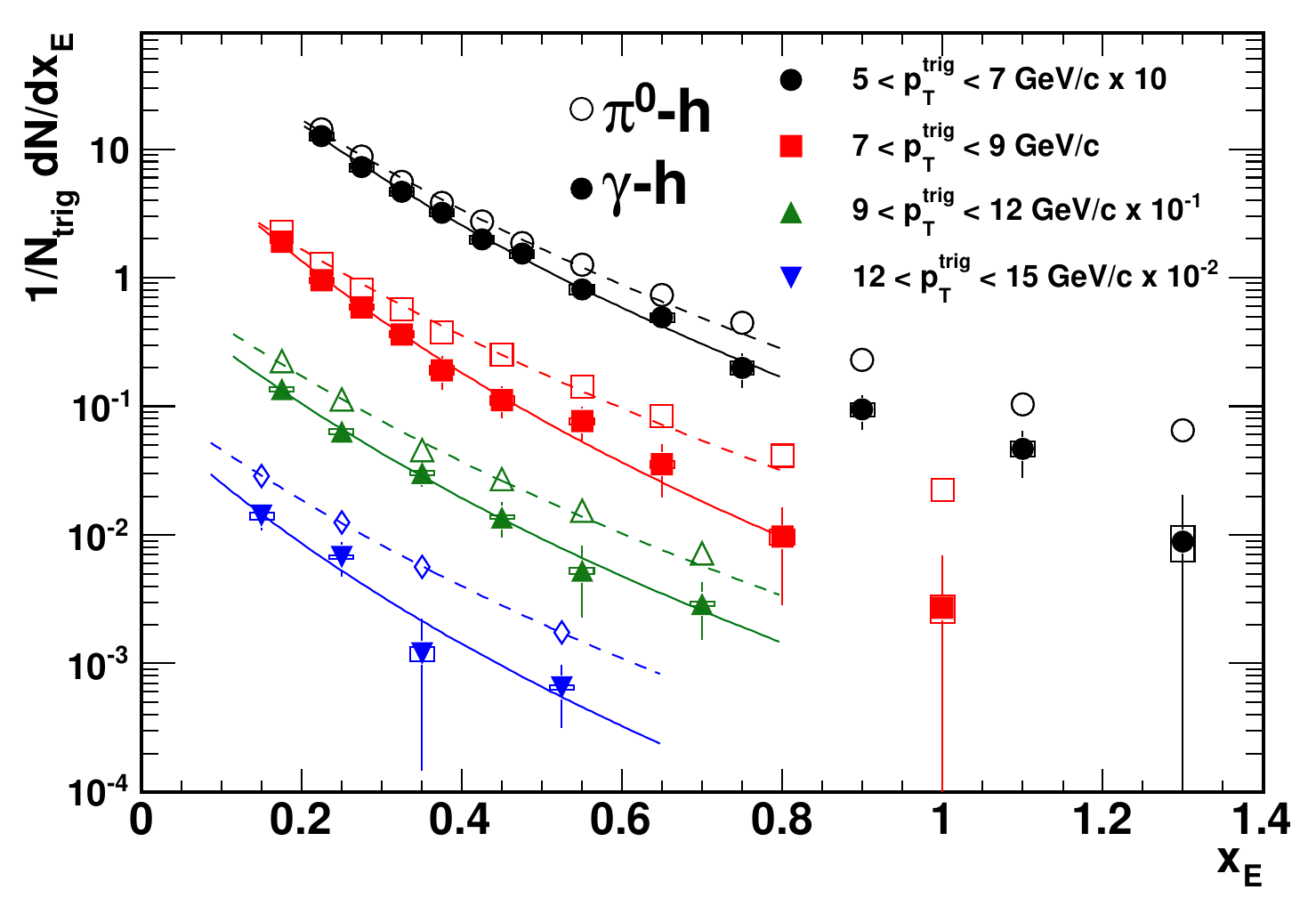}
b)\includegraphics[width=0.47\linewidth]{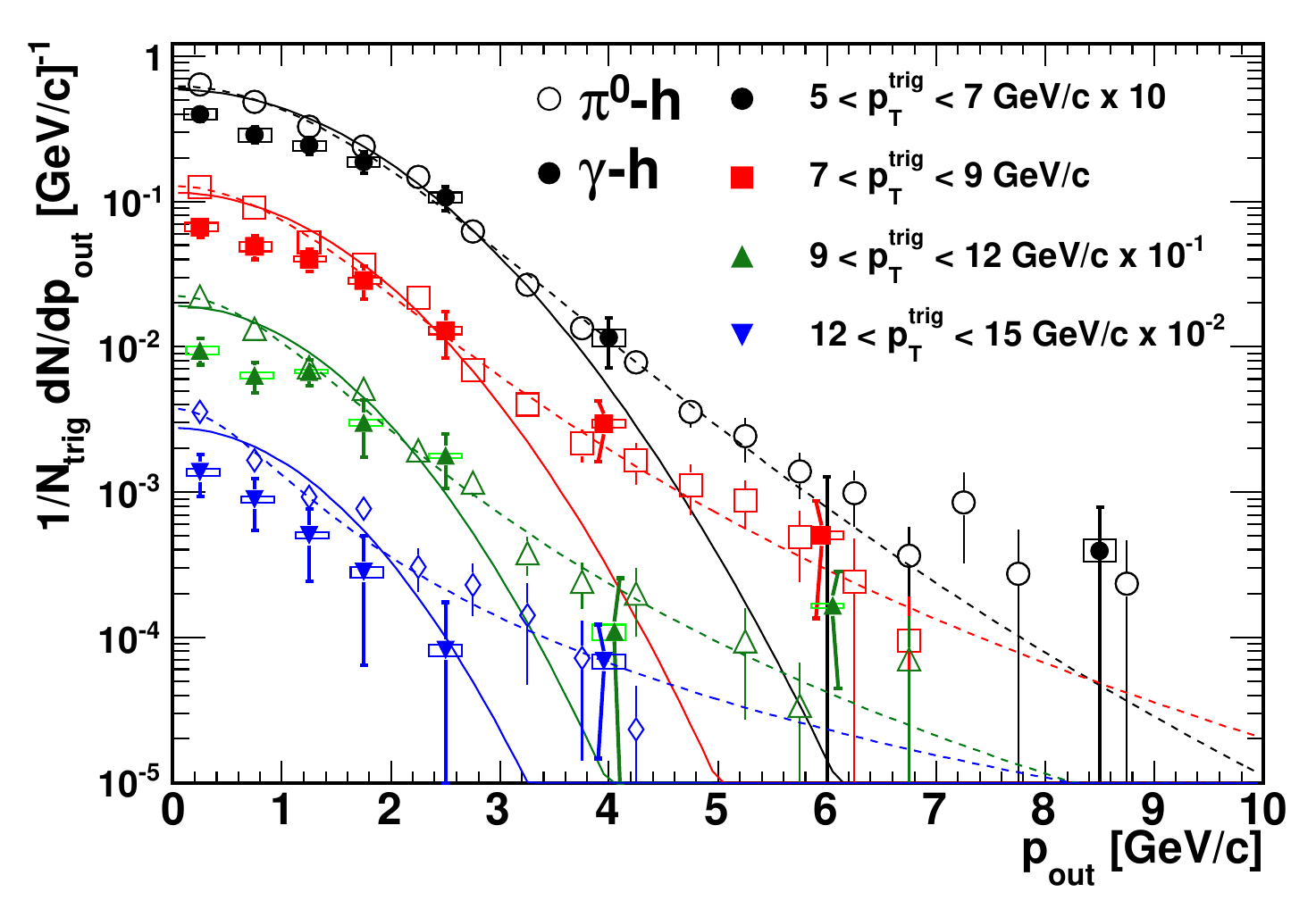}
\end{center}
\caption[]{a) $x_E$ and  b) $p_{\rm out}$ distributions from $\pi^0$-h and isolated direct-$\gamma$-h correlations~\cite{ppg095} }
\label{fig:pi0gamhxE}
\end{figure}

    The $x_E$ and $p_{\rm out}$ distributions in p-p collisions from $\pi^0$-h and direct-$\gamma$-h correlations from new data at RHIC~\cite{ppg095} are shown in Fig.~\ref{fig:pi0gamhxE}. For these measurements of direct-$\gamma$-h correlations, an isolation cut has been made to improve the systematic uncertainty in subtracting the background from $\pi^0$-h correlations.
The $x_E$ distributions show a flatter slope of the $\pi^0$-h distribution compared to the direct-$\gamma$-h distribution, as discussed for Fig.~\ref{fig:mjt-ccorazi}d (which is for jet-h vs $\pi^0$-h correlations)~\cite{FFF}, showing that the $\gamma$ plays the expected role of a jet for correlation measurements, albeit with a much higher precision measurement of $p_T$ and a much lower rate. Of course, the $x_E$ distribution from the isolated-direct-$\gamma$-h measurement is actually the fragmentation function. The $p_{\rm out}$ distributions for $\pi^0$-h and isolated-direct-$\gamma$-h correlations (Fig.~\ref{fig:pi0gamhxE}) appear to be nearly identical and both show the features of a Gaussian distribution for $p_{\rm out}< 3$ GeV/c, which is the $k_T$ effect, now thought to be due to resummation of soft gluons~\cite{Aurenche}, and a power-law tail from NLO hard-gluon emission. This is the first time that the $k_T$ effect has been measured with direct-$\gamma$; and the $\sqrt{\mean{k_T^2}}$ is essentially identical in $\pi^0$ and direct-$\gamma$ production.  

\subsection{The fragmentation function in p-p collisions}
\begin{figure}[h]
\begin{center}
\includegraphics[width=0.50\linewidth]{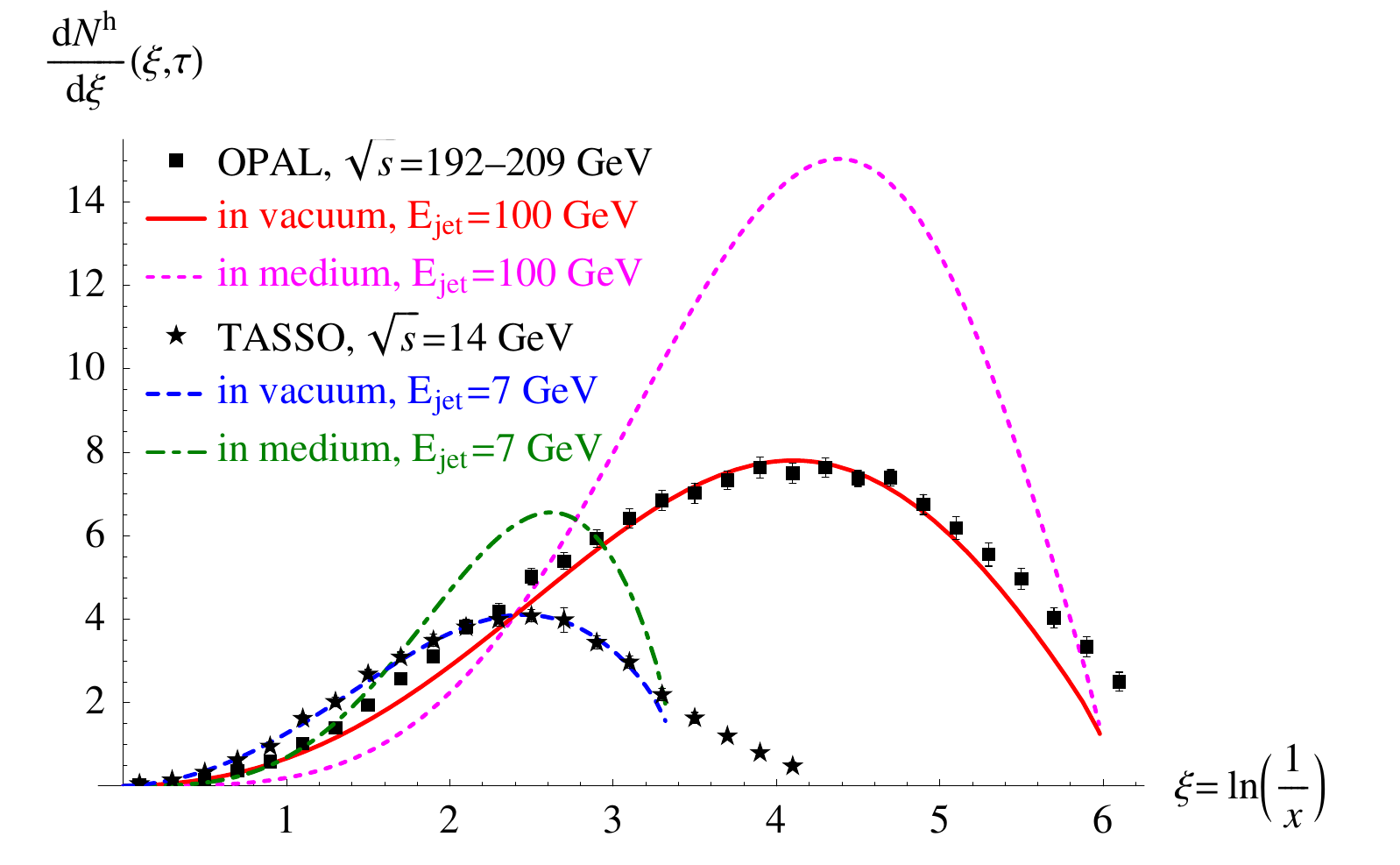}
\includegraphics[width=0.47\linewidth]{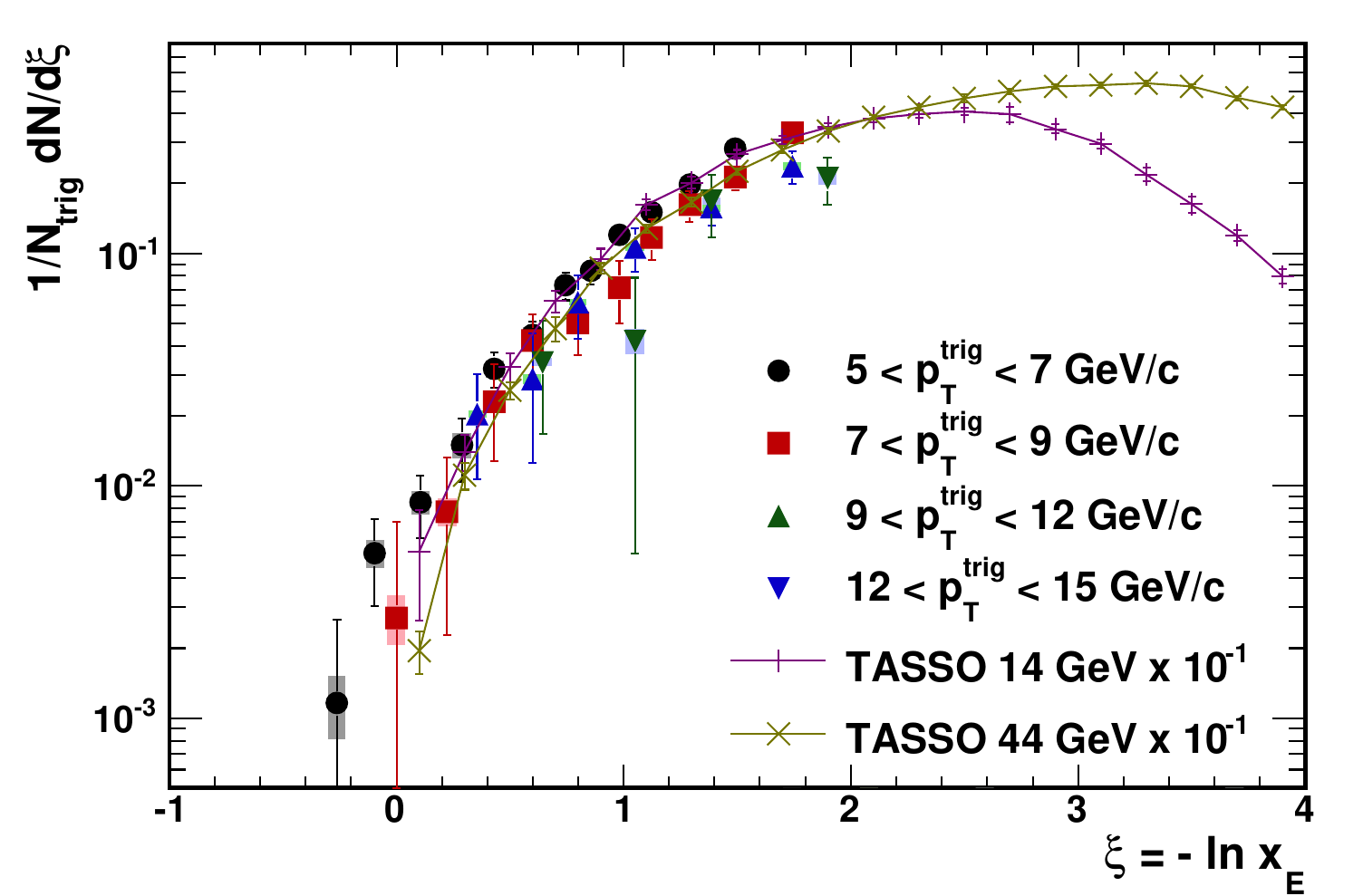}
\end{center}
\caption[]{a) Predicted $\xi$ distributions in vacuum and in medium for two jet energies~\cite{BW06}, together with measurements from $e^+ e^-$ collisions~\cite{TASSO,OPAL}   b) $\xi=\ln 1/x_E$ distributions for PHENIX isolated direct-$\gamma$ data for all $p_{T_t}$ ranges combined, compared to $e^+ e^-$ collisions at $\sqrt{s}=14$ and 44 GeV~\cite{TASSO}.}
\label{fig:BorgWied}
\end{figure}

   From the discussion above, the $x_E$ ($z_T$) distribution for direct-$\gamma$-h (Fig.~\ref{fig:pi0gamhxE}a) should be a  measurement of the $u$ quark fragmentation function. To make this more apparent, we follow the approach of
Borghini and Wiedemann (Fig.~\ref{fig:BorgWied}a)~\cite{BW06} who proposed using the hump-backed or $\xi=\ln(1/z)$ 
   distribution of jet fragments, which is a signature of QCD coherence~\cite{MLLA} for small values of  particle momentum fraction, $z=p/E_{\rm jet}$, to explore the medium-modification of jets in heavy ion collisions. The use of the $\xi$ variable would emphasize the increase in the emission of fragments at small $z$ due to the medium induced depletion of the number of fragments at large $z$. The jet energy must be known for this measurement so that it was presumed that full jet reconstruction would be required. 
However, the direct-$\gamma$-h correlation is (apart from the low rate) actually better for this measurement since both the energy and identity of the jet (8/1 $u$-quark, maybe 8/2 if the $\bar{q}+q\rightarrow \gamma+g$ channel is included) are known to high precision. The PHENIX $x_E$ distributions (Fig.~\ref{fig:pi0gamhxE}a) converted to the $\xi$ distributions are shown in Fig.~\ref{fig:BorgWied}b in quite excellent agreement with the dominant $u$-quark fragmentation functions measured in $e^+ e^-$ collisions at $\sqrt{s}/2=7$ and 22 GeV~\cite{TASSO}, which cover a comparable range in jet energy. Also, Fig.~\ref{fig:BorgWied}b is plotted on a log scale for $dN/d\xi$ which allows the full range of the fragmentation function, notably for $z>0.2$, $\xi<1.6$ to be visible.  
 
\subsection{``The golden channel'', direct-$\gamma$-h in Au+Au collisions}
\begin{figure}[h]
\begin{center}
\includegraphics[width=0.50\linewidth]{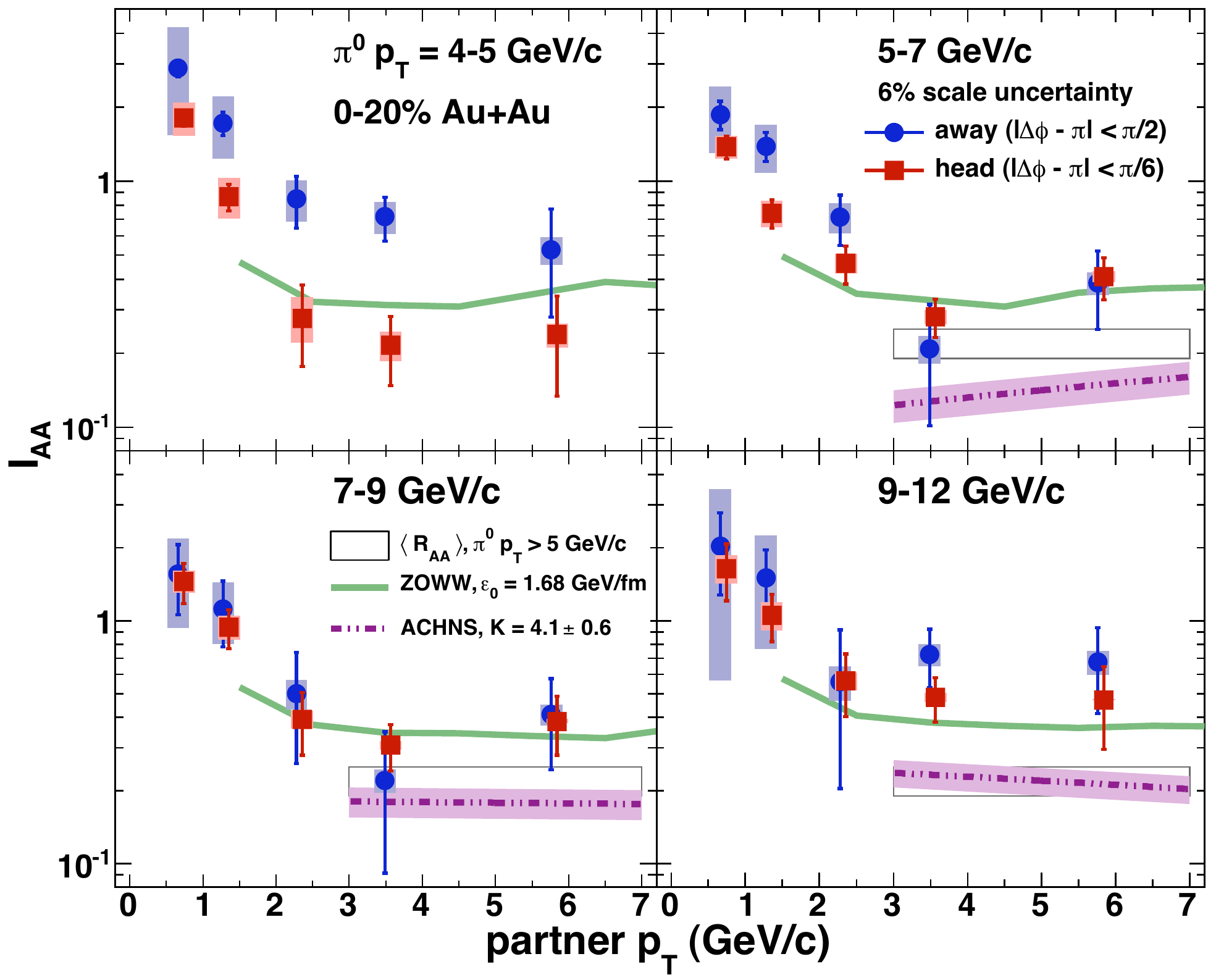}

\end{center}
\caption[]{a) PHENIX~\cite{ppg106} Away-side $I_{\rm AA}$ for a ``head-only'' (HO) $|\Delta\phi-\pi|<\pi/6$ selection (solid squares)
and the entire away-side (H+S), $|\Delta\phi-\pi|<\pi/2$ (solid circles) as a function of partner-h $p_T$ for various $p_{T_t}$. Calculations from two different predictions are shown for
    the head region in applicable \mbox{partner-h} $p_T$ ranges. $R_{AA}(p_T)$ of $\pi^0$ for $p_{T_t}>5$ GeV/c~\cite{ppg080} are included for comparison.  }
\label{fig:IAApi0h}
\end{figure}

Before discussing direct-$\gamma$-h correlations in Au+Au collisions, it is worth reviewing the results for $\pi^0$-h correlations, presented in Fig.~\ref{fig:IAApi0h}~\cite{ppg106} as $I_{AA}$, the ratio of the $p_{T}$ distributions of the partner-h in Au+Au central collisions to p-p collisions, for several values of trigger $\pi^0$ $p_{T_t}$ at $\sqrt{s_{NN}}=200$ GeV. 
 All the distributions show the characteristic shape of an exponentially falling $I_{AA}$ for lower values of $p_{T_a}$ due to energy loss of the away-jet and a flat distribution for larger values of $p_{T_a}$ due to partons that punch-through the medium with no energy loss and then fragment normally as in p-p collisions. Two other important points are evident from Fig.~\ref{fig:IAApi0h}: a) the H+S and HO data become the same for $p_{T_t}>7$ GeV/c because the away-jets in Au+Au are no longer anomalously wide and are consistent with the shape of the p-p azimuthal correlation in both the falling and flat part of the distribution; b) In the punch-thru region, $I_{AA}>R_{AA}$ at the same value of $p_{T}$.

The expectations for the difference in behavior of $\pi^0$-h and direct-$\gamma$-h correlations for the cases of either a totally absorbing medium (surface emission only) and a translucent lossy medium (partons can emerge from the medium having lost energy) are briefly summarized:\\
\par\noindent{\bf Totally absorbing medium:}
 \begin {description}
\item [direct-$\gamma$-h:] $\gamma$ emerge from all throughout the medium. Away partons are either totally absorbed or emitted from the surface with normal fragmentation, same surface bias as for inclusive hadrons (but on the away surface). $I_{AA}(p_{T_t})=$constant=fraction not absorbed= $R_{AA}^h(p_{T_t})$.
\item [$\pi^0$-h:] For inclusive hadrons, strong surface bias: only partons emitted from the surface are seen. For di-hadrons, only tangential emission is seen. $I_{AA}=$constant=the fraction of surface emission that is tangent to the surface. Therefore $I_{AA}< R_{AA}(h)$. 
\end{description}  

\noindent{\bf Translucent lossy medium:}
 \begin {description}
 \item [direct-$\gamma$-h:] Away partons lose energy. Some punch through or are tangent. $I_{AA}$ is exponentially decreasing at small $x_E$ due to partons with energy loss, constant at larger $x_E$ due to punch-thru.  
 \item [$\pi^0$-h:] Roughly the same except away partons may lose more energy due to surface bias of the trigger, and fewer may punch through.
\end{description}  
The $\pi^0$-h data in Fig.~\ref{fig:IAApi0h} show $I_{AA}>R_{AA}$ so disfavor the totally absorbing medium. The present direct-$\gamma$-h data are less conclusive. 

\begin{figure}[h]
\begin{center}
a)\includegraphics[width=0.49\linewidth]{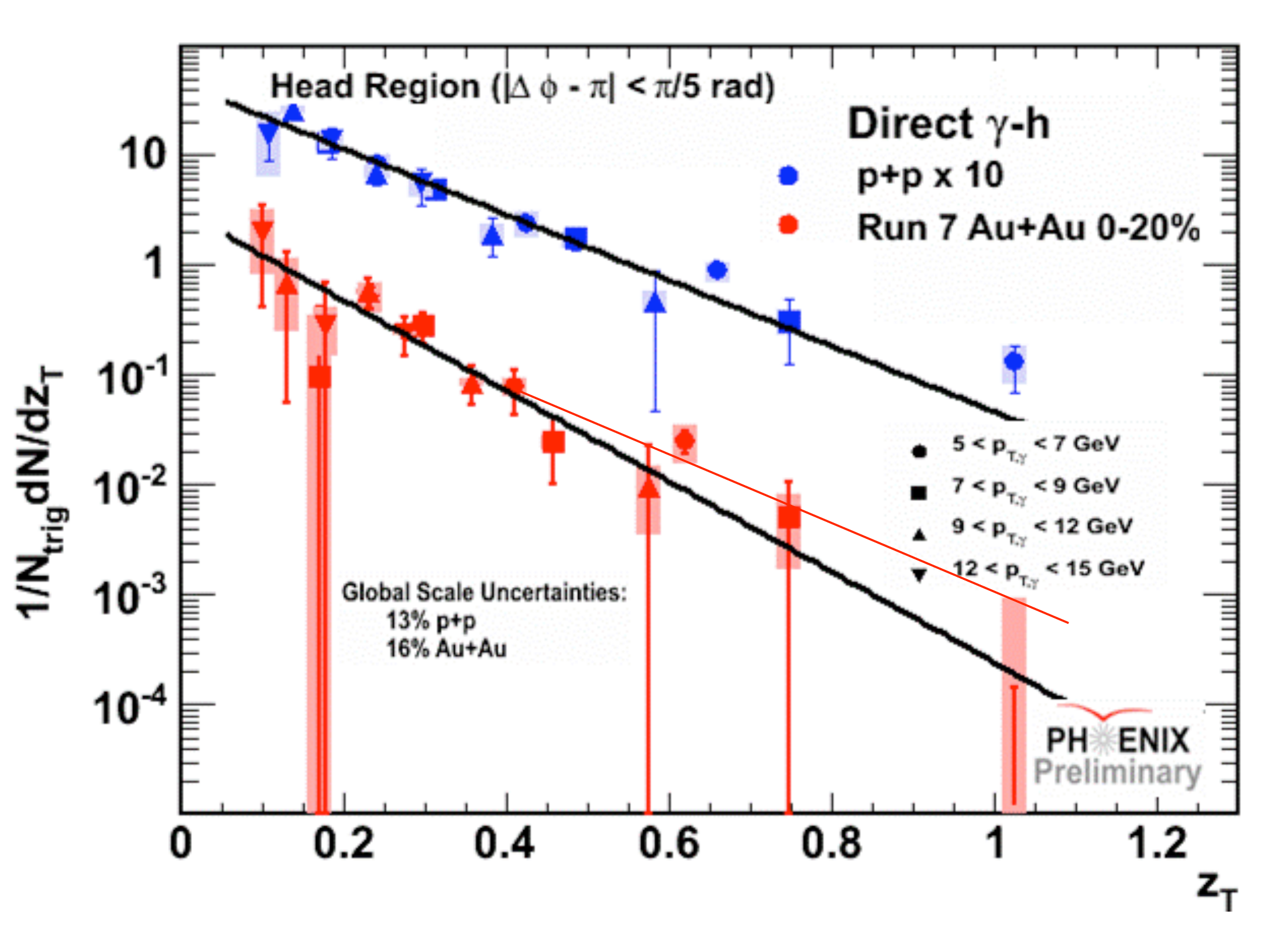}
b)\includegraphics[width=0.46\linewidth]{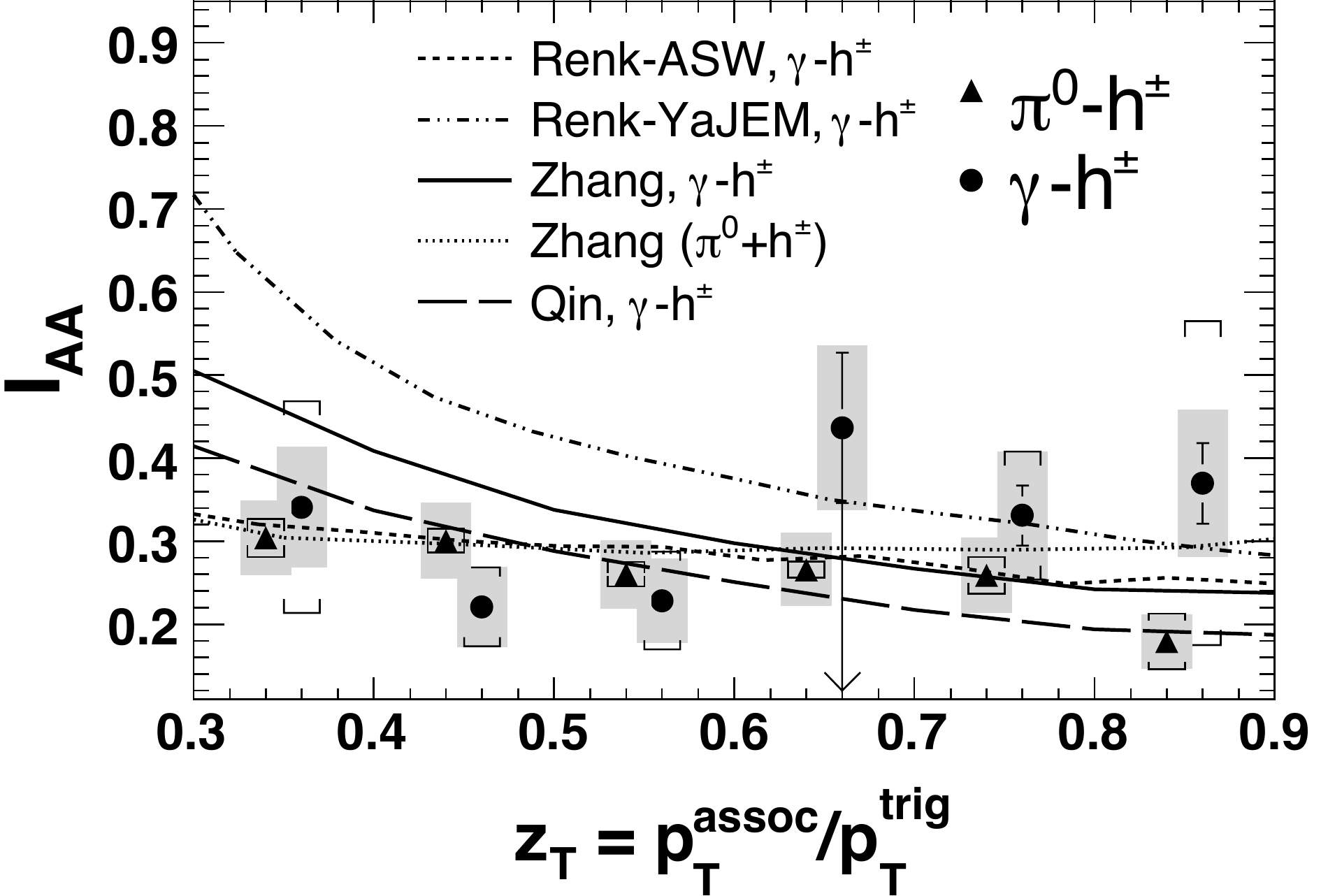}
\end{center}
\caption[]{a) PHENIX~\cite{MeganQM09} preliminary $z_T$ ($x_E$) distribution ($|\Delta\phi-\pi|<\pi/6$) for direct-$\gamma$-h, in Au+Au (red), and multiplied by a factor of 10 for p-p (blue), at $\sqrt{s_{NN}}=200$ GeV. b) STAR~\cite{Ahmed09}  away-side $I_{AA}$ vs. $z_T$ for $\pi^0$-h (triangles) and direct-$\gamma$-h (circles) with $8<p_{T_t}<16$ GeV/c at at $\sqrt{s_{NN}}=200$ GeV. For a discussion of the curves see Refs.~\cite{Ahmed09,Renk09}.}
\label{fig:IAAgam}
\end{figure}

Fig.~\ref{fig:IAAgam}a shows the PHENIX preliminary $x_E$ $(z_T)$ distributions for direct-$\gamma$-h correlations in p-p and Au+Au from QM2009~\cite{MeganQM09}. The p-p data show nice $x_E$ scaling for all $p_{T_t}$ (as would be expected for a fragmentation function) and are fit to an exponential, $e^{-b z_T}$, with slope $b=6.89\pm 0.64$. The Au+Au data also exhibit reasonable $x_E$ scaling for all $p_{T_t}$ with an exponential slope $b=9.49\pm 1.37$. The $x_E$ scaling and steeper slope in Au+Au suggest a constant fractional energy loss of the away-side parton; but the statistical significance is marginal. The thin red line shows that possibly $b\rightarrow 6.89$ for $z_T>0.4$ which would imply $I_{AA}$=constant for $x_E>0.4$ as for $\pi^0$-h. 
The STAR~\cite{Ahmed09} measurement of $\pi^0$-h and direct-$\gamma$-h $I_{AA}$ as a function of $z_T$ for $8<p_{T_t}<16$ GeV/c are shown in Fig.~~\ref{fig:IAAgam}b and show equal and constant $I_{AA}\simeq 0.3$ for $0.3\leq z_T\leq 0.9$ for both the $\pi^0$-h and direct-$\gamma$ correlations. There is no evidence in either case for an exponential rise of $I_{AA}$ towards $z_T=0$, possibly because the $z_T>0.3$ range is too high. Also the STAR value of $I_{AA}\simeq 0.3$ for $\pi^0$-h is smaller than the PHENIX value, $I_{AA}=0.50\pm 0.08$, in the same $p_{T_t}$ range.

 In summary, the results of measurements of the ``golden'' (but difficult) channel, direct-$\gamma$-h, are at too early a stage to tell whether this channel will live up to its promise to understand the {\QGP}, and {\QCD} in a medium; i.e. whether it will be a lasting paradigm, or go the way of the $J/\psi$.

\end{document}